\documentstyle[12pt]{article}
\input epsf.sty

\newcommand{\eqn}[1]{(\ref{#1})}
\newcommand{\ft}[2]{{\textstyle\frac{#1}{#2}}}
\newcommand{\tr}{\mbox{tr}}

\newcommand{\g}{{\sl g}}

\newcommand{\cD}{{\cal D}}
\newcommand{\cK}{{\cal K}}

\newcommand{\cO}{{\cal O}}

\newcommand{\cM}{{\cal M}}

\newcommand{\cN}{{\cal N}}
\newcommand{\cR}{{\cal R}}

\newcommand{\cL}{{\cal L}}

\newcommand{\mG}{{\mit\Gamma}}
\newcommand{\mP}{{\mit\Psi}}
\newcommand{\mS}{{\mit\Sigma}}
\newcommand{\phan}[1]{{\textstyle\phantom{#1}}}
\font\cmss=cmss12 
\def\1{\hbox{{1}\kern-.25em\hbox{l}}}
\def\bfZ{\relax{\hbox{\cmss Z\kern-.4em Z}}}
\begin{document}
\begin{titlepage}
\begin{center}
\hfill YITP-SB-00-08 \\
\vskip 30mm

{\Large\bf YANG-MILLS- AND D-INSTANTONS}

\vskip 10mm

{\bf A.V. Belitsky, S. Vandoren and P. van Nieuwenhuizen}

\vskip 8mm

\centerline{\it C.N.\ Yang Institute for Theoretical Physics}
\centerline{\it State University of New York at Stony Brook}
\centerline{\it NY 11794-3840, Stony Brook, USA}

\vskip 5mm

{\tt belitsky,vandoren,vannieu@insti.physics.sunysb.edu}

\vskip 8mm

\end{center}

\vskip .2in

\begin{center}
{\bf Abstract }
\end{center}
\begin{quotation}\noindent

In these lectures, which are written at an elementary and pedagogical
level, we discuss general aspects of (single) instantons in $SU(N)$
Yang-Mills theory, and then specialize to the case of $\cN = 4$
supersymmetry and the large $N$ limit. We show how to determine the
measure of collective coordinates and compute instanton corrections to
certain correlation functions. We then relate this to
D-instantons in type IIB supergravity. By taking the D-instantons to
live in an $AdS_5\times S^5$ background, we perform explicit checks
of the AdS/CFT correspondence.

\end{quotation}

\end{titlepage}

\tableofcontents

%%%%%%%%%%%%%%%%%%%%%%%%%%%%%%%%%%%%%%%%%%%%%%%%%%%%%%%%%%%%%%%%%%%
\section{Introduction.}
%%%%%%%%%%%%%%%%%%%%%%%%%%%%%%%%%%%%%%%%%%%%%%%%%%%%%%%%%%%%%%%%%%%

In the last decade we have seen an enormous progress in understanding
non-perturbative effects both in supersymmetric field theories and
superstring theories. When we talk about non-perturbative effects, we
usually mean solitons and instantons, whose masses and actions, respectively,
are inversely proportional to the coupling constant. Therefore, these
effects become important in the strongly coupled regime. The typical examples
of solitons are the kink and the magnetic monopole in field theory, and the
D-branes in supergravity or superstring theories. In the context of
supersymmetry, these solutions preserve one half of the supersymmetry and
are therefore BPS. As for instantons, we have the Yang-Mills (YM) instantons
\cite{Belavin}, and there are various kinds of instantons in string theory,
of which the D-instantons \cite{Dinst} are the most important for these
lectures.

In more general terms, without referring to supersymmetry, instantons are
solutions to the field equations in euclidean space with finite action, and
describe tunneling processes in Minkowski space-time from one vacuum at time
$t_1$ to another vacuum at time $t_2$. The
simplest model to consider is a quantum mechanical system with a double
well potential having two vacua. Classically there is no trajectory for
a particle to interpolate between the two vacua, but quantum mechanically
tunneling occurs. The tunneling amplitude can be computed in the WKB
approximation, and is typically exponentially suppressed. In the euclidean
picture, after performing a Wick rotation, the potential is turned upside down,
and it is possible for a particle to propagate between the two vacua, as
described by the classical solution to the equations of motion (see e.g.\
\cite{Rev}).

Also in YM theories, instantons are known to describe tunneling processes
between different vacua, labeled by an integer winding number, and lead
to the introduction of the CP-violating $\theta$-term \cite{CDG,JR}. It
was hoped that instantons could shed some light on the mechanism of quark
confinement. Although this was successfully shown in three-dimensional
gauge theories \cite{Pol}, the role of instantons in relation to confinement
in four dimensions is much more obscure. Together with the non-perturbative
chiral $U(1)$ anomaly in the instanton background, which led to baryon
number violation and the solution to the $U(1)$ problem \cite{Hooft1,Hooft2},
instantons have shown their relevance to phenomenological models like QCD
and the Standard Model. To avoid confusion, note that the triangle chiral
anomalies in perturbative field theories in Minkowski space-time are
canceled by choosing suitable multiplets of fermions. There are, however,
also chiral anomalies at the non-perturbative level. It is hard to compute
the non-perturbative terms in the effective action which lead to a breakdown
of the chiral symmetry by using methods in Minkowski space-time. However,
by using instantons in euclidean space, one can relatively easy determine
these terms. As we shall see, due to the presence of instantons there are
fermionic zero modes (and also bosonic zero modes) which appear in the path
integral measure. One must saturate these integrals and this leads to
correlation functions of composite operators with fermion fields which do
violate the chiral $U (1)$ symmetry. The new non-perturbative terms are
first computed in euclidean space, but then continued to Minkowski space
where they give rise to new physical effects \cite{Hooft2}. They have the
following form in the effective action
\begin{eqnarray}
\label{EffAct}
S_{\rm eff} \propto \exp
\left\{- \frac{8 \pi^2}{\g^2} \left( 1 + \cO (\g^2) \right) + i \theta
\right\}
\left( \lambda \lambda \right)^n  \ ,
\end{eqnarray}
where $n$ depends on the number of fermionic zero modes. The prefactor is
due to the classical instanton action and is clearly non-perturbative. The
terms indicated by $\cO (\g^2)$ are due to standard radiative corrections
computed by using Feynman graphs in an instanton background. The
term $\left( \lambda \lambda \right)^n$ involving the chiral spinor
$\lambda$ comes from saturating the integration in the path integral over
the fermionic collective coordinates and violates in general the chiral
symmetry. On top of \eqn{EffAct} we have to add the contributions from
anti-instantons, generating $\left( \bar\lambda\bar\lambda \right)^n$ terms
in the effective action. As we shall discuss, in euclidean space the chiral
and anti-chiral spinors are independent, but in Minkowski space-time they
are related by complex conjugation, and one needs the sum of instanton and
anti-instanton contributions to obtain a hermitean action.

In these lectures we will mainly concentrate on supersymmetric YM theories,
especially on the $\cN = 4$ $SU(N)$ SYM theory, and its large $N$ limit.
Instantons in $\cN = 1,2$ models have been extensively studied in the
past, and still are a topic of current research. For the $\cN = 1$ models,
one is mainly interested in the calculation of the superpotential and the
gluino condensate \cite{n=1}. In some specific models, instantons also
provide a mechanism for supersymmetry breaking \cite{susy-break}, see
\cite{S-V} for a recent review on these issues. In the case of $\cN = 2$,
there are exact results for the prepotential \cite{SW}, which acquires
contributions from all multi-instanton sectors. These predictions were
successfully tested in the one-instanton sector in \cite{test-SW}, and
for two-instantons in \cite{2inst-SW}.

Our interest in $\cN = 4$ SYM is twofold. On the one hand, since this
theory is believed to be S-dual \cite{M-O}, one expects that the complete
effective action, including all instanton and anti-instanton effects,
organizes itself into an $SL(2,{\bf Z})$ invariant expression. It would be
important to test this explicitly using standard field-theoretical
techniques. On the other hand, not unrelated to the previous, we are
motivated by the AdS/CFT correspondence
\cite{ads/cft}. In this picture, D-instantons in type IIB supergravity are
related to YM-instantons in the large $N$ limit \cite{BG,BGKR}. By making
use of the work by Green et al.\ on D-instantons \cite{GG}, definite
predictions come out for the large $N$ SYM theory, which were successfully
tested to leading order in the coupling constant, in the one-instanton
sector in \cite{DorKhoMatVan98}, and for multi-instantons in
\cite{DorHolKhoMatVan99}. Although these calculations are already
sufficiently complicated, it is nevertheless desirable to go beyond
leading order, such that one can obtain exact results for certain
correlation functions at the non-perturbative level. This will be the
guideline for our subsequent investigation.

The lectures are set up as follows. In section 2, we discuss the bosonic
YM instanton solution for $SU(N)$ and relate the counting of bosonic
collective coordinates to the index of the Dirac operator. Section 3
deals with fermionic collective coordinates, parametrizing the solutions
of the Dirac equation in the background of an instanton. We write down
explicit formulae for these solutions in the one-instanton sector and
elaborate further on the index of the Dirac operator. For
multi-instantons, one must use the ADHM construction \cite{ADHM}, which
is beyond the scope of these lectures. There are already comprehensive
reviews on this topic \cite{KMS,DorHolKhoMatVan99}. Section 4 gives a
treatment of the zero modes and the one-instanton measure on the moduli
space of collective coordinates. We explain in detail the normalization
of the zero modes since it is crucial for the construction of the
measure. In section 5 we discuss the one-loop determinants in the
background of an instanton, arising from integrating out the quantum
fluctuations. We then apply this to supersymmetric theories, and show
that the determinants for all supersymmetric YM theories cancel
each other \cite{dada-divecchia}.

Starting from section 6, we apply the formalism to $\cN = 4$ SYM theory.
We explicitly construct the euclidean action, and discuss in detail the
reality conditions on the bosonic and fermionic fields. In section 7, we
set up an iteration procedure in Grassmann collective coordinates to solve
the equations of motion. For the gauge group $SU(2)$ this iteration
amounts to applying successive ordinary and conformal supersymmetry
transformations on the fields. However, in the case of $SU(N)$ not all
solutions can be obtained by means of supersymmetry, and we solve
the equations of motions explicitly. Then, in section 8, we show how to
compute correlation functions, and discuss the large $N$ limit. Finally,
in section 9, we briefly discuss D-instantons in type IIB supergravity,
both in flat space and in an $AdS_5 \times S^5$ curved background,
and perform checks on the AdS/CFT correspondence.

After a short outlook we present a few appendices where we set up our
conventions and give a detailed derivation of some technical results
in order to make the paper self-contained.

%%%%%%%%%%%%%%%%%%%%%%%%%%%%%%%%%%%%%%%%%%%%%%%%%%%%%%%%%%%%%%%%%%%
\section{Classical euclidean solutions and collective coordinates.}
\setcounter{equation}{0}
%%%%%%%%%%%%%%%%%%%%%%%%%%%%%%%%%%%%%%%%%%%%%%%%%%%%%%%%%%%%%%%%%%%

%%%%%%%%%%%%%%%%%%%%%%%%%%%%%%%%%%%%%%%%%%%%%%%%%%%%%%%%%%%%%%%%%%%
\subsection{Generalities.}
%%%%%%%%%%%%%%%%%%%%%%%%%%%%%%%%%%%%%%%%%%%%%%%%%%%%%%%%%%%%%%%%%%%

We start with some elementary facts about instantons in $SU (N)$
Yang-Mills theories. The action, continued to euclidean space, is
\begin{equation}
\label{YMaction}
S = - \frac{1}{2 \g^2} \int\, d^4 x\, \tr\, F_{\mu\nu} F_{\mu\nu} \ .
\end{equation}
We have chosen traceless anti-hermitean $N$ by $N$ generators satisfying
$\left[ T_a, T_b \right] = f_{abc} T_c$ with real structure constants
and ${\rm tr} \left\{ T_a T_b \right\} = - \ft12 \delta_{ab}$. For
instance, for $SU(2)$, one has $T_a = - \ft{i}2 \tau_a$, where
$\tau_a$ are the Pauli matrices. Notice that the action is positive.
Further conventions are $\cD_\mu Y = \partial_\mu Y + [ A_\mu, Y ]$
for any Lie algebra valued field $Y$, and $F_{\mu\nu} = \partial _\mu
A_\nu - \partial_\nu A_\mu + [ A_\mu, A_\nu ]$, such that $F_{\mu\nu} =
[\cD_\mu, \cD_\nu]$. The euclidean metric is $\delta_{\mu\nu} =
{\rm diag} (+,+,+,+)$. In \eqn{YMaction}, the only appearance of the
coupling constant is in front of the action.

By definition, a Yang-Mills instanton is a solution to the euclidean
equations of motion with finite action. The equations of motion read
\begin{equation}
\label{eom}
\cD_\mu F_{\mu\nu} = 0 \ .
\end{equation}
To find solutions with finite action, we require that the field strength
tends to zero at infinity, hence the gauge fields asymptotically approaches
a pure gauge~\footnote{\label{Comp} Another way of satisfying the finite
action requirement is to first formulate the theory on a compactified
${\bf R}^4$, by adding and identifying points at infinity. Then the topology
is that of the four-sphere, since ${\bf R}^4 \cup \infty \simeq S^4$. The
stereographic map from ${\bf R}^4 \cup \infty$ to $S^4$ preserves the
angles, and is therefore conformal. Also the YM action is conformally
invariant, implying that the field equations on ${\bf R}^4 \cup \infty$
are the same as on $S^4$. The finiteness requirement is satisfied when
the gauge potentials can be smoothly extended from ${\bf R}^4$ to $S^4$.
The action is then finite because $S^4$ is compact and $A_\mu$ is smooth
on the whole of the four-sphere.}
\begin{equation}
\label{puregauge}
A_\mu \stackrel{x^2\rightarrow \infty}{=} U^{-1} \partial_\mu U \ ,
\end{equation}
for some $U\in SU(N)$.

There is actually a way of classifying fields which satisfy this boundary
condition. It is known from homotopy theory (the Pontryagin class) that
all gauge fields with vanishing field strength at infinity can be classified
into sectors characterized by an integer number
\begin{equation}
\label{winding}
k = - \frac{1}{16\pi^2} \int\, d^4 x\, \tr\, F_{\mu\nu}{^*\!F_{\mu\nu}} \ ,
\end{equation}
where ${^*\!F_{\mu\nu}} = \ft12 \epsilon_{\mu\nu\rho\sigma} F_{\rho\sigma}$
is the dual field strength, and $\epsilon_{1234} = 1$. The derivation of
this result can be found in Appendix \ref{Winding}. As a part of the proof,
one can show that the integrand in \eqn{winding} is the divergence of the
current
\begin{equation}
K_\mu = - \frac{1}{8\pi^2} \epsilon_{\mu\nu\rho\sigma}
\tr\, A_\nu \left( \partial_\rho A_\sigma + \ft23 A_\rho A_\sigma \right) \ .
\end{equation}
The four-dimensional integral in \eqn{winding} then reduces to an integral
over a three-sphere at infinity, and one can use \eqn{puregauge} to
show that the integer $k$ counts how many times this sphere covers the gauge
group three-sphere $S^3 \approx SU(2) \subset SU(N)$. In more mathematical
terms, the integer $k$ corresponds to the third homotopy group $\pi_3(SU(2))
= {\bf Z}$.

Since we require instantons to have finite action, they satisfy the
above boundary conditions at infinity, and hence they are classified by
an integer number $k$, called the instanton number or topological charge.
Gauge potentials leading to field strengths with different instanton number
can not be related by gauge transformations. This follows from the fact
that the instanton number is a gauge invariant quantity.

We now show that, in a given topological sector, there is a unique solution
to the field equations, the (anti-)instanton, that minimizes the action.
This is the field configuration which has (anti-)selfdual field strength
\begin{equation}
F_{\mu\nu} = \pm {^*\!F_{\mu\nu}}
= \pm \ft12 \epsilon_{\mu\nu\rho\sigma} F_{\rho\sigma} \ .
\end{equation}
This equation is understood in euclidean space, where $(^*)^2 = 1$. In
Minkowski space there are no solutions to the selfduality equations since
$(^*)^2 = - 1$. So, as seen from \eqn{winding}, instantons (with selfdual
field strength) have $k>0$ whereas anti-instantons (with anti-selfdual
field strength) have $k < 0$. To see that this configuration is
indeed the unique minimum of the action, we perform a trick similar to the
one used for deriving the BPS bound for solitons:
\begin{eqnarray}
S &=& - \frac{1}{2\g^2} \int\, d^4x \, \tr\, F^2
= - \frac{1}{4\g^2} \int\, d^4 x \, \tr\, (F \mp {^*\!F})^2
\mp \frac{1}{2\g^2} \int\, d^4x \, \tr\, F{^*F} \nonumber\\
&\geq& \mp \frac{1}{2\g^2} \int\, d^4x \, \tr\, F{^*\!F}
= \frac{8\pi^2}{\g^2} |k| \ ,
\end{eqnarray}
where the equality is satisfied if and only if the field strength is
(anti-)selfdual. The action is then $S_{\rm cl} = (8\pi^2/\g^2)|k|$, and
has the same value for the instanton as well as for the anti-instanton.
However, in euclidean space, we can also add a theta-angle term to the
action, which reads
\begin{equation}
S_\theta =- i \frac{\theta}{16\pi^2} \int\,
d^4x \, \tr \, F_{\mu\nu}{^*\!F}^{\mu\nu} = \pm i\theta |k| \ .
\end{equation}
The plus or minus sign corresponds to the instanton and anti-instanton
respectively, so the theta angle distinguishes between them.

It is worth mentioning that the energy-momentum tensor for a selfdual
field strength is always zero
\begin{equation}
{\mit\Theta}_{\mu\nu} = \frac{2}{\g^2}
\tr \left\{ F_{\mu\rho}F_{\nu\rho} - \ft14 \delta_{\mu\nu}
F_{\rho\sigma} F_{\rho\sigma} \right\} = 0 \ .
\end{equation}
This follows from the observation that the instanton action
$\int d^4 x\, F^2 = \int d^4 x\, {^\ast\! F}F$ is metric independent.
The vanishing of the energy-momentum tensor is consistent with the fact
that instantons are topological in nature. It also implies that instantons
do not curve euclidean space, as follows from the Einstein equations.

Note that we have not shown that all the solutions of \eqn{eom} with
finite action are given by instantons, i.e.\ by selfdual field strengths.
In principle there could be configurations which are local minima of the
action, but are neither selfdual nor anti-selfdual. No such examples of
exact solutions have been found in the literature so far.

%%%%%%%%%%%%%%%%%%%%%%%%%%%%%%%%%%%%%%%%%%%%%%%%%%%%%%%%%%%%%%%%%%%
\subsection{The $k = 1$ instanton in $SU(2)$.}
%%%%%%%%%%%%%%%%%%%%%%%%%%%%%%%%%%%%%%%%%%%%%%%%%%%%%%%%%%%%%%%%%%%

An explicit construction of finite action solutions of the euclidean
classical equations of motion was given by Belavin et al. \cite{Belavin}.
The gauge configuration for one-instanton ($k = 1$) in $SU(2)$ is
\begin{equation}
\label{1inst}
A_\mu^a (x; x_0,\rho)
= 2 \frac{\eta^a_{\mu\nu} (x - x_0)_\nu}{( x - x_0 )^2 + \rho^2}\ ,
\end{equation}
where $x_0$ and $\rho$ are arbitrary parameters called collective
coordinates. They correspond to the position and the size of the instanton.
The above expression solves the selfduality equations for any value of the
collective coordinates. Notice that it is regular for $x = x_0$, as long
as $\rho \neq 0$. The antisymmetric eta-symbols are defined as (see Appendix
\ref{HooftSpinor} for more of their properties)
\begin{eqnarray}
\eta_{\mu\nu}^a = \epsilon^a_{\mu\nu}
\qquad &\mu,\nu = 1, 2, 3 \, , & \qquad
\eta^a_{\mu 4} = - \eta^a_{4\mu} = \delta^a_\mu\ ,\nonumber\\
{\bar \eta}_{\mu\nu}^a = \epsilon^a_{\mu\nu}
\qquad &\mu,\nu = 1, 2, 3 \, , & \qquad
{\bar \eta}^a_{\mu 4} = - {\bar \eta}^a_{4\mu} = -\delta^a_\mu\ .
\end{eqnarray}
The $\eta$ and $\bar\eta$-tensors are selfdual
and anti-selfdual respectively, for each index $a$. They form a basis
for the antisymmetric four by four matrices, and we have listed their
properties in Appendix \ref{HooftSpinor}. The gauge transformation
corresponding to \eqn{puregauge} is simply $U (x) = \sigma_\mu
x_\mu/{\sqrt {x^2}}$, where the sigma matrices are given by
$\sigma_\mu = ({\vec \tau},i)$.

The field strength corresponding to this gauge potential is
\begin{equation}
F^a_{\mu\nu}
= - 4 \eta^a_{\mu\nu} \frac{\rho^2}{[( x - x_0 )^2 + \rho^2]^2} \ ,
\end{equation}
and it is selfdual. Notice that the special point $\rho = 0$, called
zero size instantons, leads to zero field strength and corresponds to pure
gauge. This point must be excluded from the instanton moduli space of
collective coordinates, since it is singular. Finally one can compute the
action
by integrating the density
\begin{equation}
\label{inst-dens}
\tr\, F_{\mu\nu}F^{\mu\nu} = - 96\,
\frac{\rho^4}{[( x - x_0 )^2 + \rho^2]^4} \ .
\end{equation}
Using the integral given at the end of Appendix \ref{HooftSpinor}, one finds
that this indeed corresponds to $k = 1$.

One can also consider the instanton in singular gauge, for which
\begin{equation}
\label{inst-sing}
A^a_\mu = 2\, \frac{\rho^2{\bar \eta}^a_{\mu\nu} (x - x_0)_\nu}
{( x - x_0 )^2 [( x - x_0 )^2 + \rho^2]}
= - \bar\eta^a_{\mu\nu} \partial_\nu\, \ln
\left\{ 1 + \frac{\rho^2}{( x - x_0 )^2} \right\} \ .
\end{equation}
This gauge potential is singular for $x = x_0$, where it approaches
a pure gauge
configuration $A_\mu \stackrel{x\rightarrow x_0}{=}
U \partial_\mu U^{- 1}$ with
$U(x - x_0)$ given before. Moreover this gauge group transformation relates
the regular gauge instanton \eqn{1inst} to the singular one \eqn{inst-sing}
at all points. The field strength in singular gauge is then (taking the
instanton position zero, $x_0 = 0$, otherwise replace $x \rightarrow x -
x_0$)
\begin{equation}
F^a_{\mu\nu} =
- \frac{4\rho^2}{( x^2 + \rho^2)^2}
\left\{ \bar\eta^a_{\mu\nu}
- 2 \bar\eta^a_{\mu\rho} \frac{x_\rho x_\nu}{x^2}
+ 2 \bar\eta^a_{\nu\rho} \frac{x_\rho x_\mu}{x^2}
\right\} \ .
\end{equation}
Notice that, despite the anti-selfdual eta-tensors, the field strength is
still selfdual, as can be seen by using the properties of the eta-tensors
given in Appendix \ref{HooftSpinor}. Singular gauge is frequently used,
because, as we
will see later, the instanton measure can be computed most easily in this
gauge. One can compute the winding number again in singular gauge. Then one
finds that there is no contribution coming from infinity. Instead, all the
winding is coming from the singularity at the origin.

At first sight it seems there are five collective coordinates. There are
however extra collective coordinates corresponding to the gauge
orientation. In fact, one can act with an $SU (2)$ matrix on the solution
\eqn{1inst} to obtain another solution,
\begin{equation}
\label{1inst-angles}
A_\mu \left( x; x_0, \rho, \vec\theta \right)
= U \left( \vec\theta \right) \, A_\mu(x; x_0,\rho)\,
U^\dagger \left( \vec\theta \right) \ ,
\qquad U \in SU(2) \ .
\end{equation}
One might think that, since this configuration is gauge equivalent to the
expression given above, it should not be considered as a new solution.
This is not true however, the reason is that, after we fix the gauge, we
still have left a rigid $SU(2)$ symmetry which acts as in \eqn{1inst-angles}.
So in total there are eight collective coordinates, also called moduli.

In principle, one could also act with the (space-time) rotation matrices
$SO(4)$ on the instanton solution, and construct new solutions. However,
as was shown by Jackiw and Rebbi \cite{JackReb}, these rotations can be
undone by suitably chosen gauge transformations. If one puts together the
instanton and anti-instanton in a four by four matrix, the gauge group
$SU(2)$ can be extended to $SO(4) = (SU(2)\times SU(2))/ Z_2$, which
is the same
as the euclidean rotation group. A similar analysis holds for the other
generators of the conformal group. In fact, Jackiw and Rebbi showed that
for the (euclidean) conformal group $SO(5,1)$, the subgroup $SO(5)$
consisting of rotations and combined special conformal transformation with
translations
($R^\mu \equiv K^\mu + P^\mu$), leaves the instanton invariant, up to
gauge transformations. This leads to a 5 parameter instanton moduli
space $SO(5,1)/SO(5)$, which is the euclidean version of the
five-dimensional anti-de Sitter space $AdS_5$. The coordinates on this
manifold correspond to the four positions and the size $\rho$ of the
instanton. On top of that, there are still three gauge orientation
collective coordinates, yielding a total of eight moduli.

%%%%%%%%%%%%%%%%%%%%%%%%%%%%%%%%%%%%%%%%%%%%%%%%%%%%%%%%%%%%%%%%%%%
\subsection{The $k = 1$ instanton in $SU(N)$.}
%%%%%%%%%%%%%%%%%%%%%%%%%%%%%%%%%%%%%%%%%%%%%%%%%%%%%%%%%%%%%%%%%%%

Instantons in $SU(N)$ can be obtained by embedding $SU(2)$ instantons
into $SU(N)$. For instance, a particular embedding is given by the
following $N$ by $N$ matrix
\begin{equation}
\label{1instSUN}
A_\mu^{SU(N)} =
\left(
\begin{array}{cc}
0 & 0             \\
0 & A_\mu^{SU(2)}
\end{array}
\right) \ .
\end{equation}
Of course this is not the most general solution, as one can choose
different embeddings. One could act with a general $SU(N)$ element on
the solution \eqn{1instSUN} and obtain a new one. Some of them
correspond to a different embedding~\footnote{There are
also embeddings which can not be obtained by $SU(N)$ or any other
similarity transformations. They are completely inequivalent, but
correspond to higher instanton numbers $k$ \cite{Wilczek}. Since we
do not cover multi-instantons in these lectures, these embeddings are
left out of the discussion here.} inside $SU(N)$. Not all elements of
$SU(N)$ generate a new solution. There is a stability group that
leaves \eqn{1instSUN} invariant, acting only on the zeros, or commuting
trivially with the $SU(2)$ embedding. Such group elements should be
divided out, so we consider, for $N > 2$,
\begin{equation}
A_\mu^{SU(N)}
= U\,
\left(
\begin{array}{cc}
0 & 0             \\
0 & A_\mu^{SU(2)}
\end{array}
\right)
\, U^\dagger, \qquad
U \in \frac{SU(N)}{SU(N - 2) \times U(1)} \ .
\label{coset}
\end{equation}
One can now count the number of collective coordinates. From counting
the dimension of the coset space in \eqn{coset}, one finds there are
$4 N - 5$ angles. Together with the position and the scale of the
$SU(2)$ solution, we find in total $4N$ collective coordinates.

It is instructive to work out the example of $SU(3)$. Here we use the
eight Gell-Mann matrices $\{\lambda_\alpha\}, \alpha = 1,\dots,8$. The
first three $\lambda_a, a = 1, 2, 3$, form an $SU(2)$ algebra and are
used to define the $k = 1$ instanton by contracting \eqn{1inst} with
$\lambda_a$. The generators $\lambda_4,\dots,\lambda_7$ form two doublets
under this $SU(2)$, and can be used to generate new solutions. Then
there is $\lambda_8$, which is a singlet, corresponding to the $U(1)$
factor in \eqn{coset}. It commutes with the $SU(2)$ subgroup spanned
by $\lambda_a$, and so it belongs to the stability group leaving the
instanton invariant. For $SU(3)$ there are seven gauge orientation
zero modes.

The question then arises whether or not these are all the solutions. To
find this out, one can study deformations of the solution \eqn{1instSUN},
$A_\mu + \delta A_\mu$, and
see if they preserve selfduality.
 Expanding to first
order in the deformation, this leads to the condition
\begin{equation}
\label{SDdef}
\cD_\mu \delta A_\nu - \cD_\nu \delta A_\mu
= {^*\!\left( \cD_\mu \delta A_\nu - \cD_\nu \delta A_\mu \right)} \ ,
\end{equation}
where the covariant derivative depends only on the original classical
solution.

In addition we require that the new solution is not related to the old
one by a gauge transformation. This can be achieved by requiring that
the deformations are orthogonal to any gauge transformation $\cD_\mu
\Lambda$, for any function $\Lambda$, i.e.
\begin{equation}
\int\,\, d^4x \, \tr\, \cD_\mu \Lambda \delta A_\mu = 0 \ .
\end{equation}
After partial integration the orthogonality requirement leads to the
usual background field gauge condition
\begin{equation}
\label{backgr-gauge}
\cD_\mu \delta A^\mu = 0 \ .
\end{equation}
Bernard et al.\ in \cite{Bern-et-al} have studied the solutions of
\eqn{SDdef} subject to the condition \eqn{backgr-gauge} using the
Atiyah-Singer index theorem. Index theory turns out to be a useful
tool when counting the number of solutions to a certain linear
differential equation of the form $\hat D T = 0$, where $\hat D$ is
some differential operator and $T$ is a tensor. We will elaborate
on this in the next subsection and also when studying fermionic
collective coordinates. The ultimate result of \cite{Bern-et-al} is
that there are indeed $4Nk$ solutions, leading to the above constructed
$4N$ (for $k=1$) collective coordinates. An assumption required to
apply index theorems is that the space has to be compact. One must therefore
compactify euclidean space to a four-sphere $S^4$, as was also mentioned in
footnote \ref{Comp}.

%%%%%%%%%%%%%%%%%%%%%%%%%%%%%%%%%%%%%%%%%%%%%%%%%%%%%%%%%%%%%%%%%%%
\subsection{Bosonic collective coordinates and the Dirac operator.}
%%%%%%%%%%%%%%%%%%%%%%%%%%%%%%%%%%%%%%%%%%%%%%%%%%%%%%%%%%%%%%%%%%%

In this section we will make more precise statements on how to count
the number of solutions to the selfduality equations, by relating it
to the index of the Dirac operator. A good reference on this topic
is \cite{Brown}.

The problem is to study the number of solutions to the (anti-)selfduality
equations with topological charge $k$. As explained in the last subsection,
we study deformations of a given classical solution $A_\mu^{\rm cl} +
\delta A_{\mu}$. Let us denote $\phi_\mu \equiv \delta A_{\mu}$ and
$f_{\mu\nu} \equiv \cD_{\mu} \phi_\nu - \cD_\nu \phi_\mu$. The covariant
derivative here contains only $A_\mu^{\rm cl}$. The constraints on the
deformations of an anti-instanton~\footnote{\label{Convent}{\bf Note on
conventions}: We are switching here, and in the remainder, from
instantons to anti-instantons. The reason has to do with the conventions
for the $\sigma_\mu$ and ${\bar \sigma}_\mu$ matrices as defined in the
text (see also Appendix \ref{HooftSpinor}).
Our conventions are different from  most of the instanton literature, but
are in agreement with the literature on supersymmetry. Due to this, we
obtain the somewhat unfortunate result that $\sigma_{\mu\nu}$ is
anti-selfdual, while $\eta^a_{\mu\nu}$ is selfdual. For this reason,
we will choose to study anti-instantons.} can then be written as
\begin{equation}
{\eta}^a_{\mu\nu} f_{\mu\nu} = 0 \ , \qquad \cD_\mu \phi_\mu = 0 \ .
\end{equation}
The first of this equation says that the selfdual part of $f_{\mu\nu}$
must vanish. It also means that the deformation cannot change the
anti-instanton into an instanton. It will prove convenient to use
quaternionic notation
\begin{equation}
\Phi^{\alpha\beta'} = \phi_\mu \sigma_\mu^{\alpha\beta'} \ ,
\end{equation}
with $\sigma_\mu = (\vec{\tau}, i)$, and $\bar\sigma_\mu = (\vec\tau, -i)$.
These sigma matrices satisfy $\sigma_\mu \bar\sigma_\nu + \sigma_\nu
\bar\sigma_\mu = 2\delta_{\mu\nu}$ and
\begin{equation}
\label{sigmamunu}
\sigma_{\mu\nu} \equiv \ft12
\left( \sigma_\mu \bar\sigma_\nu
- \sigma_\nu \bar\sigma_\mu \right)
= i \bar\eta^a_{\mu\nu} \tau^a, \qquad
\bar\sigma_{\mu\nu} \equiv \ft12
\left( \bar\sigma_\mu \sigma_\nu
- \bar\sigma_\nu \sigma_\mu \right)
= i \eta^a_{\mu\nu} \tau^a\ ,
\end{equation}
so $\sigma_{\mu\nu}$ and $\bar\sigma_{\mu\nu}$ are anti-selfdual and
selfdual respectively. The constraints on the deformations can then be
rewritten as a quaternion valued Dirac equation
\begin{equation}
\label{Dirac-quat}
\not\!\!{\bar \cD} \Phi = 0 \ ,
\end{equation}
with $\not\!\!\bar\cD = {\bar \sigma}^\mu \cD_\mu$. We can represent the
quaternion by
\begin{equation}
\Phi =
\left( \begin{array}{cc} a & -b^* \\ b & a^* \end{array} \right) \ ,
\end{equation}
with $a$ and $b$ complex adjoint-valued functions. Then
\eqn{Dirac-quat} reduces to two adjoint spinor equations, one for
\begin{equation}
\lambda = \left( {a \atop b} \right)
\qquad
\not\!\!\bar\cD {\lambda} = 0 \ ,
\label{Dirac-spinor}
\end{equation}
and one for $- i \sigma^2 \lambda^*$. Conversely, for each spinor solution
$\lambda$ to the Dirac equation, one shows that also $-i\sigma^2\lambda^*$
is a solution. Therefore, the number of solutions
for $\Phi$ is twice the number of solutions for a single two-component
adjoint spinor. So, the problem of counting the number of bosonic
collective coordinates is now translated to the computation of the Dirac
index, which we will discuss in the next section.

%%%%%%%%%%%%%%%%%%%%%%%%%%%%%%%%%%%%%%%%%%%%%%%%%%%%%%%%%%%%%%%%%%%
\section{Fermionic collective coordinates and the index theorem.}
\setcounter{equation}{0}
%%%%%%%%%%%%%%%%%%%%%%%%%%%%%%%%%%%%%%%%%%%%%%%%%%%%%%%%%%%%%%%%%%%

Both motivated by the counting of bosonic collective coordinates, as
argued in the last section, and by the interest of coupling YM theory
to fermions, we study the Dirac equation in the presence of an
anti-instanton. We start with a Dirac fermion $\psi$, in an arbitrary
representation (adjoint, fundamental, etc) of the gauge group, and
consider the Dirac equation in the presence of an anti-instanton background
\begin{equation}
\gamma_\mu \cD_\mu^{\rm cl} \psi \, = \, \not\!\!\cD^{\rm cl} \psi = 0 \ .
\end{equation}
The Dirac spinor can be decomposed into its chiral and anti-chiral
components
\begin{equation}
\lambda \equiv \ft12 \left( 1 + \gamma^5 \right) \psi \ ,
\qquad {\bar \chi} \equiv \ft12 \left( 1 - \gamma^5 \right) \psi \ .
\end{equation}
A euclidean representation for the Clifford algebra is given by~\footnote{In
euclidean space the Lorentz group decomposes according to $SO(4) = SU(2)
\times SU(2)$. The spinor indices $\alpha$ and $\alpha'$ correspond to the
doublet representations of these two $SU(2)$ factors. As opposed to
Minkowski space, $\alpha$ and $\alpha'$ are not in conjugate
representations.}
\begin{equation}
\label{gamma-sigma}
\gamma^\mu =
\left(
\begin{array}{cc}
0                               & -i \sigma^{\mu\, \alpha \beta'} \\
i \bar\sigma^\mu_{\alpha'\beta} & 0
\end{array}
\right) \ ,
\qquad
\gamma^5 = \gamma^1\gamma^2\gamma^3\gamma^4
= \left( \begin{array}{cc} 1 & 0 \\ 0 & - 1 \end{array} \right) \ .
\end{equation}
The Dirac equation then becomes
\begin{equation}
\label{DiracEq}
\not\!\!\bar\cD^{\rm cl} \lambda = 0 \ ,
\qquad \not\!\!\cD^{\rm cl} \bar\chi = 0 \ ,
\end{equation}
where $\not\!\!\cD$ is a two by two matrix.

%%%%%%%%%%%%%%%%%%%%%%%%%%%%%%%%%%%%%%%%%%%%%%%%%%%%%%%%%%%%%%%%%%%
\subsection{The index of the Dirac operator.}
%%%%%%%%%%%%%%%%%%%%%%%%%%%%%%%%%%%%%%%%%%%%%%%%%%%%%%%%%%%%%%%%%%%

We now show that in the presence of an anti-instanton (recall the
footnote \ref{Convent}), \eqn{DiracEq} has solutions for $\lambda$, but
not for $\bar\chi$. Conversely, in the background of an instanton,
${\not\!\!\cD}$ has zero modes, but ${\not\!\!\bar\cD}$ has not. The
argument goes as follows.
Given a zero mode $\bar\chi$ for ${\not\!\!\cD}$, it also satisfies
${\not\!\!\bar\cD \not\!\!\cD \bar\chi = 0}$. In other words, the kernel
${\rm ker} \not\!\!\cD \subset {\rm ker} \left\{ \not\!\!\bar\cD
\not\!\!\cD \right\}$. Now we evaluate
\begin{equation}
\not\!\!\bar\cD \not\!\!\cD
= \bar\sigma_\mu \sigma_\nu \cD_\mu \cD_\nu
= \cD^2 + \bar\sigma_{\mu\nu} F_{\mu\nu} \ ,
\end{equation}
where we have used $\bar\sigma_\mu \sigma_\nu + \bar\sigma_\nu \sigma_\mu
= 2 \delta_{\mu\nu}$, and $\bar\sigma_{\mu\nu}$ is defined as in
\eqn{sigmamunu}. But notice that the anti-instanton field strength is
anti-selfdual whereas the tensor $\bar\sigma_{\mu\nu}$ is selfdual, so the
second term vanishes. From this it follows that $\bar\chi$ satisfies
$\cD^2 \bar\chi = 0$. Now we can multiply with its conjugate $\bar\chi^*$
and integrate to get, after partial integration and assuming that the
fields go to zero at infinity, $\int\,d^4x \left| \cD_\mu \bar\chi
\right|^2 = 0$. From this it follows that $\bar\chi$ is covariantly
constant, and so $ F_{\mu\nu}^{a,{\rm cl}}T_a \bar\chi = 0$. This implies
that $\eta^a_{\mu\nu}T_a{\bar \chi}=0$, and hence $T_a{\bar \chi}=0$ for
all $T_a$. We conclude that $\bar\chi = 0$. Stated differently, $- \cD^2$
is a positive definite operator and has no zero modes (with vanishing
boundary conditions). This result is independent of the representation of
the fermion.

For the $\lambda$-equation, we have $\not\!\!\cD \not\!\!\bar\cD
\lambda = 0$, i.e. ${\rm ker} \not\!\!\bar\cD \subset {\rm ker}
\left\{ \not\!\!\cD \not\!\!\bar\cD \right\}$, and compute
\begin{equation}
\label{DbarD}
\not\!\!\cD \not\!\!\bar\cD
= \cD^2 + \sigma_{\mu\nu} F_{\mu\nu} \ .
\end{equation}
This time the second term does not vanish in the presence of an
anti-instanton, so zero modes are possible. Knowing that ${\not\!\!\cD}$
has no zero modes, one easily concludes that ${\rm ker} {\not\!\!\bar\cD}
= {\rm ker} \left\{ \not\!\!\cD \not\!\!\bar\cD \right\}$. Now we can
count the number of solutions using index theorems. The index of the
Dirac operator is defined as
\begin{equation}
{\rm Ind} \not\!\!\bar\cD
= {\rm dim} \left\{ {\rm ker} \not\!\!\bar\cD \right\}
- {\rm dim} \left\{ {\rm ker} \not\!\!\cD  \right\} \ .
\end{equation}
This index will give us the relevant number, since the second term is
zero. There are several ways to compute its value, and we represent it
by
\begin{equation}
\label{Ind-M}
{\rm Ind} \not\!\!\bar\cD
= \tr\, \left\{ \frac{M^2}{-\not\!\!\cD \not\!\!\bar\cD + M^2}
- \frac{M^2}{ -\not\!\!\bar\cD \not\!\!\cD + M^2} \right\} \ ,
\end{equation}
where $M$ is an arbitrary parameter. The trace stands for a sum over
group indices, spinor indices, and includes an integration over space-time.
We can in fact show that this expression is independent of $M$. The
reason is that the operators ${\not\!\!\cD \not\!\!\bar\cD}$ and
${\not\!\!\bar\cD \not\!\!\cD}$ have the same spectrum for non-zero
eigenvalues. Indeed, if $\psi$ is an eigenfunction of
${\not\!\!\bar\cD \not\!\!\cD}$, then ${\not\!\!\cD} \psi$ is an
eigenfunction of ${\not\!\!\cD \not\!\!\bar \cD}$ with the same eigenvalue
and ${\not\!\!\cD}\psi$ does not vanish. Conversely, if $\psi$ is an
eigenfunction of ${\not\!\!\cD \not\!\!\bar \cD}$, then ${\not\!\!\bar\cD}
\psi$ does not vanish and is an eigenfunction of ${\not\!\!\bar\cD
\not\!\!\cD}$ with the same eigenvalue. This means that there is a pairwise
cancellation in \eqn{Ind-M} coming from the sum over eigenstates with
non-zero eigenvalues. So the only contribution is coming from the zero
modes, for which the first term simply gives one for each zero mode, and
the second term vanishes because there are no zero modes. The result is
then clearly independent of $M$, and moreover, it is an integer, namely
${\rm dim} \left\{ {\rm ker} \not\!\!\bar\cD \right\}$.

In the basis of the four-dimensional Dirac matrices, the index can be
written as
\begin{equation}
{\rm Ind} \not\!\!\bar\cD
= \tr\, \left\{ \frac{M^2}{- \not\!\!\cD^2 + M^2 }\,\gamma_5 \right\} \ .
\end{equation}
Because this expression is independent of $M^2$, we might as well evaluate
it in the large $M^2$ limit. The calculation is then identical to the
calculation of the chiral anomaly, and we will not repeat it here. The
results are well known, and depend on the representation of the generators,
\begin{equation}
{\rm Ind} \not\!\!\bar\cD
= -\frac{1}{16\pi^2}\int\,d^4x\,F^a_{\mu\nu}{}^*F^b_{\mu\nu}\tr\,(T_aT_b) \ ,
\end{equation}
which yields
\begin{eqnarray}
{\rm Ind}_{\rm adj} \not\!\!\bar\cD &=& 2 N k, \qquad \mbox{adjoint} \ ,
\nonumber\\
{\rm Ind}_{\rm fund} \not\!\!\bar\cD &=& k, \qquad \mbox{fundamental} \ .
\end{eqnarray}
This also proves the fact that there are $4 N k$ bosonic collective
coordinates, as mentioned in the last subsection.

%%%%%%%%%%%%%%%%%%%%%%%%%%%%%%%%%%%%%%%%%%%%%%%%%%%%%%%%%%%%%%%%%%%
\subsection{Construction of the fermionic instanton.}
%%%%%%%%%%%%%%%%%%%%%%%%%%%%%%%%%%%%%%%%%%%%%%%%%%%%%%%%%%%%%%%%%%%

In this subsection we will construct the solutions to the Dirac equation
explicitly. Because we only know the gauge field for $k=1$ explicitly,
we can only construct the fermionic zero modes for the single
anti-instanton case. For an $SU(2)$ adjoint fermion, there are 4 zero
modes, and these can be written as \cite{Shi83}
\begin{equation}
\label{xi-eta}
\lambda^\alpha = -\ft12 \sigma_{\mu\nu\ \, \beta}^{\phan{ii}\alpha}
\left(
\xi^\beta
- \bar\eta_{\gamma'} \bar\sigma_\rho^{\gamma' \beta} (x - x_0)^\rho
\right)
F_{\mu\nu} \ .
\end{equation}
Actually, this expression also solves the Dirac equation for higher
order $k$, but there are additional solutions, $4k$ in total for $SU(2)$.
The four fermionic collective coordinates are denoted by $\xi^\alpha$
and $\bar\eta_{\gamma'}$, where $\alpha, \gamma' = 1,2$ are spinor
indices in euclidean space. They can somehow be thought of as the
fermionic partners of the translational and dilatational collective
coordinates in the bosonic sector. These solutions take the same form
in any gauge, one just takes the corresponding gauge for the field
strength. For $SU(N)$, there are a remaining of $2 \times (N - 2)$ zero
modes, and their explicit form depends on the chosen gauge. In regular
gauge, with color indices $u,v = 1,\dots,N$ explicitly written, the
gauge field is (setting $x_0=0$, otherwise replace $x \rightarrow x -
x_0$)
\begin{equation}
{{A_\mu}}^u{}_v = A_\mu^a \left( T_a \right)^u{}_v
= - \frac{ \sigma_{\mu\nu\ \, v}^{\phan{ii}u} x_\nu}{x^2 + \rho^2} \ ,
\qquad
\sigma_{\mu\nu\ \, v}^{\phan{ii}u}
= \left(
\begin{array}{cc}
0 & 0 \\
0 & \sigma_{\mu\nu\ \, \beta}^{\phan{ii}\alpha}
\end{array}
\right)\ .
\end{equation}
Then the corresponding fermionic instanton reads
\begin{equation}
{\lambda^{\alpha}}^u{}_v
= \frac{\rho}{\sqrt{ ( x^2 + \rho^2 )^3}}
\left( \mu^u {\delta^\alpha{}_v}
+ \epsilon^{\alpha u} \bar\mu_v \right) \ .
\end{equation}
Here we have introduced Grassmann collective coordinates (GCC)
\begin{equation}
\mu^u = (\mu^1, \dots , \mu^{N - 2}, 0, 0) \ ,
\qquad
\epsilon^{\alpha u} = ( 0, \dots , 0, \epsilon^{\alpha\beta'})
\quad\mbox{with}\quad
N - 2 + \beta' = u \ ,
\end{equation}
and similarly for $\bar\mu$. The canonical dimension of $\mu$ is chosen
to be $- 1/2$.

In singular gauge, the gauge field is
\begin{equation}
A_{\mu\, u}{}^v
= - \frac{\rho^2}{x^2(x^2 + \rho^2)} \bar\sigma_{\mu\nu\, u}{}^v x_\nu\ .
\end{equation}
Notice that the position of the color indices is different from that
in regular gauge. This is due to the natural position of indices on
the sigma matrices. The fermionic anti-instanton in singular gauge reads
\cite{Cor79}
\begin{equation}
\label{ferm-angles}
{\lambda^{\alpha}}_u{}^v
= \frac{\rho}{\sqrt{x^2 (x^2 + \rho^2)^3}}
\left( \mu_u x^{\alpha v} + {x^\alpha}_u \bar\mu^v \right) \ ,
\end{equation}
where for fixed $\alpha$, the $N$-component vectors $\mu_u$
and $x^{\alpha v}$ are given by
\begin{equation}
\mu_u = \left( \mu_1 , \dots , \mu_{N - 2} , 0 , 0 \right) \ ,
\qquad
x^{\alpha v}
= \left( 0 , \dots , 0 , x^\mu \sigma_\mu^{\alpha\beta'} \right)
\quad\mbox{with}\quad
N - 2 + \beta' = v \ .
\end{equation}
Further, ${x^\alpha}_u = x^{\alpha v} \epsilon_{vu}$ and $\bar\mu^v$
also has $N - 2$ nonvanishing components. The particular choice of zeros
in the last two entries corresponds to the chosen embedding
of the $SU(2)$ instanton inside $SU(N)$. Notice that the adjoint field
$\lambda$ is indeed traceless in its color indices. This follows from the
observation that $\lambda$ only has non-zero entries on the off-diagonal
blocks inside $SU(N)$.

Depending on whether or not there is a reality condition on $\lambda$ in
euclidean space, the $\mu$ and $\bar\mu$ are related by complex conjugation.
We will illustrate this in a more concrete example when we discuss instantons
in $\cN = 4$ SYM theory.

We should also mention that while the bosonic collective coordinates are
related to the rigid symmetries of the theory, this is not obviously true
for the fermionic collective coordinates, although, as we will see later,
the $\xi$ and $\bar\eta$ collective coordinates can be obtained by
supersymmetry and superconformal transformations in SYM theories.

A similar construction holds for a fermion in the fundamental
representation. Now there is only one fermionic collective coordinate,
which we denote by $\cK$. The explicit expression, in singular gauge, is
\begin{equation}
({\lambda^\alpha})_u
= \frac{\rho}{\sqrt{x^2 (x^2 + \rho^2)^3}}\, {x^\alpha}_u\, \cK \ .
\end{equation}

%%%%%%%%%%%%%%%%%%%%%%%%%%%%%%%%%%%%%%%%%%%%%%%%%%%%%%%%%%%%%%%%%%%
\section{Zero modes and the measure.}
\setcounter{equation}{0}
%%%%%%%%%%%%%%%%%%%%%%%%%%%%%%%%%%%%%%%%%%%%%%%%%%%%%%%%%%%%%%%%%%%

In the following two sections we will show how to set up and do
(one-loop) perturbation theory around an (anti)-instanton. As a
first step, in this section, we will discuss the zero mode structure
and show how to reduce the path integral measure over instanton field
configurations to an integral over the moduli space of collective
coordinates, closely following \cite{Bernard}. In the next section, we
compute the fluctuations around an anti-instanton background.

%%%%%%%%%%%%%%%%%%%%%%%%%%%%%%%%%%%%%%%%%%%%%%%%%%%%%%%%%%%%%%%%%%%
\subsection{Normalization of the zero modes.}
%%%%%%%%%%%%%%%%%%%%%%%%%%%%%%%%%%%%%%%%%%%%%%%%%%%%%%%%%%%%%%%%%%%

In order to construct the zero modes and discuss perturbation theory, we
first decompose the fields into a background part and quantum fields
\begin{equation}
\label{pert}
A_\mu = A_\mu^{\rm cl} (\gamma) + A_\mu^{\rm qu} \ .
\end{equation}
Here $\gamma_i$ denote a set of collective coordinates, and, for
gauge group $SU(N)$, $i = 1, \dots, 4Nk$. Before we make the expansion
in the action, we should also perform gauge fixing and introduce ghosts,
$c$, and anti-ghosts, $b$. We choose the background gauge
condition
\begin{equation}
\cD_\mu^{\rm cl} A_{\mu}^{\rm qu} = 0 \ .
\end{equation}
Then the action, expanded up to quadratic order in the quantum fields,
is of the form
\begin{equation}
\label{gf-action}
S = \frac{8\pi^2}{\g^2} + \frac{1}{\g^2}
\tr\, \int\, d^4 x \left\{
A_\mu^{\rm qu}\, M_{\mu\nu}^{\rm cl}\, A_\nu^{\rm qu}
- 2 b\, M^{\rm gh}\, c \right\}\ ,
\end{equation}
with $M^{\rm gh} = \cD^2$ and
\begin{eqnarray}
M_{\mu\nu} &=& \cD^2 \delta_{\mu\nu} + 2 F_{\mu\nu}\nonumber\\
&=& \left( \cD^2 \delta_{\mu\nu} - \cD_\nu \cD_\mu + F_{\mu\nu} \right)
+ \cD_\mu \cD_\nu \equiv M_{\mu\nu}^1 + M_{\mu\nu}^2 \ ,
\end{eqnarray}
where we have dropped the subscript ${\rm cl}$. Here, $M^1$ stands for the
quadratic operator coming from the classical action, and $M^2$
corresponds to the gauge fixing term.

In making an expansion as in \eqn{gf-action}, we observe the existence of
zero modes (i.e.\ eigenfunctions of the operator $M_{\mu\nu}$ with zero
eigenvalues),
\begin{equation}
\label{zeromode}
Z^{(i)}_\mu \equiv \frac{\partial A_\mu^{\rm cl}}{\partial \gamma_i}
+ \cD_\mu^{\rm cl} \Lambda^i \ ,
\end{equation}
where the $\Lambda^i$-term is chosen to keep $Z_\mu$ in the background
gauge, so that
\begin{equation}
\label{BGC}
\cD_\mu^{\rm cl} Z_\mu^{(i)} = 0 \ .
\end{equation}
The first term in \eqn{zeromode} is a zero mode of $M^1$, as follows from
taking the derivative with respect to $\gamma_i$ of the field equation.
The $\cD_\mu \Lambda$ term is also a zero mode of $M^1$, since it is a
pure gauge transformation. The sum of the two terms is also a zero mode
of $M^2$, because $\Lambda$ is chosen such that $Z_\mu$ is in the
background gauge.

Due to these zero modes, we cannot integrate the quantum fluctuations,
since the corresponding determinants would give zero and yield divergences
in the path integral. They must therefore be extracted from the quantum
fluctuations, in a way we will describe in a more general setting in the
next subsection. It will turn out to be important to compute
the matrix of inner products
\begin{equation}
\label{norms}
U^{ij} \equiv \langle Z^{(i)} | Z^{(j)} \rangle
\equiv - \frac{2}{\g^2}
\int \, d^4x \, \tr\, \left\{ Z_\mu^{(i)} Z^{\mu (j)} \right\} \ .
\end{equation}
We now evaluate this matrix for the anti-instanton. For the four
translational zero modes, one can easily keep the zero mode in the
background gauge by choosing $\Lambda^\mu = A_\mu^{\rm cl}$.
Indeed,
\begin{equation}
Z_{\mu}^{(\nu)} =
\frac{\partial A_\mu^{\rm cl}}{\partial {x_0}_\nu}
+ \cD_\mu A_{\nu}^{\rm cl}
= - \partial_\nu A_\mu^{\rm cl} + \cD_\mu A_{\nu}^{\rm cl}
= F_{\mu\nu}^{\rm cl} \ ,
\end{equation}
which satisfies the background gauge condition. The norm of this
zero mode is
\begin{equation}
U^{\mu\nu} = \frac{8\pi^2|k|}{\g^2} \delta^{\mu\nu}
= S_{\rm cl}\, \delta^{\mu\nu}\ .
\end{equation}
This result actually holds for any $k$, and arbitrary gauge group.

Now we consider the dilatational zero mode corresponding to $\rho$
and limit ourselves to $k = 1$. Taking the derivative with respect to
$\rho$ leaves the zero mode in the background gauge, so we can set
$\Lambda^\rho = 0$. In singular gauge we have
\begin{equation}
Z_\mu^{(\rho)}
= - 2 \,\frac{\rho\, \bar\sigma_{\mu\nu} \, x_\nu}{(x^2 + \rho^2)^2} \ .
\end{equation}
Using the integral given in Appendix \ref{HooftSpinor}, one easily
computes that
\begin{equation}
U^{\rho \rho} = \frac{16\pi^2}{\g^2} = 2 S_{\rm cl} \ .
\end{equation}

The gauge orientation zero modes can be obtained from \eqn{1inst-angles}. By
expanding\footnote{Note the factor 2 in the exponent. This is to make the
normalization the same as in \cite{Bernard}. In that paper, the generators
are normalized as $\tr T_aT_b = - 2 \delta_{ab}$ (versus $-1/2$ in our
conventions), and there is no factor of two in the exponent. If we leave
out the factor of 2 in the exponent, then subsequent formula for the norms
of the gauge orientation zero modes will change, but this would eventually
be compensated by the integration over the angles $\theta^a$, such that the
total result remains the same. See Appendix \ref{Volume} for more details.}
$U(\theta) = \exp( - 2 \theta^a T_a)$ infinitesimally in \eqn{1inst-angles}
we get
\begin{equation}
\label{angle-zm}
\frac{\partial A_\mu}{\partial \theta^a} = 2 \left[ A_\mu, T_a \right] \ ,
\end{equation}
which is not in the background gauge. To satisfy \eqn{BGC} we have to add
appropriate gauge transformations, which differ for different generators of
$SU(N)$. First, for the $SU(2)$ subgroup corresponding to the instanton
embedding, we add
\begin{equation}
\label{L-su2}
\Lambda^a = - 2 \, \frac{\rho^2}{x^2+\rho^2} \, T_a \ ,
\end{equation}
and find that
\begin{equation}
\label{su2zm}
Z_{\mu}^{(a)} = 2 \cD_\mu \left[
\frac{x^2}{x^2 + \rho^2} T_a \right] \ .
\end{equation}
We have given the zero mode by working infinitesimally in $\theta^a$. One
should be able to redo the analysis for finite $\theta$, and we expect the
result to be an $SU(2)$ rotation on \eqn{su2zm}, which drops out under the
trace in the computation of the zero mode norms. One can now show, using
\eqn{eta-eta3} that the zero mode \eqn{su2zm} is in the background gauge,
and its norm reads
\begin{equation}
U^{ab} = \delta^{ab} 2 \rho^2 S_{\rm cl} \ .
\end{equation}

It is also fairly easy to prove that there is no mixing between
the different modes, i.e. $U^{\mu(\rho)} = U^{\mu a} = U^{(\rho) a} = 0$.

The matrix $U^{ij}$ for $SU(2)$ is eight by eight, with non-vanishing entries
\begin{equation}
U^{ij}=\pmatrix{\delta^{\mu\nu}S_{\rm cl}&&\cr &2S_{\rm cl} &\cr &&2
\delta^{ab}\rho^2S_{\rm cl}}_{[8] \times [8]}\ ,
\end{equation}
 and the square root of
the determinant is
\begin{equation}
{\sqrt U} = 2^2 S_{\rm cl}^4 \rho^3 = \frac{2^{14} \pi^8\rho^3}{\g^8} \ .
\end{equation}

Now we consider the remaining generators of $SU(N)$ by first analyzing the
example of $SU (3)$. There are seven gauge orientation zero modes, three
of which are given in
\eqn{su2zm} by taking for $T_a$ ($-i/2$) times the first three Gell-Mann
matrices $\lambda_1,\lambda_2,\lambda_3$. For the other four zero modes,
corresponding to $\lambda_4,\dots,\lambda_7$, the formula \eqn{angle-zm}
still holds, but we have to change the gauge transformation in order to
keep the zero mode in background gauge,
\begin{equation}
\label{L-su3}
\Lambda^k
= 2 \left[ \sqrt{\frac{x^2}{x^2 + \rho^2}} - 1 \right]
T_k \ , \qquad k = 4, 5, 6, 7 \ ,
\end{equation}
with $T_k = (-i/2) \lambda_k$. The difference in $x$-dependence of the gauge
transformations \eqn{L-su2} and \eqn{L-su3} is due to the change in
commutation relations. Namely, $\sum_{a = 1}^{3} [\lambda_a, [\lambda_a,
\lambda_\beta]] = - (3/4) \lambda_\beta$ for $\beta = 4, 5, 6, 7$, whereas
it is $-2 \lambda_\beta$ for $\beta = 1, 2, 3$. As argued before, there is
no gauge orientation zero mode associated with $\lambda_8$, since it
commutes with the $SU(2)$ embedding. The zero modes are then
\begin{equation}
\label{su3zm}
Z_\mu^{(k)} = 2 \cD_\mu
\left[ \sqrt{\frac{x^2}{x^2 + \rho^2}} T_k \right] \ ,
\qquad  k = 4, 5, 6, 7 \ ,
\end{equation}
with norms
\begin{equation}
U^{kl} = \delta^{kl} \rho^2 S_{\rm cl} \ ,
\end{equation}
and are orthogonal to \eqn{su2zm}, such that $U^{ka} = 0$.

This construction easily generalizes to $SU (N)$. One first chooses an
$SU (2)$ embedding, and this singles out 3 generators. The other generators
can then be split into $2 (N - 2)$ doublets under this $SU (2)$ and the
rest are singlets. There are no zero modes associated with the singlets,
since they commute with the chosen $SU(2)$. For the doublets, each
associated zero mode has the form as in \eqn{su3zm}, with the same norm
$\rho^2 S_{\rm cl}$. This counting indeed leads to $4 N - 5$ gauge
orientation zero modes. Straightforward calculation for the square-root
of the complete determinant then yields
\begin{equation}
{\sqrt U} = \frac{ 2^{6 N + 2}}{\rho^5}
\left( \frac{\pi \rho}{\g} \right)^{4N} \ .
\end{equation}
This ends the discussion about the (bosonic) zero mode normalization.

%%%%%%%%%%%%%%%%%%%%%%%%%%%%%%%%%%%%%%%%%%%%%%%%%%%%%%%%%%%%%%%%%%%
\subsection{Measure of collective coordinates.}
%%%%%%%%%%%%%%%%%%%%%%%%%%%%%%%%%%%%%%%%%%%%%%%%%%%%%%%%%%%%%%%%%%%

We now construct the measure on the moduli space of collective coordinates,
and show how the matrix $U$ plays the role of a Jacobian. We first
illustrate the idea for a generic system without gauge invariance, with
fields $\phi^A$, and action $S[\phi]$. We expand around the instanton
solution
\begin{equation}
\phi^A (x) = \phi^A_{\rm cl} \left( x, \gamma \right)
+ \phi^A_{\rm qu} \left( x,\gamma \right) \ .
\end{equation}
The collective coordinate is denoted by $\gamma$ and, for notational
simplicity, we assume there is only one. At this point the fields
$\phi^A_{\rm qu}$ can still depend on the collective coordinate, as it
can include zero modes. The action, up to quadratic terms in the quantum
fields is
\begin{equation}
\label{ActionExpand}
S = S_{\rm cl} + \ft12
\phi^A_{\rm qu} M_{AB} \left( \phi_{\rm cl} \right) \phi^B_{\rm qu} \ .
\end{equation}
The operator $M$ has zero modes given by
\begin{equation}
Z^A = \frac{\partial \phi^A_{\rm cl}}{\partial \gamma}\ ,
\end{equation}
since $MZ$ is just the derivative of the field equation with respect to
the collective coordinate. More generally, if the operator $M$ is
hermitean, it has a set of eigenfunctions $F_\alpha$ with eigenvalues
$\epsilon_\alpha$,
\begin{equation}
MF_\alpha = \epsilon_\alpha F_\alpha\ .
\end{equation}
One of the solutions is of course the zero mode $Z=F_0$ (we are suppressing
the index $A$) with $\epsilon_0=0$. Any function can be written in the
basis of eigenfunctions, in particular the quantum fields,
\begin{equation}
\phi^A_{\rm qu} = \sum_\alpha \xi_\alpha F^A_\alpha \ ,
\end{equation}
with some coefficients $\xi_\alpha$. The eigenfunctions have norms,
determined by their inner product
\begin{equation}
\langle F_\alpha | F_\beta \rangle
= \int\, d^4x\, F_\alpha (x) F_\beta (x) \ .
\end{equation}
The eigenfunctions can always be chosen orthogonal, such that
$\langle F_\alpha | F_\beta \rangle = \delta_{\alpha\beta} u_\alpha$.
The action then becomes
\begin{equation}
\label{S}
S = S_{\rm cl} + \ft12 \sum_\alpha
\xi_\alpha \xi_\alpha \epsilon_\alpha u_\alpha\ .
\end{equation}
If there would be a coupling constant in the action \eqn{ActionExpand},
then we rescale the inner product with the coupling, such that
\eqn{S} still holds. This is done in \eqn{norms}, where also a factor
of 2 is brought in, but it cancels after taking the trace. The measure is
defined as
\begin{equation}
\left[ d\phi \right] \equiv
\prod_{\alpha = 0} \sqrt{\frac{u_\alpha}{2\pi}}\, d \xi_\alpha \ .
\end{equation}
We now perform the gaussian integration over the
$\xi_\alpha$ and get
\begin{equation}
\int\, \left[ d\phi \right] \, e^{ - S[\phi]}
= \int\, \sqrt{\frac{u_0}{2\pi}} d \xi_0\, e^{-S_{\rm cl}}
({\det}'M)^{-1/2} \ .
\end{equation}
One sees that if there would be no zero modes, it produces the correct
result, namely the determinant of $M$. In the case of zero modes, the
determinant of $M$ is zero, and the path integral would be ill-defined.
Instead, we must leave out the zero mode in $M$, take the amputated
determinant (denoted by $\det'$), and integrate over the mode $\xi_0$.

The next step is to convert the $\xi_0$ integral to an integral over the
collective coordinate $\gamma$. This can be done by inserting unity into
the path integral \cite{GerSakTom}. Consider the identity
\begin{equation}
1 = \int\, d \gamma\, \delta \left( f (\gamma) \right)
\frac{\partial f}{\partial \gamma} \ ,
\end{equation}
which holds for any (invertible) function $f(\gamma)$. Taking $f(\gamma)
= - \langle \phi - \phi_{\rm cl}(\gamma)| Z \rangle$, (recall that
the original field $\phi$ is independent of $\gamma$), we get
\begin{equation}
1 = \int\, d \gamma\,
\left(
u_0 -
\left\langle
\phi_{\rm qu}
\left|
\frac{\partial Z}{\partial \gamma}
\right\rangle\right.
\right)
\delta
\Big( \langle \phi_{\rm qu} | Z \rangle \Big)
= \int\, d \gamma\,
\left( u_0
- \left\langle \phi_{\rm qu} \left| \frac{\partial Z}{\partial \gamma}
\right\rangle\right. \right) \delta \Big( \xi_0 u_0 \Big) \ .
\end{equation}
This trick
is somehow similar to the Faddeev-Popov trick for gauge fixing. In the
semiclassical approximation, the second term between the brackets is
subleading and we will neglect it\footnote{It will appear however
as a two loop contribution. To see this, one first writes this term in
the exponential, where it enters without $\hbar$, so it is at least a
one loop effect. Then, $\phi_{\rm qu}$ has
a part proportional to the zero mode, which drops out by means of the delta
function insertion. The other part of $\phi_{\rm qu}$ is genuinely quantum
and contains a power of $\hbar$ (which we have suppressed). Therefore, it
contributes at two loops \cite{TwoLoop1,TwoLoop2}.}. This leads to
\begin{equation}
\int\, \left[ d\phi \right]
\, e^{-S} = \int\, d \gamma\, \sqrt{\frac{u_0}{2\pi}} e^{-S_{\rm cl}}
\left( {\det}'M \right)^{-1/2} \ .
\end{equation}
For a system with more zero modes $Z^i$ with norms-squared $U^{ij}$, the
result is
\begin{equation}
\label{measure}
\int\, \left[ d\phi \right] \, e^{-S} = \int \,
\prod_{i = 1} \frac{d\gamma^i}{\sqrt{2\pi}}
\left( \det\, U \right)^{1/2} e^{-S_{\rm cl}}
\left( {\det}' M \right)^{-1/2} \ .
\end{equation}
Notice that this result is invariant under rescalings of $Z$, which can
be seen as rescalings on the collective coordinates. More generally, the
matrix $U^{ij}$ can be interpreted as the metric on the moduli space of
collective coordinates. The measure is then invariant under general
coordinate transformations on the moduli space.

This expression for the measure also generalizes to systems with fermions.
The only modifications are dropping the factors of ${\sqrt {2\pi}}$
(because gaussian integration over fermions does not produce this factor),
and inverting the determinants.

One can repeat the analysis for gauge theories to show that
\eqn{measure} also holds for Yang-Mills instantons in singular gauge. For
regular gauges, there are some modifications due to the fact that
the gauge orientation zero mode functions $\Lambda^a$ do not fall off
fast enough at infinity. This is explained in \cite{Bernard}, and we will
not repeat it here. For this reason, it is more convenient to work in
singular gauge.

The collective coordinate measure for $k = 1$ $SU(N)$ YM theories, without
the determinant from integrating out the quantum fluctuations which will
be analyzed in the next section, is now
\begin{equation}
\label{bos-meas}
\frac{2^{4 N + 2} \pi^{4 N - 2}}{(N - 1)!(N - 2)!}
\frac{1}{\g^{4N}} \int\, d^4 x_0 \, \frac{d\rho}{\rho^5}
\rho^{4N}\ .
\end{equation}
This formula contains the square-root of the determinant of $U$, $4N$
factors of $1/\sqrt{2\pi}$, and we have also integrated out the gauge
orientation zero modes. This may be done only if we are evaluating gauge
invariant correlation functions. The result of this integration follows
from the volume of the coset space
\begin{equation}
\label{VolCoset}
{\rm Vol} \left\{
\frac{SU(N)}{SU(N - 2) \times U(1)} \right\}
= \frac{\pi^{2 N - 2}}{(N - 1)!(N - 2)!}\ .
\end{equation}
The derivation of this formula can be found in Appendix \ref{Volume},
and is based on \cite{Bernard}.

%%%%%%%%%%%%%%%%%%%%%%%%%%%%%%%%%%%%%%%%%%%%%%%%%%%%%%%%%%%%%%%%%%%
\subsection{The fermionic measure.}
%%%%%%%%%%%%%%%%%%%%%%%%%%%%%%%%%%%%%%%%%%%%%%%%%%%%%%%%%%%%%%%%%%%

Finally we construct the measure on the  moduli space of fermionic
collective coordinates. For the $\xi$ zero modes \eqn{xi-eta}, one finds
\begin{equation}
Z^\alpha_{(\beta)} = \frac{\partial \lambda^\alpha}{\partial \xi^\beta}
= - \ft12 \sigma_{\mu\nu\ \, \beta}^{\phan{ii}\alpha} F_{\mu\nu} \ .
\end{equation}
The norms of these two zero modes are given by
\begin{equation}
{(U_\xi)_\beta}^\gamma = - \frac{2}{\g^2} \int \, d^4 x \,
\tr\, \left\{ Z^\alpha_{(\beta)} Z^{(\gamma)}_\alpha \right\}
= 4 S_{\rm cl} {\delta_\beta}^\gamma \ ,
\end{equation}
where we have used the expression \eqn{norms}. This produces a term
in the measure~\footnote{Sometimes one finds in the literature
that $U_\xi=2 S_{\rm cl}$. This is true when one uses the conventions for
Grassmann integration $\int d^2\xi\,\xi^\alpha\xi^\beta=\ft12
\epsilon^{\alpha\beta}$. In our conventions $d^2\xi\equiv d\xi^1d\xi^2$.}
\begin{equation}
\int \, d \xi^1 d \xi^2\, \left( 4 S_{\rm cl} \right)^{-1}\ ,
\label{xi-meas}
\end{equation}
So the Jacobian
for the $\xi$ zero modes is given by $U_\xi = 4 S_{\rm cl}$, and the result
\eqn{xi-meas} actually holds  for any $k$.

For the $\bar\eta$ zero modes, we obtain, using some algebra for the
$\sigma$-matrices,
\begin{equation}
(U_{\bar \eta})_{\alpha'}{}^{\beta'}
= 8 S_{\rm cl} \delta_{\alpha'}{}^{\beta'}\ ,
\end{equation}
so that the corresponding measure is
\begin{equation}
\int\,d{\bar \eta}_1d{\bar \eta}_2\,(8S_{\rm cl})^{-1}\ ,
\end{equation}
which only holds for $k = 1$.

Finally we compute the Jacobian for the fermionic ``gauge orientation'' zero
modes.For convenience, we take the solutions in regular gauge (the Jacobian
is gauge invariant anyway), and find
\begin{equation}
{\left( Z^\alpha_{({\mu^w})} \right)^u}_v
= \frac{\rho}{\sqrt{(x^2 + \rho^2)^3}}
\, {\delta^\alpha}_v \, {\Delta^u}_w\ ,
\qquad
{\left( Z^\alpha_{({\bar \mu}_w)} \right)^u}_v
= \frac{\rho}{\sqrt{(x^2 + \rho^2)^3}} \,
\epsilon^{\alpha u} {\Delta^w}_v\  ,
\end{equation}
where the $N$ by $N$ matrix $\Delta$ is the unity matrix in the $(N-2)$
by $(N-2)$ upper diagonal block, and zero elsewhere. The norms of
$Z_{\mu}$ and $Z_{\bar\mu}$ are easily seen to be zero, but the
nonvanishing inner product is
\begin{equation}
(U_{\mu\bar \mu})^u{}_v = - \frac{2}{\g^2} \int\, d^4 x \,
\tr\, Z^\alpha_{(\bar\mu_u)}
Z_{\alpha\, (\mu_v)} = \frac{2\pi^2}{\g^2} \Delta^u{}_v\ ,
\end{equation}
where we have used the integral \eqn{Integral}. It also follows from the
index structure that the $\xi$ and $\bar\eta$ zero modes are orthogonal
to the $\mu$ zero modes, so there is no mixing in the Jacobian.

Putting everything together, the fermionic measure for $\cN$ adjoint
fermions coupled to $SU(N)$ YM theory, with $k=1$, is
\begin{equation}
\label{ferm-meas}
\int\, \prod_{A = 1}^{\cN} d^2 \, \xi^A
\left( \frac{\g^2}{32\pi^2} \right)^\cN
\int\, \prod_{A = 1}^{\cN} d^2\, \bar\eta^A
\left( \frac{\g^2}{64\pi^2\rho^2} \right)^\cN
\int\, \prod_{A = 1}^{\cN} \prod_{u = 1}^{N - 2} \,
d \mu^{A,u} \, d \bar\mu_u^A
\left( \frac{\g^2}{2\pi^2} \right)^{\cN (N - 2)} \ .
\end{equation}
Similarly, one can include fundamental fermions, for which the Jacobian
factor is
\begin{equation}
U_{\cK} \equiv \frac{1}{\g^2}\int\,d^4x\,Z^\alpha{}_u Z_\alpha{}^u
= \frac{\pi^2}{\g^2}\ ,
\end{equation}
for each specie.

%%%%%%%%%%%%%%%%%%%%%%%%%%%%%%%%%%%%%%%%%%%%%%%%%%%%%%%%%%%%%%%%%%%
\section{One loop determinants.}
\setcounter{equation}{0}
%%%%%%%%%%%%%%%%%%%%%%%%%%%%%%%%%%%%%%%%%%%%%%%%%%%%%%%%%%%%%%%%%%%

After having determined the measure on the collective coordinate moduli
space, we now compute the determinants that arise after Gaussian
integration over the quantum fluctuations. Before doing so, we extend
the model by adding real scalar fields in the adjoint representation.
The action is
\begin{equation}
S = - \frac{1}{\g^2} \int \, d^4 x\, \tr\,
\left\{\ft12 F_{\mu\nu} F_{\mu\nu} +\left( \cD_\mu \phi \right)
\left( \cD_\mu \phi \right) - i \bar\lambda \not\!\!\bar\cD \lambda
- i \lambda \not\!\!\cD \bar\lambda \right\} \ .
\label{scal-ferm}
\end{equation}
Here, $\lambda$ is a two-component Weyl spinor which we take in the
adjoint representation. Generalization to fundamental fermions is
straightforward. In Minkowski space, $\bar\lambda$ belongs to the
conjugate representation of the Lorentz group, but in euclidean space
it is unrelated to $\lambda$.

The anti-instanton solution which we will expand around is
\begin{equation}
\label{cl-conf}
A_\mu^{\rm cl}\ ,
\qquad \phi_{\rm cl} = 0\ ,
\qquad \lambda_{\rm cl}=0\ ,
\qquad \bar\lambda_{\rm cl} = 0\ ,
\end{equation}
where $A_{\mu}^{\rm cl}$ is the anti-instanton. Although this background
represents an exact solution to the field equations, it does not include
the fermionic zero modes, which are the solutions to the Dirac equation.
In this approach, one should treat these zero modes in perturbation theory.
As will become clearer in later sections, we would like to include the
fermionic zero modes in the classical anti-instanton background and treat
them exactly. This is also more compatible with supersymmetry and the ADHM
construction for (supersymmetric) multi-instantons. But then one would have
to redo the following analysis, which, to our knowledge, has not been done
so far. We comment on this issue again at the end of this section.

After expanding $A_\mu = A_\mu^{\rm cl} + A_\mu^{\rm qu}$, and similarly
for the other fields, we add gauge fixing terms
\begin{equation}
S_{\rm gf} = \frac{1}{\g^2} \int\, d^4x \,\tr\,
\left\{ \left( \not\!\!\cD_\mu^{\rm cl} \, A_\mu^{\rm qu} \right)^2
- 2\, b \, \cD^2_{\rm cl} \, c \right\}\ ,
\end{equation}
such that the total gauge field action is given by \eqn{gf-action}.
The integration over $A_\mu$ gives
\begin{equation}
\label{det-gauge}
\left[ {\det}' \Delta_{\mu\nu} \right]^{-1/2} \ ,
\qquad
\Delta_{\mu\nu} = - \cD^2 \delta_{\mu\nu} - 2 F_{\mu\nu}\ ,
\end{equation}
where the prime stands for the amputated determinant, with zero eigenvalues
left out. We have suppressed the subscript `cl' and Lie algebra indices.

Integration over the scalar fields results in
\begin{equation}
\left[ \det \Delta_\phi \right]^{-1/2}\ ,
\qquad \Delta_\phi = - \cD^2\ ,
\end{equation}
and the ghost system yields similarly
\begin{equation}
\left[ \det \Delta_{\rm gh} \right]\ ,
\qquad \Delta_{\rm gh} = - \cD^2\ .
\end{equation}
For the fermions $\lambda$ and $\bar\lambda$, we give a bit more
explanation. Since neither ${\not\!\!\cD}$ nor ${\not\!\!\bar\cD}$ is
hermitean, we can not evaluate the determinant in terms of its
eigenvalues. But both products
\begin{equation}
\Delta_- = - \not\!\!\cD \not\!\!\bar\cD
= - \cD^2 - \sigma_{\mu\nu} F_{\mu\nu}\ ,
\qquad
\Delta_+ = - \not\!\!\bar\cD \not\!\!\cD = - \cD^2\ ,
\end{equation}
which still have unwritten spinor indices, are hermitean. Hence we can
expand $\lambda$ in terms of eigenfunctions $F_i$ of $\Delta_-$ with
coefficients $\xi_i$, and $\bar\lambda$ in terms of eigenfunctions
$\bar F_i$ of $\Delta_+$ with coefficients $\bar\xi_i$. We have seen
in section 3 that both operators have the same spectrum of non-zero
eigenvalues, and the relation between the eigenfunctions is $\bar F_i
= \not\!\!\bar\cD F_i$. Defining the path integral over $\lambda$ and
$\bar\lambda$ as the integration over $\xi_i$ and $\bar\xi_i$, one gets
the determinant over the nonzero eigenvalues. The result for the
integration over the fermions gives
\begin{equation}
\left[ {\det}' \Delta_- \right]^{1/4}
\left[ \det \Delta_+ \right]^{1/4} \ .
\end{equation}
As stated before, since all the eigenvalues of both $\Delta_-$ and
$\Delta_+$ are the same, the determinants are formally equal. This
result can also be obtained by writing the spinors in terms of Dirac
fermions, the determinant we have to compute is then
\begin{equation}
\left[ {\det}' \, \Delta_D^2 \right]^{1/2}\ ,
\qquad
\Delta_D =
\left(
\begin{array}{cc}
0               & \not\!\!\cD \\
\not\!\!\bar\cD & 0
\end{array}
\right)\ .
\end{equation}
Now we notice that the determinants for the bosons are related to the
determinants of $\Delta_-$ and $\Delta_+$. For the ghosts and adjoint
scalars this is obvious,
\begin{equation}
\det \, \Delta_{\phi} = \det \, \Delta_{\rm gh}
= \left[ \det \, \Delta_+ \right]^{1/2}\ .
\end{equation}
For the vector fields, we use the identity
\begin{equation}
\Delta_{\mu\nu}
= \ft12 \tr \left\{ \bar\sigma_\mu \Delta_- \sigma_\nu \right\}
= \ft12 \bar\sigma_{\mu\, \alpha'\beta}\, {\Delta_-}^\beta{}_\gamma\,
{\delta^{\alpha'}}_{\delta'} \, \sigma_\nu^{\gamma\delta'}\ ,
\end{equation}
to prove that
\begin{equation}
{\det}' \Delta_{\mu\nu} = \left[ {\det}' \Delta_- \right]^2\ .
\end{equation}
Now we can put everything together. The determinant for a YM system
coupled to $n$ adjoint scalars and $\cN$ Weyl spinors is
\begin{equation}
\left[ {\det}' \Delta_- \right]^{- 1 + \cN/4}
\left[ {\det}' \Delta_+ \right]^{\ft14 ( 2 + \cN - n)}\ .
\end{equation}
This expression simplifies to the ratio of the determinants when
$\cN - \ft{n}2 = 1$. Particular cases are
\begin{eqnarray}
&&\cN = 1 \quad n = 0 \quad \rightarrow \quad
\left[ \frac{\det \Delta_+}{{\det}' \Delta_-} \right]^{3/4}\ ,
\nonumber\\
&&\cN = 2 \quad n = 2 \quad \rightarrow \quad
\left[ \frac{\det \Delta_+}{{\det}' \Delta_-} \right]^{1/2}\ ,
\nonumber\\
&&\cN = 4 \quad n = 6 \quad \rightarrow \quad
\left[ \frac{\det \Delta_+}{{\det}' \Delta_-} \right]^{0}\ .
\end{eqnarray}
These cases correspond to supersymmetric models with $\cN$-extended
supersymmetry. Notice that for $\cN = 4$, the determinants between
bosons and fermions cancel, so there is no one-loop contribution. For
$\cN = 1, 2$, the determinants give formally unity since the non-zero
eigenvalues are the same. As we will explain below, however, we must
first regularize the theory to define the determinants properly and this
may yield different answers. In all other cases, we will not get this
ratio of these particular determinants.

All of the above manipulations are a bit formal. We know that as soon as
we  do perturbation theory, one must first choose a regularization scheme
in order to define the quantum theory. After that, the renormalization
procedure must be carried out and counterterms must be added. The
counterterms are the same as in the theory without instantons and their
finite parts must be specified by physical normalization conditions. The
ratios of products of non-zero eigenvalues have the meaning of a mass
correction to the instanton (seen as a five-dimensional soliton). One
can write this ratio as the exponent of the difference of two infinite
sums
\begin{equation}
\frac{\det \Delta_+}{\det^\prime \Delta_-}
= \exp \left( \sum_n \omega_n^{(+)} - \sum_n \omega_n^{(-)} \right) \ ,
\end{equation}
where the eigenvalues $\lambda_n = \exp \omega_n$. The frequencies
$\omega_n^{(+)}$ and $\omega_n^{(-)}$ are discretized by putting the
system in a box of size $R$ and imposing suitable boundary conditions
on the quantum fields at $R$ (for example, $\phi (R) = 0$, or
$\frac{d}{dR} \phi (R) = 0$, or a combination of thereof \cite{Hooft1}).
These boundary conditions may be different for different fields.
The sums over $\omega_n^{(+)}$ and $\omega_n^{(-)}$ are divergent;
their difference is still divergent (although less divergent than
each sum separately) but after adding counterterms $\Delta S$ one obtains
a finite answer. The problem is that one can combine the terms in
both series in different ways, possibly giving different answers.
By combining $\omega_n^{(+)}$ with $\omega_n^{(-)}$ for each fixed $n$,
one would find that the ratio $\left(\det\Delta_+/\det'\Delta_-\right)$
equals unity. However, other values could result by using different ways
to regulate these sums. It is known that in field theory the results
for the effective action due to different regularization schemes
differ at most by a local finite counterterm. In the background field
formalism we are using, this counterterm must be background gauge
invariant, and since we consider only vacuum expectation values
of the effective action, only one candidate is possible: it is
proportional to the gauge action $\int d^4 x\, \tr\, F^2$ and multiplied
by the one-loop beta-function for the various fields which can run
in the loop,
\begin{equation}
\Delta S \propto \beta (\g) \int d^4 x\, \tr\, F^2 \
\ln \frac{\mu^2}{\mu_0^2} \ .
\end{equation}
The factor $\ln \left( \mu^2/ \mu_0^2 \right)$ parametrizes the freedom
in choosing different renormalization schemes.

A particular regularization scheme used in \cite{Hooft1} is Pauli-Villars
regularization. In this case 't Hooft used first $x$-dependent regulator
masses to compute the ratios of the one-loop determinants $\Delta$ in
the instanton background and $\Delta^{(0)}$ in the trivial vacuum. Then he
argued that the difference between using the $x$-dependent masses and
using the more usual constant masses, was of the form $\Delta S$ given
above. The final result for pure YM $SU(N)$ in the $k = 1$ sector is
\cite{Hooft1,Bernard}
\begin{equation}
\left[ \frac{{\det}' \Delta_-}{\det \Delta_-^{(0)}} \right]^{-1}
\left[ \frac{\det \Delta_+}{\det \Delta_+^{(0)}} \right]^{1/2}
= \mu^{4N}
\exp \left\{
- \ft13 N\, \ln (\mu \rho) - \alpha(1)
- 2 ( N - 2 ) \alpha \left( \ft12 \right) \right\}\ .
\end{equation}
Here we have normalized the determinants against the vacuum, indicated
by the superscript $(0)$. Note that Pauli-Villars regulator fields
contribute one factor of $\mu$ for each zero mode of the original
fields. The numerical values of the function $\alpha(t)$ are related to
the Riemann zeta function, and take the values $\alpha \left( \ft12
\right) = 0.145 873$ and $\alpha(1) = 0.443 307$. Notice that this
expression for the determinant depends on $\rho$, and therefore changes
the behaviour of the $\rho$ integrand in the collective coordinate measure.
Combined with \eqn{bos-meas} one correctly reproduces the $\beta$-function
of $SU (N)$ YM theory.

Let us briefly come back to the point of expanding around a background which
includes the fermionic zero modes. Upon expanding around this classical
configuration, one finds mixed terms between the gauge field and fermion
quantum fluctuations, e.g. terms like ${\bar \lambda}_{\rm qu}A_\mu^{\rm qu}
\lambda_{\rm cl}$. Integrating out the quantum fields yields a
superdeterminant in the space of all the fields, which will in general
depend on the Grassmann collective coordinates (GCC) appearing in
$\lambda_{\rm cl}$. It remains to be seen if this superdeterminant will
still give unity in the supersymmetric cases, and if not, one would like
to find its dependence on the GCC. We hope to report on this in a future
publication.

%%%%%%%%%%%%%%%%%%%%%%%%%%%%%%%%%%%%%%%%%%%%%%%%%%%%%%%%%%%%%%%%%%%
\section{$\cN = 4$ supersymmetric Yang-Mills theory.}
\setcounter{equation}{0}
%%%%%%%%%%%%%%%%%%%%%%%%%%%%%%%%%%%%%%%%%%%%%%%%%%%%%%%%%%%%%%%%%%%

For reasons explained in the introduction, we now focus on the $\cN = 4$
model \cite{Sch77}. The action is of course well known in Minkowski space,
but instantons require, however, the formulation of the $\cN = 4$ euclidean
version. Due to absence of a real representation of Dirac matrices in
four-dimensional euclidean space, the notion of Majorana spinor is absent.
This complicates the construction of euclidean Lagrangians for
supersymmetric models \cite{Zum,Nic,Nie96}. For $\cN = 2, 4$ theories, one
can replace the Majorana condition by the so-called simplectic Majorana
condition and consequently construct real supersymmetric Lagrangians
\cite{BVV, Bla97}.

In the following subsection we write down the action in  Minkowski
space-time and discuss the reality conditions on the fields. Next we
construct the hermitean $\cN = 4$ euclidean model via the dimensional
reduction of 10D $\cN = 1$ super-Yang-Mills theory along the time direction.
Using this, we study in the consequent section the solutions of the
classical equations of motion, using an iteration procedure in the
Grassmann collective coordinates.

%%%%%%%%%%%%%%%%%%%%%%%%%%%%%%%%%%%%%%%%%%%%%%%%%%%%%%%%%%%%%%%%%%%
\subsection{Minkowskian $\cN = 4$ SYM.}
%%%%%%%%%%%%%%%%%%%%%%%%%%%%%%%%%%%%%%%%%%%%%%%%%%%%%%%%%%%%%%%%%%%

The $\cN = 4$ action in Minkowski space-time with the signature
$\eta^{\mu\nu} = {\rm diag}(-,+,+,+)$ is given by
\begin{eqnarray}
\label{action}
S\!\!\!&=&\!\!\!
\frac{1}{\g^2} \int\, d^4x \, \tr \,
\Bigg\{ \ft12 F_{\mu\nu} F^{\mu\nu}
- i \bar\lambda_A^{\dot\alpha} \not\!\!{\bar\cD}_{\dot\alpha\beta}
\lambda^{\beta, A}
- i \lambda_{\dot\alpha}^A \not\!\!\cD^{\alpha\dot\beta}
\bar\lambda_{A\dot\beta}
+ \ft12 \left( \cD_\mu {\bar \phi}_{AB} \right)
\left( \cD^\mu \phi^{AB} \right) \\
&&\!\!\!- \sqrt{2} {\bar \phi}_{AB}
\left\{ \lambda^{\alpha,A}, \lambda^B_{\alpha}\right\}
- \sqrt{2} \phi^{AB}
\left\{ \bar\lambda_A^{\dot\alpha} , \bar\lambda_{\dot\alpha, B} \right\}
+ \ft18 \left[ \phi^{AB}, \phi^{CD} \right]
\left[ \bar\phi_{AB} , \bar\phi_{CD} \right] \Bigg\}\ . \nonumber
\end{eqnarray}
The on-shell $\cN = 4$ supermultiplet consists out of a real gauge field,
$A_\mu$, four complex Weyl spinors $\lambda^{\alpha,A}$ and an
antisymmetric complex scalar $\phi^{AB}$ with labels $A,B = 1, \dots, 4$
of internal $R$ symmetry group $SU (4)$.

The reality conditions on
the components of this  multiplet are~\footnote{Unless specified otherwise,
equations which involve complex conjugation of fields will be understood
as not Lie algebra valued, i.e.\ they hold for the components $\lambda^{a,
\alpha, A}$, etc.} the Majorana conditions $\left( \lambda^{\alpha,A}
\right)^\ast = - \bar\lambda^{\dot\alpha}_A$ and $(\lambda^A_\alpha)^\ast
= \bar\lambda_{\dot\alpha,A}$ and
\begin{equation}
\label{MinkRealScal}
{\bar \phi}_{AB} \equiv \left( \phi^{AB} \right)^\ast
= \ft12 \epsilon_{ABCD} \phi^{CD} \ .
\end{equation}
The sigma matrices are
defined by $\sigma^{\mu\, \alpha\dot \beta} = (-1, \tau^i)$, ${\bar
\sigma}^\mu_{\dot\alpha\beta} = (1, \tau^i)$ and complex conjugation
gives $\left( \sigma_{\mu}^{\alpha \dot\beta} \right)^\ast =
\sigma_\mu^{\beta\dot\alpha} = \epsilon^{\dot\alpha\dot\gamma}
\epsilon^{\beta\delta} \bar\sigma_{\mu\, \dot\gamma\delta}$, with
$\epsilon^{\dot\alpha\dot\beta} = \epsilon_{\dot\alpha\dot\beta} =
- \epsilon^{\alpha\beta} = - \epsilon_{\alpha\beta}$.

Since $\phi^{AB}$ is antisymmetric, one can express it in a basis
spanned by the eta-matrices
 (see Appendix \ref{HooftSpinor})
\begin{equation}
\phi^{AB} = \frac{1}{\sqrt 2}
\left\{ S^i \eta^{iAB} + i P^i \bar\eta^{iAB} \right\},
\qquad
\bar\phi_{AB} = \frac{1}{\sqrt 2}
\left\{ S^i \eta^{i}_{AB} - i P^i \bar\eta^{i}_{AB} \right\} \ ,
\end{equation}
in terms of real scalars $S^i$ and pseudoscalars $P^i$, $i = 1, 2, 3$,
so that reality condition is automatically fulfilled. Then the
kinetic terms for the $(S,P)$ fields take the standard form.

The action (\ref{action})
is invariant under the supersymmetry transformation laws
\begin{eqnarray}
\label{susy-Mink}
\delta A_\mu
\!\!\!&=&\!\!\! - i {\bar\zeta}^{\dot\alpha}_A
\bar\sigma_{\mu\, \dot\alpha\beta} \lambda^{\beta, A}
+ i \bar\lambda_{\dot\beta, A} \sigma_\mu^{\alpha\dot\beta} \zeta^A_\alpha\ ,
\nonumber\\
\delta \phi^{AB}
\!\!\!&=&\!\!\! \sqrt{2} \Big( \zeta^{\alpha,A} \lambda^B_\alpha
- \zeta^{\alpha,B} \lambda^A_\alpha
+ \epsilon^{ABCD} \bar\zeta^{\dot\alpha}_C \bar\lambda_{\dot\alpha,D}
\Big)\ ,
\nonumber\\
\delta \lambda^{\alpha,A}
\!\!\!&=&\!\!\! - \ft12
\sigma^{\mu\nu\, \alpha}_{\phan{mn}\beta}
F_{\mu\nu} \zeta^{\beta,A}
- i \sqrt{2}
\bar\zeta_{\dot\alpha,B} \not\!\!\cD^{\alpha\dot\alpha} \phi^{AB}
+ \left[ \phi^{AB}, \bar\phi_{BC} \right] \zeta^{\alpha,C} \ ,
\end{eqnarray}
which are consistent with the reality conditions.
Let us turn now to the discussion of the euclidean version of this model
and discuss the differences with the Minkowski theory.

%%%%%%%%%%%%%%%%%%%%%%%%%%%%%%%%%%%%%%%%%%%%%%%%%%%%%%%%%%%%%%%%%%%%%
\subsection{Euclidean $\cN = 4$ SYM.}
\label{EuclidN4}
%%%%%%%%%%%%%%%%%%%%%%%%%%%%%%%%%%%%%%%%%%%%%%%%%%%%%%%%%%%%%%%%%%%%%

To find out the $\cN = 4$ supersymmetric YM model in euclidean $d = (4, 0)$
space, we follow the same procedure as in \cite{Sch77}. We start
with the $\cN = 1$ SYM model in $d = (9, 1)$ Minkowski space-time, but
contrary to the original paper we reduce it on a six-torus with one time
and five space coordinates \cite{BVV,Bla97}. As opposed to action
(\ref{action}) with the $SU(4)=SO(6)$ $R$-symmetry group, this reduction
leads to the internal non-compact $SO(5, 1)$ $R$-symmetry group in euclidean
space. As we will see, the reality conditions on bosons and fermions
will both use an internal metric for this non-compact
internal symmetry group.

The $\cN = 1$ Lagrangian with $d = (9,1)$ reads
\begin{equation}
\cL_{10} = \frac{1}{\g^2_{10}} {\rm tr}
\left\{
\ft12 F_{MN} F^{MN} + \bar \mP \mG^M \cD_M \mP
\right\} \ ,
\end{equation}
with the field strength $F_{MN} = \partial_M A_N - \partial_N A_M +
\left[ A_M, A_N \right]$ and the Majorana-Weyl spinor $\mP$ defined
by the conditions
\begin{equation}
\label{MajoranaWeyl}
\mG^{11} \mP = \mP \ ,
\qquad
\mP^T C^-_{10} = \mP^\dagger i \mG^0 \equiv \bar \mP \ .
\end{equation}
Here the hermitean matrix $\mG^{11} \equiv \ast\!\mG$ is a product of
all Dirac matrices, $\mG^{11} = \mG^0 \dots \mG^9 $, normalized to $\left(
\ast\!\mG \right)^2 = + 1$. The $\mG$-matrices obey the Clifford algebra
$\left\{ \mG^M, \mG^N \right\} = 2 \eta^{MN}$ with metric $\eta^{MN} =
{\rm diag} (-, +, \dots, +)$. The Lagrangian is a density under the
standard transformation rules
\begin{equation}
\delta A_M = \bar \zeta \mG_M \mP \ ,
\qquad
\delta {\mit\Psi} = - \ft12 F_{MN} \mG^{MN} \zeta \ ,
\end{equation}
with $\mG^{MN} = \ft12 [\mG^M \mG^N - \mG^M \mG^N]$ and $\bar\zeta =
\zeta^T C^-_{10} = \zeta^\dagger i \mG^0$.

To proceed with the dimensional reduction we choose a particular
representation of the gamma matrices in $d = (9, 1)$ , namely
\begin{equation}
\label{10Gammas}
\mG^M
= \left\{ \hat\gamma^a \otimes \gamma^5, \1_{[8] \times [8]}
\otimes \gamma^\mu \right\},
\qquad
\mG^{11} = \mG^0 \dots \mG^9 = \hat\gamma^7 \otimes \gamma^5 ,
\end{equation}
where the $8 \times 8$ Dirac matrices $\hat\gamma^a$ and $\hat\gamma^7$
of $d = (5,1)$ with $a = 1, \dots, 6$ can be conveniently defined by
means of 't Hooft symbols as follows
\begin{equation}
\hat\gamma^a
= \left(
\begin{array}{cc}
0              & \mS^{a, AB} \\
\bar\mS^a_{AB} & 0
\end{array}
\right) , \qquad
\hat\gamma^7 = \hat\gamma^1 \dots \hat\gamma^6
= \left(
\begin{array}{cc}
1 & 0 \\
0 & - 1
\end{array}
\right) \ ,
\end{equation}
with the notations $\mS^{a, AB} = \left\{ - i \eta^{1,AB},
\eta^{2,AB}, \eta^{3,AB}, i \bar\eta^{k,AB} \right\}$, $\bar\mS^a_{AB}
= \left\{ i \eta^1_{AB}, - \eta^2_{AB}, - \eta^3_{AB}, i \bar\eta_{AB}^k
\right\}$ so that $\ft12 \epsilon_{ABCD} \mS^{a\, CD} = - \bar\mS^a_{AB}$.
Meanwhile $\gamma^\mu$ and $\gamma^5$ are the usual of $d = (4, 0)$
introduced in \eqn{gamma-sigma}. Note that in this construction we
implicitly associated one of the Dirac matrices, namely $\eta^1$, in $6$
dimensions with the time direction and thus it has square $- 1$; all other
(as well as all $d = (4,0)$) are again hermitean with square $+ 1$.

Let us briefly discuss the charge conjugation matrices in $d = (9, 1)$,
$d = (5, 1)$ and $d = (4, 0)$. One can prove by means of finite group
theory \cite{Nie81} that all their properties are representation
independent. In general there are two charge conjugation matrices $C^+$
and $C^-$ in even dimensions, satisfying $C^\pm \mG^\mu = \pm \left(
\mG^\mu \right)^T C^\pm$, and $C^+ = C^- \ast\!\mG$.  These charge
conjugation matrices do not depend on the signature of space-time and
obey the relation $C^- \ast\!\mG = \pm \left(\ast\!\mG\right)^T C^-$
with $-$ sign in $d = 10,\, 6$ and $+$ sign in $d=4$. The
transposition depends on the dimension and leads
 to $\left( C^\pm \right)^T = \pm
C^\pm$ in $d = 10$, $\left( C^\pm \right)^T = \mp C^\pm$ in $d = 6$,
and finally $\left( C^\pm \right)^T = - C^\pm$ for $d = 4$. Explicitly,
the charge conjugation matrix $C^-_{10}$ is given by $C_{6}^- \otimes
C_{4}^-$ where
\begin{equation}
C_4^- = \gamma^4 \gamma^2
= \left(
\begin{array}{cc}
\epsilon_{\alpha\beta} & 0                      \\
0                      & \epsilon^{\alpha'\beta'}
\end{array}
\right),
\qquad
C_6^- = i \hat\gamma^4 \hat\gamma^5 \hat\gamma^6
= \left(
\begin{array}{cc}
0                  & \delta_{A}^{\phan{A} B}  \\
\delta^{A}_{\phan{A} B} & 0
\end{array}
\right) \ .
\end{equation}

Upon compactification to euclidean $d = (4, 0)$ space the 10-dimensional
Lorentz group $SO(9,1)$ reduces to $SO(4) \times SO(5,1)$ with compact
space-time group $SO(4)$ and $R$-symmetry group $SO(5,1)$. In these
conventions a $32$-component chiral Weyl spinor $\mP$ decomposes as follows
into $8$ and $4$ component chiral-chiral and antichiral-antichiral
spinors
\begin{equation}
\mP
= \left(
\begin{array}{c}
1 \\
0
\end{array}
\right)
\otimes
\left(
\begin{array}{c}
\lambda^{\alpha, A} \\
0
\end{array}
\right)
+
\left(
\begin{array}{c}
0 \\
1
\end{array}
\right)
\otimes
\left(
\begin{array}{c}
0 \\
\bar\lambda_{\alpha', A}
\end{array}
\right) \ ,
\end{equation}
where $\lambda^{\alpha, A}$ ($\alpha = 1,2$) transforms only under
the first $SU(2)$ in $SO(4) = SU(2) \times SU(2)$, while
$\bar\lambda_{\alpha', A}$ changes only under the second $SU (2)$.
Furthermore, $\bar\lambda_{\alpha',A}$ transforms in the complex
conjugate of the $SO(5,1)$ representation of $\lambda^{\alpha,A}$,
namely, $\left( \lambda^\ast \right)^{\alpha,B} \eta^1_{BA}$
transforms like $\bar\lambda_{\alpha,A}$, and the two spinor
representation of $SO(5,1)$ are pseudoreal, i.e.\ $[\hat\gamma^a ,
\hat\gamma^b]^\ast_L \eta^1 = \eta^1 [\hat\gamma^a , \hat\gamma^b]_R$
where $L$ ($R$) denotes the upper (lower) $4$-component spinor.

Substituting these results, the Lagrangian reduces to
\begin{eqnarray}
\label{N4susy}
\cL_E^{\cN = 4}\!\!\! &=&\!\!\! \frac{1}{\g^2} {\rm tr}\
\Bigg\{
\ft12 F_{\mu\nu} F_{\mu\nu}
- i \bar\lambda^{\alpha'}_{A} \not\!\!{\bar\cD}_{\alpha'\beta}
\lambda^{\beta, A}
- i \lambda_{\alpha}^{A} \not\!\!\cD_{\mu}^{\alpha\beta'}
\bar\lambda_{\beta', A}
+ \ft12 \left( \cD_\mu \bar\phi_{AB} \right)
\left( \cD_\mu \phi^{AB} \right)
\nonumber\\
&-&\!\!\! \sqrt{2}
\bar\phi_{AB} \left\{ \lambda^{\alpha, A}, \lambda_{\alpha}^{B} \right\}
- \sqrt{2}
\phi^{AB}
\left\{ \bar\lambda^{\alpha'}_{A}, \bar\lambda_{\alpha', B} \right\}
+ \ft18 \left[ \phi^{AB}, \phi^{CD} \right]
\left[ \bar\phi_{AB}, \bar\phi_{CD} \right]
\Bigg\}\ ,
\end{eqnarray}
where we still use the definition for $\bar\phi_{AB} \equiv \ft12
\epsilon_{ABCD} \phi^{CD}$. These scalars come from the ten-dimensional
gauge field, and can be grouped into $\phi^{AB} = \ft1{\sqrt{2}}
\mS^{a\, AB} A_a $, where $A_a$ are the first six real components of
the ten dimensional gauge field $A_M$. Writing the action in
terms of these 6 scalars, one finds however that one of the fields,
say $A_0$, has a different sign in the kinetic term, which reflects the
$SO(5,1)$ symmetry of the theory.
In the basis with the $\phi^{AB}$
fields, we obtain formally the same action for the Minkowski case
by reducing on a torus with $6$ space coordinates, but the difference
hides in the reality conditions which we will discuss in the next
subsection.

The action is invariant under the dimensionally reduced supersymmetry
transformation rules
\begin{eqnarray}
\delta A_\mu
\!\!\!&=&\!\!\! - i {\bar\zeta}^{\alpha'}_A
\bar\sigma_{\mu\, \alpha'\beta} \lambda^{\beta, A}
+ i \bar\lambda_{\beta', A} \sigma_\mu^{\alpha\beta'} \zeta^A_\alpha\ ,
\nonumber\\
\delta \phi^{AB}
\!\!\!&=&\!\!\! \sqrt{2} \Big( \zeta^{\alpha,A} \lambda^B_\alpha
- \zeta^{\alpha,B} \lambda^A_\alpha
+ \epsilon^{ABCD} \bar\zeta^{\alpha'}_C \bar\lambda_{\alpha',D} \Big)\ ,
\nonumber\\
\delta \lambda^{\alpha,A}
\!\!\!&=&\!\!\! - \ft12
\sigma^{\mu\nu\, \alpha}_{\phan{mn}\beta}
F_{\mu\nu} \zeta^{\beta,A}
- i \sqrt{2}
\bar\zeta_{\alpha',B} \not\!\!\cD^{\alpha\alpha'} \phi^{AB}
+ \left[ \phi^{AB}, \bar\phi_{BC} \right] \zeta^{\alpha,C}\ ,
\nonumber\\
\delta \bar\lambda_{\alpha',A}
\!\!\!&=&\!\!\! - \ft12
\bar\sigma^{\mu\nu\phan{ii}\beta'}_{\phan{m}\alpha'}
F_{\mu\nu} \bar\zeta_{\beta',A}
+ i \sqrt{2}
\zeta^{\alpha,B} \not\!\!{\bar\cD}_{\alpha'\alpha} \bar\phi_{AB}
+ \left[ \bar\phi_{AB}, \phi^{BC} \right] \bar\zeta_{\alpha',C}\ .
\end{eqnarray}
Again, these rules are formally the same as in \eqn{susy-Mink}.

%%%%%%%%%%%%%%%%%%%%%%%%%%%%%%%%%%%%%%%%%%%%%%%%%%%%%%%%%%%%%%%%%%%%%
\subsection{Involution in euclidean space.}
%%%%%%%%%%%%%%%%%%%%%%%%%%%%%%%%%%%%%%%%%%%%%%%%%%%%%%%%%%%%%%%%%%%%%

The Majorana-Weyl condition (\ref{MajoranaWeyl}) on $\mP$ leads in
4D euclidean space to reality conditions on $\lambda^\alpha$
which are independent of those on $\bar\lambda_{\alpha'}$, namely,
\begin{equation}
\label{RealityConds1}
\left( \lambda^{\alpha, A} \right)^\ast
= - \lambda^{\beta, B} \epsilon_{\beta\alpha} \eta^1_{BA} \ ,
\qquad
\left( \bar\lambda_{\alpha', A} \right)^\ast
= - \bar\lambda_{\beta', B} \epsilon^{\beta'\alpha'} \eta^{1, BA} \ .
\end{equation}
These reality conditions are consistent and define a
simplectic Majorana spinor in euclidean space. The $SU (2) \times SU (2)$
covariance of (\ref{RealityConds1}) is obvious from the pseudoreality of
the ${\bf 2}$ of $SU (2)$, but covariance under $SO(5,1)$ can also be
checked (use $[\eta^a, \bar\eta^b] = 0$). Since the first $\mS$ matrix
has an extra factor $i$ in order that $\left( \mG^0 \right)^2 = - 1$,
see (\ref{10Gammas}), the reality condition on $\phi^{AB}$ involves
$\eta^1_{AB}$
\begin{equation}
\label{RealityConds2}
\left( \phi^{AB} \right)^\ast
= \eta^1_{AC} \phi^{CD} \eta^1_{DB}\ .
\end{equation}
The euclidean action in (\ref{N4susy}) is hermitean under the reality
conditions in (\ref{RealityConds1}) and (\ref{RealityConds2}). For
the $\sigma$-matrices, we have under complex conjugation
\begin{equation}
\left( \sigma_{\mu}^{\alpha\beta'} \right)^\ast
= \sigma_{\mu\, \alpha\beta'} \ ,
\qquad
\left( \bar\sigma_{\mu\, \alpha'\beta} \right)^\ast
= \bar\sigma_{\mu}^{\alpha'\beta}\ .
\end{equation}
Obviously due to the nature of the Lorentz group the involution cannot
change one type of indices into another as opposed to the minkowskian
case.

%%%%%%%%%%%%%%%%%%%%%%%%%%%%%%%%%%%%%%%%%%%%%%%%%%%%%%%%%%%%%%%%%%%%%
\section{Instantons in $\cN = 4$ SYM.}
\setcounter{equation}{0}
%%%%%%%%%%%%%%%%%%%%%%%%%%%%%%%%%%%%%%%%%%%%%%%%%%%%%%%%%%%%%%%%%%%%%

One can easily derive the euclidean equations of motion from
(\ref{N4susy})
\begin{eqnarray}
\label{EqOfMot}
&& \cD_\nu F_{\nu\mu} - i
\left\{ \bar\lambda^{\alpha'}_A \bar\sigma_{\mu\, \alpha'\beta} ,
\lambda^{\beta, A} \right\}
- \ft12 \left[ \bar\phi_{AB}, \cD_\mu \phi^{AB} \right] = 0\ ,
\nonumber\\
&& \cD^2 \phi^{AB}
+ \sqrt{2} \left\{ \lambda^{\alpha, A}, \lambda_{\alpha}^B \right\}
+ \ft1{\sqrt{2}} \epsilon^{ABCD}
\left\{ \bar\lambda^{\alpha'}_C, \bar\lambda_{\alpha', D} \right\}
- \ft12 \left[ \bar\phi_{CD}, \left[ \phi^{AB}, \phi^{CD} \right] \right]
= 0\ ,
\nonumber\\
&& \not\!\!{\bar\cD}_{\alpha'\beta} \lambda^{\beta, A}
+ i \sqrt{2} \left[ \phi^{AB}, \bar\lambda_{\alpha', B} \right] = 0 ,
\quad
\not\!\!\cD^{\alpha\beta'} \bar\lambda_{\beta',A}
- i \sqrt{2} \left[ \bar\phi_{AB}, \lambda^{\alpha, B} \right] = 0\ .
\end{eqnarray}
An obvious solution is the configuration ${\mit\Phi} =
\{ A_\mu = A^{\rm cl}_\mu, \phi^{AB} = \lambda^{\alpha, A} =
\bar\lambda_{\alpha', A} = 0 \}$. However, as we have seen in previous
sections, in the background of an anti-instanton the Dirac operator has
zero eigenvalues $\lambda$ satisfying the Dirac equation
$\not\!\!{\bar\cD} \lambda = 0$. There are two
equivalent though formally different approaches to account for these new
configurations.

According to the first, one starts with the above mentioned purely bosonic
configuration ${\mit\Phi}$. Then one must treat the fermionic collective
coordinates in perturbation theory, because they would appear in the
quantum fluctuations as the zero eigenvalue modes of the Dirac operator.
Although it is legitimate to do so, it is inconvenient for the reason
that usually in perturbation theory, one restricts to quadratic order in
the fluctuations (Gaussian approximation). This would however be
insufficient for the fermionic collective coordinates, as we want to
construct its effective action to all orders. In other words, we want to
treat them exactly, and not in perturbation theory.  The second approach
is to include the fermionic instanton in the classical configuration. Doing
this, we automatically treat them exactly and to all orders, as
long as we can find exact solutions to the equations of motion. This
procedure is also more consistent with supersymmetry and the ADHM
construction for multi-instantons. For these reasons, we choose
the second approach.

Now we describe the procedure for constructing the solution to
the classical equations of motion. It is obvious that for a system with
uncoupled scalars and fermions \eqn{scal-ferm}, the configuration
$\{A_\mu^{\rm cl}, \lambda^{\rm cl}, {\bar \lambda}^{\rm cl}
= \phi^{\rm cl} = 0 \}$, with $\lambda^{\rm cl}$ a solution of the Dirac
equation, is an exact solution of the field equations.
As soon as the Yukawa couplings are turned on, the situation changes
drastically as it happens for the case at hand with the system
(\ref{EqOfMot}). The point we want to emphasize here is that the above
instanton configuration with non-zero fermion mode no longer satisfies the
field equations. Indeed, since $\lambda^A$ is turned on, by looking at the
equation for the scalar field in (\ref{EqOfMot}), we conclude that
$\phi^{AB}$ cannot be taken to be zero at quadratic order in Grassmann
collective coordinates (GCC) due to nonvanishing $\left\{
\lambda^{\alpha, A}, \lambda_{\alpha}^B \right\}$. Knowing this, then
also ${\bar \lambda}$ is turned on at cubic order in GCC, as follows from
its field equation. This leads to an iteration procedure which yields a
solution as an expansion in the GCC which stops (for finite $N$) after a
finite number of steps. We will now demonstrate this more explicitly,
first in the case of $SU(2)$, and in subsection 7.2 for $SU(N)$.

%%%%%%%%%%%%%%%%%%%%%%%%%%%%%%%%%%%%%%%%%%%%%%%%%%%%%%%%%%%%%%%%%%%%%
\subsection{Iterative solution in case of $SU (2)$ group.}
%%%%%%%%%%%%%%%%%%%%%%%%%%%%%%%%%%%%%%%%%%%%%%%%%%%%%%%%%%%%%%%%%%%%%

Let us consider first the gauge group $SU (2)$. This is an exceptional
case in the sense that all fermionic zero modes can be generated by means
of supersymmetric and superconformal transformation. E.g.\ if we denote,
suppressing indices, by $Q$ the supersymmetry generators, then a new
solution is given by
\begin{equation}
{\mit\Phi} (\xi) = e^{i \xi \cdot Q}\, {\mit\Phi}\, e^{- i \xi \cdot Q}\ .
\end{equation}
We start generating solutions of the above equations of motion iteratively
in Grassmann parameters from the purely bosonic anti-instanton configuration
${\mit\Phi} = \{ A_\mu = A^{\bar I}_\mu, \phi^{AB} = \lambda^{\alpha, A}
= \bar\lambda_{\alpha', A} = 0 \}$. The exact solution can be expanded as
\begin{equation}
{\mit\Phi} (\xi) = \sum_{n = 0}^{\infty} \frac{1}{n!} \delta^n {\mit\Phi}\ .
\end{equation}
Explicitly we produce from the anti-instanton potential
\begin{equation}
{^{(0)}\!\!A_{\mu\, u}^{\phan{iii}v}}
= {A_{\mu\, u}^{\bar I \phan{ii}v}}
= - \frac{\rho^2}{ x^2 \left( x^2 + \rho^2 \right)}
\bar\sigma^{\mu\nu\phan{n}v}_{\phan{m}u} x_\nu \ ,
\end{equation}
the already known fermionic zero mode
\begin{equation}
{^{(1)}\!\lambda^{\alpha, A}} = - \ft12
\sigma_{\mu\nu\ \, \beta}^{\phan{ii}\alpha}
\xi^{\beta, A}\, {^{(0)}\!F}_{\mu\nu}\ .
\end{equation}
It is obvious that we can only use the $\zeta$ supersymmetry
transformation rules, because ${\bar \zeta}$ leaves the bosonic
anti-instanton invariant. The left superscript on each field indicates
how many GCC it contains. Given this solution for $\lambda$,
we use the supersymmetry transformations, first on $\phi^{AB}$, then
on ${\bar \lambda}_A$, to determine
\begin{eqnarray}
&&{^{(2)}\!\phi^{AB}} = \ft1{\sqrt{2}}
\Big(\xi^A\sigma_{\mu\nu}
\xi^{B}\Big) {^{(0)}\!F}_{\mu\nu}\ , \\
&&{^{(3)}\!\bar\lambda_{\alpha', A}} = \frac{i}{6} \epsilon_{ABCD}
\bar\sigma_{\nu\, \alpha'\beta} \xi^{\beta,B}
\left( \xi^C
\sigma_{\rho\sigma}
\xi^{ D} \right) {^{(0)}\!\cD_\nu} {^{(0)}\!F}_{\rho\sigma}\ .
\end{eqnarray}
When we suppress spinor indices, we understand that they appear in their
natural position, dictated by the sigma-matrices. One can check that these
are indeed the solutions of the field equations to the required order in
the Grassmann parameters, i.e.\
\begin{equation}
\label{LOeom}
\begin{array}{lccl}
{^{(0)}\!\cD_\nu} {^{(0)}\!F_{\nu\mu}} = 0\ ,
& & &
{^{(0)}\!\cD}^2\, {^{(2)}\!\phi^{AB}}
+ \sqrt{2}
\left\{ {^{(1)}\!\lambda^{\alpha, A}},
{^{(1)}\!\lambda_{\alpha}^B} \right\} = 0\ , \\
{^{(0)}\!\!\not\!\!{\bar\cD}}_{\alpha'\beta}
{^{(1)}\!\lambda^{\beta, A}} = 0\ ,
& & &
{^{(0)}\!\!\not\!\!\cD}^{\alpha\beta'}
{^{(3)}\!\bar\lambda_{\beta', A}}
- i \sqrt{2} \left[ {^{(2)}\!\bar\phi_{AB}},
{^{(1)}\!\lambda^{\alpha, B}} \right] = 0\ .
\end{array}
\end{equation}
To check, for example that the field equation for ${\bar \lambda}$ is
indeed satisfied, one can use the field equation for the gauge field,
the Bianchi identity and the Fierz relation for two-component spinors
\begin{equation}
\label{Fierz}
\xi^\alpha_{(1)} \xi_{(2)\beta}
= \frac{1}{2} \delta^\alpha_{\phan{\alpha}\beta}
\left( \xi^\gamma_{(1)} \xi_{(2)\gamma} \right)
- \frac{1}{8}
\sigma_{\mu\nu\ \beta}^{\phan{n}\alpha}
\left( \xi_{(1)\gamma}
\sigma_{\mu\nu\ \, \delta}^{\phan{ii}\gamma}
\xi^{\delta}_{(2)} \right)\ ,
\end{equation}
together with the self-duality properties of the sigma-matrices, like
e.g. $\epsilon_{\nu\theta\rho\sigma} \sigma_{\mu\theta} =
- \delta_{\mu\nu} \sigma_{\rho\sigma}
- \delta_{\mu\rho} \sigma_{\sigma\nu}
- \delta_{\mu\sigma} \sigma_{\nu\rho}$ stemming from the
anti-selfduality of $\sigma_{\mu\nu}$.

Proceeding along these line, we start generating corrections to already
non-vanishing fields. To fourth order in $\xi$, we find for the gauge field
\begin{equation}
{^{(4)}\!A_{\mu}} = \frac{1}{24} \epsilon_{ABCD}
\left( \xi^A
\sigma_{\mu\nu}
\xi^{ B} \right)
\left( \xi^C
\sigma_{\rho\sigma}
\xi^{ D} \right) {^{(0)}\!\cD_\nu} {^{(0)}\!F}_{\rho\sigma}\ .
\end{equation}
From this it follows that the field strength constructed from
${^{(0)}\!A_{\mu}} + {^{(4)}\!A_{\mu}}$ is no longer selfdual.

It is obvious that due to algebraic nature of the procedure one can
easily continue it to higher orders in $\xi$, but for present
purposes (to compute the correlation function of four stress-tensors) it is
sufficient to stop at fourth order. Using the superconformal supersymmetry
transformation laws, one can similarly construct a solution in  $\bar\eta$,
with equation (\ref{xi-eta}) being the first term in the expansion.

It is instructive to compare this method, which is sometimes called
the sweeping procedure, with the explicit solution of the above equations
of motion. For instance, for $\phi^{AB}$ one finds by direct integration
of (\ref{LOeom})
\begin{equation}
{^{(2)}\!\phi^{AB}}_u{}^v
= - \left\{
\frac{8 \sqrt{2}}{(x^2 + \rho^2)^2}
+ \frac{C_1}{x^2 (x^2 + \rho^2)}
+ \frac{C_2}{\rho^4}
\left( 1 + \frac{x^4}{3\rho^2(x^2 + \rho^2)}\right)
\right\}
\xi_u^{[A} \xi^{B],v}\ ,
\end{equation}
where $C_1$ and $C_2$ are integration constants,
$\xi_u=(\xi_1,\xi_2)$, and antisymmetrization is done with weight
one, i.e. $[A,B]=\ft12(AB-BA)$.
The SUSY procedure generates only the first term here. The second one
is a solution everywhere except for origin, so it requires a delta
function source. Since we require regularity of the solution everywhere,
this term must be dropped. The last term has a rising
asymptotic solution in the infrared region. Thus this sweeping procedure
gives only solutions with well defined asymptotical properties.

%%%%%%%%%%%%%%%%%%%%%%%%%%%%%%%%%%%%%%%%%%%%%%%%%%%%%%%%%%%%%%%%%%%%%
\subsection{Extension to $SU (N)$ group.}
%%%%%%%%%%%%%%%%%%%%%%%%%%%%%%%%%%%%%%%%%%%%%%%%%%%%%%%%%%%%%%%%%%%%%

Let us now turn to the construction of the solution to the equation of
motion in the theory with $SU (N)$ gauge group. On top of the $\xi^A$
and $\bar\eta^A$, we have $8 (N - 2)$ extra zero modes denoted by $\mu^A_u$
and $\bar\mu^{Au}$. As we know the Dirac operator develops zero modes
in the anti-instanton background
\begin{equation}
\label{MUmodes}
{^{(1)}\!\lambda^{\alpha,A}}_u{}^v
= \frac{\rho}{\sqrt{x^2 (x^2 + \rho^2)^3}}
\left(
\mu^A_u\ x^{\alpha v} + x^\alpha_{\phan{n}u}\ \bar\mu^{A, v}
\right)\ .
\end{equation}
Recall that in writing down this expression, we have chosen a particular
embedding of the $SU (2)$ singular gauge anti-instanton in the right bottom
corner of the $SU (N)$ matrix.

We want to find the solution of the equations of motion
as an expansion in the Grassmann collective coordinates. In the previous
subsection this was done making use of the broken supersymmetry
transformations for the $\xi$ collective coordinates. Unfortunately, there
is no known symmetry that generates $\mu$ and $\bar\mu$ fermionic zero modes.
Thus we are forced to change our strategy and solve the equations explicitly.
After having obtained the fermionic zero mode \eqn{MUmodes}, the next step is
to solve the scalar field equation of motion.

By computing the fermion bilinear term in the scalar equation (\ref{LOeom}),
we find that the scalar field takes the form
\begin{equation}
{^{(2)}\!\phi^{AB}}_u{}^v
= f (x^2, \rho^2)
\left(
\mu^{[A}_u\ \bar\mu^{B],v}
- \ft12 \left( \mu^{[A}_w\ \bar\mu^{B],w} \right)
\tilde\delta_u^{\phan{u}v}
\right)\ ,
\end{equation}
where we have denoted
\begin{equation}
\tilde\delta_u^{\phan{u}v}
= \left(
\begin{array}{cc}
0 & 0 \\
0 & \1_{[2]\times[2]}
\end{array}
\right)\ .
\end{equation}
This ansatz can now be plugged into the equation of motion for $\phi^{AB}$
and leads to a second order differential equation for the function $f$,
\begin{equation}
x^2 f'' + 2 f' = - \frac{{\sqrt2}\rho^2}{(x^2+\rho^2)^3}\ ,\label{diff-f}
\end{equation}
where the prime stands for the derivative w.r.t.\ $x^2$.
Note that in the covariant derivative in (\ref{LOeom})
the connection drops out since in the $SU(2)$ subspace the colour structure
of $\phi^{AB}$ is simply a unit matrix.
The solution of \eqn{diff-f} is given by
\begin{equation}
f (x^2, \rho^2) = \frac{1}{\sqrt{2}} \frac{1}{x^2 + \rho^2}
+ \frac{C_1}{\rho^2} + \frac{C_2}{x^2}\ , \label{sol-f}
\end{equation}
where $C_1$ and $C_2$ are integration constants corresponding to the
homogeneous solutions. The third term is however not a solution in the
origin, it is rather the scalar Green's function since it satisfies
$\partial^2 \frac{1}{x^2} = - 4 \pi^2 \delta^{(4)} (x)$. We
should therefore drop it for the anti-instanton configuration. The second
term involves the constant  $C_1$, which at this point is not specified.

To demonstrate further the procedure, we
solve for ${^{(3)}\!\bar\lambda}$
and ${^{(4)}\!\!A_\mu}$. To this order, the equations to be solved read
\begin{eqnarray}
\label{barlambdaCorr}
&&{^{(0)}\!\!\not\!\!\cD}^{\alpha\beta'}
{^{(3)}\!\bar\lambda_{\beta', A}}
= i \sqrt{2} \left[
{^{(2)}\!\bar\phi_{AB}}, {^{(1)}\!\lambda^{\alpha, B}}
\right]\ , \nonumber\\
\label{aCorr}
&&{^{(0)}\!\cD}^2\, {^{(4)}\!\!A_\mu}
- {^{(0)}\!\cD_\mu} {^{(0)}\!\cD_\nu} {^{(4)}\!\!A_\nu}
- 2 \left[ {^{(0)}\!F_{\nu\mu}}, {^{(4)}\!\!A_\nu} \right]
= i \left\{ {^{(3)}\!\bar\lambda^{\alpha'}_A}
\bar\sigma_{\mu\, \alpha'\beta} , {^{(1)}\!\lambda^{\beta, A}} \right\}\ .
\end{eqnarray}
In the second equation, we have made use of the fact that the commutator of
the scalar field with its derivative is zero. Calculating the remaining
commutators shows the following structure of the solutions in collective
coordinates,
\begin{eqnarray}
\label{bar-lambda-3}
&&{^{(3)}\!\bar\lambda_{\alpha',A}\,}_u{}^v
= i g (x^2, \rho^2)\
\epsilon_{ABCD} \left( \mu^C_w\ \bar\mu^{D,w} \right)
\left(
\mu_u^B \delta_{\alpha'}^{\phan{\alpha}v}
- \epsilon_{\alpha' u}\bar\mu^{B, v} \right)\ , \\
&& {^{(4)}\!\!A_{\mu\, u}}{}^v
= h (x^2, \rho^2)\
\epsilon_{ABCD}
\left( \mu^A_w\ \bar\mu^{B, w} \right)
\left( \mu^C_z\ \bar\mu^{D, z} \right)
\bar\sigma_{\mu\nu\, u}^{\phan{mm}v} x_\nu\ .
\end{eqnarray}
Notice that ${\bar \lambda}$ is off-diagonal in colour space and
the correction to the anti-instanton gauge field lives in the $SU (2)$
lower diagonal block only.
Introducing $\tilde g = \rho^3 g$, depending
only on the dimensionless variable $y = x^2/\rho^2$ we find the
following differential equation
\begin{equation}
2 \tilde g^\prime (y)
+ \frac{3}{y (1 + y)} \tilde g (y)
= \frac{3}{4}\frac{1}{\sqrt {y(1+y)^5}}
+ \frac{3}{4} C_1 \frac{1}{\sqrt {y(1+y)^3}}\ .
\end{equation}
From this we see that the homogeneous solution for the scalar field
(corresponding
to $C_1$) now enters in the inhomogeneous part for the fermion equation.
This equation can be easily solved,
\begin{equation}
{\tilde g}= - \frac{1}{16} \frac{1 + 3 y}{\sqrt{y^3 (1 + y)^3}}
- \frac{3}{16} C_1\frac{1 + 2y}{\sqrt{y^3(1+y)}}
+ C_3 \left( \frac{1 + y}{y} \right)^{3/2} .
\end{equation}
$C_3$ is the new integration constant. In order to have a solution without
delta function singularities (which corresponds to a behaviour of $y^{-3/2}$
at the origin), we have to choose the integration constant $C_3
= \ft{1}{2^4}(1 + 3 C_1)$. Then the solution reduces to
\begin{equation}
{\tilde g} = \frac{1}{16}
\left\{ (3+y){\sqrt {\frac{y}{(1+y)^3}}}
+ 3C_1\sqrt{\frac{y}{(1+y)}} \right\}\ .
\end{equation}
To complete our analysis, we determine the function $h$ appearing in the
gauge field. Defining
$\tilde h = \rho^4 h$, we find it must satisfy
\begin{equation}
4 y \tilde h^{\prime\prime} (y)
+ 12 \tilde h^\prime (y)
+ \frac{24}{(1 + y)^2} \tilde h (y)
= \frac{2{\tilde g}}{\sqrt {y(1+y)^3}}\ .
\end{equation}
The solution reads
\begin{eqnarray}
\tilde h (y) &=&
- \frac{1}{128} \frac{1 + 4 y}{y^2 ( 1 + y )^2}
- \frac{C_1 \left(2 + 10 y - 3 y^2 - 6 y^2 \ln y \right)}{
64 y^2 ( 1 + y )^2}
+ \frac{ C_3 \left(1 + 6 y - 7 y^2 - 2 y^3 - 6 y^2 \ln y \right)}{
4 y^2 ( 1 + y )^2} \nonumber\\
&+& \frac{C_4}{( 1 + y )^2}
- \frac{ C_5 \left( 1 + 8 y - 8 y^3 - y^4 - 12 y^2 \ln y \right)}{
y^2 ( 1 + y )^2} \ .
\end{eqnarray}
Since the differential equation is of second order, there appear two new
integration constants, $C_4$ and $C_5$. We have still written the
constant $C_3$ explicitly, but its value is related to $C_1$ as
discussed above. In order to get rid of unwanted singularity at the
origin, coming from the $y^{-2}$ dependence, we have to choose
$C_5 = \frac{1 + 2 C_1}{2^7}$. As a surprise, both the next-to-leading
$y^{-1}$-asymptotic terms and the logarithms vanish. Taking the above
values for the constants, the total solution reduces to
\begin{equation}
{\tilde h}(y) = - \frac{1}{128}
\frac{1}{(1 + y)^2}
\Bigg( 2 (7 + 18 C_1 - 64 C_4) - 4 (1 + C_1)y - (1 + 2C_1) y^2 \Bigg)\ .
\end{equation}
It is actually easy to find the structure of the all-order solutions
in the colour space. Simple analysis reveals that
\begin{eqnarray}
&A_\mu
= \left(
\begin{array}{cc}
0 & 0 \\
0 & \star_{[2]\times[2]}
\end{array}
\right)\ , \qquad
\phi^{AB}
= \left(
\begin{array}{cc}
\star_{[N - 2]\times[N - 2]} & 0 \\
0 & -\ft12 \1_{[2]\times[2]} {\rm tr}[ \star ]
\end{array}
\right)&\ , \nonumber\\
&\lambda^{A, \alpha}
= \left(
\begin{array}{cc}
0                        & \star_{[N - 2]\times[2]} \\
\star_{[2]\times[N - 2]} & 0
\end{array}
\right)\ ,&
\end{eqnarray}
and the same for $\bar\lambda$. This means that the scalar fields are
uncharged with respect to the instanton, i.e.\ they commute trivially
with the gauge fields. At each step in the iteration, one generates
new integration constants, some of which are determined by requiring
the absence of delta function sources at the origin.
A detailed analysis of these constants and the asymptotic behaviour
of the fields will be given in a future publication \cite{BelVanNie00a}.

%%%%%%%%%%%%%%%%%%%%%%%%%%%%%%%%%%%%%%%%%%%%%%%%%%%%%%%%%%%%%%%%%%%%%
\subsection{Four-fermion instanton action.}
%%%%%%%%%%%%%%%%%%%%%%%%%%%%%%%%%%%%%%%%%%%%%%%%%%%%%%%%%%%%%%%%%%%%%

We can now evaluate the instanton action for the background solution
constructed in the previous section. First of all, we should mention that
there is no $\xi$ or ${\bar \eta}$ dependence, since these zero modes are
protected by supersymmetry and superconformal symmetry. Therefore, we only
concentrate on the $\mu$ dependence. Now we can use the field equations for
the fermions to see that their kinetic energy cancels against the Yukawa
terms, and this works to all orders in GCC. The resulting instanton action
is therefore
\begin{equation}
\label{S-inst}
\cL_{\rm inst}^{\cN = 4} = \frac{1}{\g^2} {\rm tr}\
\Bigg\{
\ft12 F_{\mu\nu} F_{\mu\nu}
+ \ft12 \left( \partial_\mu \bar\phi_{AB} \right)
\left( \partial_\mu \phi^{AB} \right)
+ \ft18 \left[ \phi^{AB}, \phi^{CD} \right]
\left[ \bar\phi_{AB}, \bar\phi_{CD} \right]
\Bigg\}\ .
\end{equation}
The first non-trivial correction to the instanton action, of order
four in GCC, comes from
\begin{equation}
\cL_{\rm quart} = \frac{1}{\g^2} {\rm tr}\
\Big\{
2\, {^{(0)}\!F_{\mu\nu}} {^{(0)}\!\cD_\mu} {^{(4)}\!A_\nu}
+ \ft12 \left( \partial_\mu {^{(2)}\!\bar\phi_{AB}} \right)
\left( \partial_\mu {^{(2)}\!\phi^{AB}} \right)
 \Big\}\ ,\label{S4}
\end{equation}
since the potential does not contribute to this order. Now we can plug in
the solutions for the gauge and scalar fields into the action
$S_{\rm quart} = - \int d^4 x \cL_{\rm quart}$. Taking into account the
integration constants, we find
\begin{equation}
\label{EffAction}
S_{\rm quart} = \left\{ -\frac{1}{4}+\frac{3}{8}(1+2C_1) \right\}
\frac{\pi^2}{\g^2 \rho^2}
\epsilon_{ABCD}
\left( \mu^A_u\ \bar\mu^{u, B} \right)
\left( \mu_v^C\ \bar\mu^{v, D} \right) \ .
\end{equation}
The first term inside the brackets is the contribution coming from the
scalar fields, and is independent of $C_1$. The second term, proportional
to $(1 + 2C_1)$, is the contribution from the gauge fields. It is
independent of $C_4$ and is entirely determined by the value of the
function $h$ at infinity. This can be seen by realizing that the first
term in \eqn{S4} can be written as a total derivative by using the field
equations for $A_\mu$, so there is only a contribution from the boundary
at spatial infinity.

For the moment, we will determine the constant $C_1$ such that there is no
contribution from the gauge fields, i.e.\ we will set $C_1 = - \ft12$. Again,
a careful analysis of these constants and a discussion about surface terms
that can contribute to the action will be given in \cite{BelVanNie00a}.
Notice that for $C_1 = - \ft12$, the total prefactor is then $- \ft14$ and
is the same as in \cite{DorKhoMatVan98} (up to a sign !).

We conclude this section by discussing the reality conditions on the
fermionic collective coordinates. These follow of course from the reality
conditions on the spinors (\ref{RealityConds1}). Straightforward
substitution of the solutions yields the following results
\begin{eqnarray}
\label{GCCreality}
\begin{array}{lccl}
\left( \xi^{A, \alpha} \right)^*
= - \xi^{\beta, B} \epsilon_{\beta\alpha} \eta^1_{BA}\ ,
& & &
\left( \bar\eta^{A}_{\alpha'} \right) ^*
= - \bar\eta^B_{\beta'} \epsilon^{\beta'\alpha'} \eta^1_{BA}\ , \\
\left( \mu^A_u \right)^*
= \bar\mu^{B, u} \eta^1_{BA}\ ,
& & &
\left( \bar\mu^{A, u} \right)^*
= - \mu^{B}_u \eta^1_{BA}\ .
\end{array}
\end{eqnarray}
The effective action (\ref{EffAction}) is then hermitean w.r.t.\ these
 relations since $\epsilon_{ABCD} \eta^1_{AA'} \eta^1_{BB'}
\eta^1_{CC'} \eta^1_{DD'} = \left( \det\, \eta^1 \right)
\epsilon_{A'B'C'D'}$ and $\det\, \eta^1_{AB} = 1$. The rules given in
\eqn{GCCreality} sometimes simplify calculations. For instance, in
\eqn{bar-lambda-3}
one can determine the ${\bar \mu}$ dependence by only computing the
$\mu$ dependence and using the reality conditions.

%%%%%%%%%%%%%%%%%%%%%%%%%%%%%%%%%%%%%%%%%%%%%%%%%%%%%%%%%%%%%%%%%%%%%
\section{Correlation functions.}
\setcounter{equation}{0}
%%%%%%%%%%%%%%%%%%%%%%%%%%%%%%%%%%%%%%%%%%%%%%%%%%%%%%%%%%%%%%%%%%%%%

Having discussed the zero mode structure, the measure of collective
coordinates and the instanton action, we can now finally turn to the
computation of correlation functions. We recall that the one-anti-instanton
measure, coming from bosonic and fermionic zero modes, for the $\cN = 4$
model is given by
\begin{eqnarray}
\label{N=4meas}
&&\int\,d\cM_{k = - 1} \equiv
\frac{2^{ 4 N + 2 } \pi^{ 4 N - 2 }}{( N - 1 )!( N - 2 )!}
\,\g^{- 4N} e^{- \left( \frac{8 \pi^2}{\g^2} - i \theta \right)}
\int d^4 x_0 \int \frac{d\rho}{\rho^5} \rho^{4N} \\
&&\times\
\int\prod_{A = 1}^4 d^2 \xi^A \, \left( \frac{\g^2}{32\pi^2} \right)^4
\int\prod_{A = 1}^4 d^2 \bar\eta^A \,
\left( \frac{\g^2}{64 \pi^2 \rho^2} \right)^4
\int\prod_{A = 1}^4 \prod_{u = 1}^{N - 2} \,
d \mu_u^A d \bar\mu^{A,u} \,
\left( \frac{\g^2}{2\pi^2} \right)^{4 ( N - 2 )}, \nonumber
\end{eqnarray}
as can easily be seen by combining \eqn{bos-meas} with \eqn{ferm-meas}.
We remind the reader that, as discussed in previous sections, there can be
extra corrections to the measure, proportional to the $\mu$ and ${\bar \mu}$
collective coordinates. These corrections, which are subleading in the
coupling constant, have to our knowledge never been computed, and are
currently under study \cite{BelVanNie00a}.

The measure \eqn{N=4meas} appears in the path integral, and in order to find
a nonvanishing answer, we must insert some fermion fields to saturate the
$\xi^A$ and $\bar\eta^A$ zero modes, otherwise these integrals would yield
zero. This is a generic feature of instanton calculations, and applies
as well to ${\cal N} = 2, 1$ and non-supersymmetric theories. The other
zero modes, $\mu^A_u$ and $\bar\mu^{A,u}$ can be saturated by bringing down
enough powers of the instanton action in the exponential $\exp\left( -
S_{\rm quart} \right)$. We will evaluate the $\mu^A, \bar\mu^A$ integration
below . The total action is $S_{\rm tot} = S_{\rm inst} + S_{\rm quart}$
with the usual one-anti-instanton contribution $S_{\rm inst} = \frac{8
\pi^2}{\g^2} - i\theta$. Higher order terms in the instanton action (starting
from eighth order in GCC) are suppressed. This is because one must bring down
less powers of the instanton action, and since the action has a $1/\g^2$ in
front, one has less powers of the coupling constant as compared with the
leading quartic term. We also repeat that we take the value $C_1 = - \ft12$,
such that
\begin{equation}
S_{\rm quart} = -
\frac{\pi^2}{4\g^2 \rho^2}
\epsilon_{ABCD}
\left( \mu^A_u\ \bar\mu^{u, B} \right)
\left( \mu_v^C\ \bar\mu^{v, D} \right) .
\end{equation}
Notice that in order to saturate the fermionic zero modes, we have to expand
the exponential up to $2N-4$ powers of the instanton action. This will bring
down a factor of $\rho^{-4N+8}$, such that the total $\rho$-dependence of the
measure is independent of $N$, namely $d\rho/\rho^5$.
Combining this with the instanton positions, this is just the measure
of a five-dimensional anti-de-Sitter space, see the next section.

We will now analyze two correlators. The first one involves the insertion of
sixteen fermion fields. This correlator was computed in \cite{DorKhoMatVan98},
and we briefly repeat it below. The second one is the four point function of
energy-momentum tensors. We show how the fermionic zero modes are saturated
and we outline the computation of the full correlator.

%%%%%%%%%%%%%%%%%%%%%%%%%%%%%%%%%%%%%%%%%%%%%%%%%%%%%%%%%%%%%%%%%%%%%
\subsection{$\langle \Lambda^{16} \rangle$ correlator.}
%%%%%%%%%%%%%%%%%%%%%%%%%%%%%%%%%%%%%%%%%%%%%%%%%%%%%%%%%%%%%%%%%%%%%

Because we have integrated out the gauge orientation zero modes in the
measure \eqn{N=4meas}, we must compute correlation functions of gauge
invariant objects. Since there are 16 fermionic zero modes protected
from lifting by super(conformal) symmetry, they have to be saturated by
inserting appropriate fermionic operators. The gluinos
$\lambda^A$ are not gauge invariant, but a suitable gauge invariant
composite operator is the expression
\begin{equation}
\label{Lambda}
\Lambda_\alpha^A\ = \frac{1}{2\g^2}
\sigma_{\mu\nu\ \, \alpha}^{\phan{ii}\beta}
\tr \left\{ F_{\mu\nu} \, \lambda_\beta^A \right\}\ .
\end{equation}
We could equally well have considered fermionic bilinears contracted
to gauge invariant Lorentz scalars as was originally done in
\cite{Hooft1,Hooft2}, and in $\cN = 4$ in \cite{BGKR,DorHolKhoMatVan99}.
We can write explicitly how this operator looks like in the anti-instanton
background.  Using \eqn{xi-eta} and \eqn{inst-dens}, we then find
\begin{equation}
\Lambda^{\alpha, A} (x)
= - \frac{96}{\g^2} \,
\left(
\xi^{\alpha, A} - \bar\eta_{\alpha'}^A
\bar\sigma_\mu^{\alpha'\alpha} \, ( x^\mu - x_0^\mu )
\right)
\frac{\rho^4}{ [ (x - x_0)^2 + \rho^2 ]^4}\ .
\end{equation}
This is actually an exact formula, and there is no contribution from the
$\mu$ GCC to all orders. This follows from the fact that the field strength
is diagonal and the gluino is off-diagonal in colour space. Due to this, all
higher point functions, with more than sixteen $\Lambda$'s, will be zero. The
only contribution to $\langle \Lambda^{16}\rangle$ comes from taking
$\Lambda$ to be linear in the $\xi$ or ${\bar \eta}$ zero mode.
Putting this all together, we now insert 16 copies of the composite
fermion, at 16 space points $x_i$, into the one-anti-instanton
measure \eqn{N=4meas}, and multiply by $\exp\left( -
S_{\rm quart} \right)$.

To evaluate this correlator, we first discuss the integration over the
lifted zero modes $\mu^A_u, \bar\mu^{Au}$. Define $I_N$ to be the
(un-normalized) contribution of the lifted modes to the correlator
\begin{equation}
\label{IN}
I_N \equiv \int\prod_{A = 1}^4 \prod_{u = 1}^{N - 2}
d\mu_u^A d\bar\mu^{A,u} \, e^{-S_{\rm quart}} .
\end{equation}
Explicit calculation for $N=3$ shows $I_3 = \frac{3\pi^4}{
\rho^4 \g^4}$. To evaluate $I_N$ for arbitrary $N$, it is helpful to
rewrite the quadrilinear term $S_{\rm quart}$ as a quadratic form. To
this end we introduce six independent auxiliary bosonic variables
$\chi_{AB} = - \chi_{BA},$ and substitute into \eqn{IN} the integral
representation \cite{DorKhoMatVan98}
\begin{equation}
\label{auxdef}
e^{- S_{\rm quart}} =
-\frac{ \rho^6 \g^6}{2^9 \pi^6}
\int\prod_{1 \le A' < B' \le 4} \, d \chi_{A'B'} \,
\exp\left(
\frac{-\rho^2 \g^2}{32 \pi^2} \epsilon_{ABCD} \chi_{AB} \chi_{CD}
+ \ft12 \, \chi_{AB} \Lambda^{AB}_N ( \mu, \bar\mu )
\right) ,
\end{equation}
where we have introduced the object $\Lambda^{AB}_N (\mu, \bar\mu)
= \frac{1}{2\sqrt{2}} \left( \mu_u^A \bar\mu^{B,u} - \mu_u^B \bar\mu^{A,u}
\right)$ to denote the fermion bilinear that couples to $\chi_{AB}$. Our
strategy is to perform only the integrations over $\mu^A_{N - 2}$ and
$\bar\mu^{A, N - 2}$, and thereby deduce a recursion relation, i.e.\ an
equation between $I_N$ and $I_{N - 1}$. Accordingly we break out these
terms from $\Lambda^N$:
\begin{equation}
\label{breakout}
\Lambda^{AB}_N (\mu, \bar\mu) =
\Lambda^{AB}_{N - 1} (\mu, \bar\mu)
+ \frac{1}{2 \sqrt{2}} \left( \mu^A_{N - 2} \bar\mu^{B, N - 2}
- \mu^B_{N - 2} \bar\mu^{A, N - 2} \right) .
\end{equation}
Now, the $\{ \mu^A_{N - 2}, \bar\mu^{A, N - 2} \}$ integration in
\eqn{auxdef} brings down a factor of $\frac{1}{64} \det\, \chi_{AB}$.
Next we exploit the fact that the determinant of a four-dimensional
antisymmetric matrix is a total square,
\begin{equation}
\det\, \chi_{AB} =
\left( \ft18 \, \epsilon_{ABCD} \chi_{AB} \chi_{CD} \right)^2 .
\end{equation}
Since the right-hand side of this determinant equation is proportional to
the square of the first term in \eqn{auxdef}, the result of the
$\{ \mu^A_{N - 2}, \bar\mu^{A, N - 2} \}$ integration can be rewritten
as a parametric second derivative relating $I_N$ to $I_{N - 1}\,$,
namely \cite{DorKhoMatVan98}
\begin{equation}
\label{recursion}
I_N = \ft14 \, \pi^4 \rho^6 \g^6
\frac{\partial^2}{\partial( \rho^2 \g^2 )^2}
\, \left( (\rho^2 \g^2)^{-3} I_{N - 1} \right) .
\end{equation}
The insertion of $( \rho^2 \g^2 )^{-3}$ inside the parentheses ensures
the derivatives to act only on the exponent of \eqn{auxdef}, and not on the
prefactor. This recursion relation, combined with the initial condition
for $I_3$, finally gives
\begin{equation}
I_N = \ft12 \, (2 N - 2)! \,
\left(
\frac{\pi^2}{2 \rho^2 \g^2} \right)^{2 N - 4} .
\end{equation}
Combining everything together, we find for the one-anti-instanton
contribution to the 16-fermion correlator in $\cN = 4$ SYM theory:
\begin{eqnarray}
\label{16corr}
\langle \Lambda_{\alpha_1}^1 (x_1) \cdots
\Lambda_{\alpha_{16}}^4(x_{16}) \rangle
&=& C_N e^{- \left( \frac{8 \pi^2}{\g^2} - i \theta \right)}
\int\, d^4 x_0 \, \frac{d\rho}{\rho^5}
\int\prod_{A = 1}^4 d^2\xi^A \, d^2 \bar\eta^A
\nonumber\\
&&\times\ \left( \xi^1_{\alpha_1} - \bar\eta_{\alpha'}^1
\bar\sigma_{\mu \ \alpha_1}^{\phan{.}\alpha'}
( x_1^\mu - x_0^\mu ) \right) K_4 (x_0, \rho; x_1, 0) \\
&&\times\cdots\times\
\left( \xi^4_{\alpha_{16}} - \bar\eta_{\alpha'}^4
\bar\sigma_{\mu \ \alpha_{16}}^{\phan{.}\alpha'}
(x_{16}^\mu - x_0^\mu ) \right) K_4 (x_0, \rho; x_{16}, 0) \nonumber \ ,
\end{eqnarray}
where we have denoted
\begin{equation}
\label{adsPropag}
K_4(x_0, \rho; x_i, 0) = \frac{\rho^4}{[(x_i - x_0)^2 + \rho^2]^4}\ ,
\end{equation}
and the overall constant $C_N$ is given by
\begin{equation}
C_N = \g^{-24} \, \frac{2^{-2 N}(2 N - 2)!}{(N - 1)! (N - 2)!} \,
2^{57} \, 3^{16} \, \pi^{-10} .
\end{equation}
The large $N$ limit gives by means of Stirling's formula
\begin{equation}
\label{16constant}
C_N \rightarrow \g^{-24} \, \sqrt{N} \,
2^{55} \, 3^{16} \, \pi^{-21/2} .
\end{equation}
In principle we would have to do the integrations over the $\xi$
and ${\bar \eta}$ zero modes. This would just give a numerical tensor
in spinor indices and sigma matrices, but we refrain from giving its
explicit expression. Also, we should do the $\rho$ integration. This
is not straightforward, but one can check by power counting that the
result is convergent and finite. The integration over $x_0$ is left
over. The final expression for the correlator can then be seen to
induce a sixteen fermion vertex in the effective Lagrangian, which is
integrated over $x_0$.

%%%%%%%%%%%%%%%%%%%%%%%%%%%%%%%%%%%%%%%%%%%%%%%%%%%%%%%%%%%%%%%%%%%%%
\subsection{$\langle {\mit\Theta}^4 \rangle$ correlator.}
%%%%%%%%%%%%%%%%%%%%%%%%%%%%%%%%%%%%%%%%%%%%%%%%%%%%%%%%%%%%%%%%%%%%%

In this section we study the four-point Green's function of energy-momentum
tensors ${\mit\Theta}_{\mu\nu}$.
The improved (traceless and symmetric) energy momentum tensor for the
model is
\begin{eqnarray}
\label{EMtensor}
{\mit\Theta}_{\mu\nu} \!\!\!&=&\!\!\! \frac{1}{\g^2} {\rm tr}
\Bigg\{
2 \left( F_{\mu \rho} F_{\nu \rho}
- \frac{1}{4} \delta_{\mu\nu} F_{\rho\sigma} F_{\rho\sigma}
\right)
- i \bar\lambda^{\alpha'}_{A} \bar\sigma_{(\mu\, \alpha'\beta}
\cD_{\nu )} \lambda^{\beta, A}
- i \lambda_{\alpha}^{A} {\sigma_{(\mu}}^{\alpha\beta'}
\cD_{\nu )} \bar\lambda_{\beta', A} \\
&+&\!\!\! \left( \cD_\mu \bar\phi_{AB} \right)
\left( \cD_\nu \phi^{AB} \right)
- \frac{1}{2} \delta_{\mu\nu} \left( \cD_\rho \bar\phi_{AB} \right)
\left( \cD_\rho \phi^{AB} \right)
- \ft18 \delta_{\mu\nu} \left[ \phi^{AB}, \phi^{CD} \right]
\left[ \bar\phi_{AB}, \bar\phi_{CD} \right] \nonumber\\
&-&\!\!\! \frac{1}{6} \left( \partial_\mu \partial_\nu
- \delta_{\mu\nu} \partial^2 \right) \bar\phi_{AB} \phi^{AB}
\Bigg\} , \nonumber
\end{eqnarray}
where the last term is an improvement \cite{Jac70} stemming from the
addendum $R \bar\phi \phi$ to the Lagrangian of $\cN = 4$ super-Yang-Mills
coupled selfconsistently to conformal supergravity \cite{BRdW}.
Symmetrization is done with weight one, $(\mu, \nu) = \ft12 \left\{ \mu
\nu + \nu \mu \right\}$. We dropped the equations of motion for fermions
and gauge fields in (\ref{EMtensor}). This tensor is conserved and
traceless upon using the equations of motion.

We now evaluate this expression in the anti-instanton background.
First we concentrate on the $\mu$-zero modes, and we will show that there
is no $\mu$-dependence, as was also the case for the field $\Lambda$ in the
previous subsection. This can be seen by the following argument. Since the
only possible tensor structure of ${\mit\Theta}_{\mu\nu}$ which may contain
the collective coordinates is the traceless tensor ${\mit\Delta}_{\mu\nu}
\equiv \frac{x^2}{4}
\delta_{\mu\nu} - x_\mu x_\nu$, it must take the form
\begin{equation}
{\mit \Theta}_{\mu\nu} = t(x) {\mit\Delta}_{\mu\nu}
\epsilon_{ABCD}(\mu^A{\bar \mu}^B)(\mu^C {\bar \mu}^D)\ .
\end{equation}
Now, from the conservation of energy-momentum
tensor we derive a differential equation for the $x$-dependent function,
$t(x)$, which is solved by $t(x) = c/x^6$, with $c$ an arbitrary
($\rho$-dependent) constant.
By looking at the explicit form of the instanton solution,
it is simple to see that such an $x$-dependence can never be produced.
Therefore, the only possibility is that $c = 0$. Explicit
calculation confirms this observation, as there is a subtle cancellation of
the $\mu$-dependence between the bosons and fermions, showing that indeed
 ${\mit\Theta}_{\mu\nu}$ is $\mu$ and $\bar\mu$ independent.

The story is different for the $\xi$ and ${\bar \eta}$ GCC, as we now can
construct different possible tensor structures which are both traceless and
conserved. There are three different classes of terms, one which has four
$\xi$'s, one with four ${\bar \eta}$'s and one with mixed $\xi$ and
${\bar \eta}$. The stress tensor is obtained by taking the derivative of
the action with respect to the metric. One expects that in curved space,
the action does depend on the $\xi$ and ${\bar \eta}$ GCC, in other words,
in a general curved space, these fermion zero modes are not protected by
supersymmetry.  An explicit calculation supports this, and using the results
of section 7.1, we find for the $\xi$ mode contribution
\begin{equation}
{\mit\Theta}^{(\xi)}_{\mu\nu}
= \frac{1}{\g^2}
\frac{3 \cdot 2^8 \cdot \rho^4}{[ (x - x_0)^2 + \rho^2 ]^6}
\epsilon_{ABCD} (x - x_0)_\rho (x - x_0)_\sigma
\bigg( \xi^A \sigma_{\mu\rho}
\xi^{B} \bigg) \bigg( \xi^C \sigma_{\nu\sigma}
\xi^{D} \bigg)\ .\label{T-xi}
\end{equation}
We repeat that when spinor indices are not explicitly written, they are in
their natural position dictated by the sigma matrices. To obtain this result,
we have made use of the Fierz relation \eqn{Fierz} and the identity
$\epsilon_{ABCD}(\xi^A \sigma_{\mu\rho}\xi^B) (\xi^C\sigma_{\rho\nu}\xi^D)=0$
which follows from anti-selfduality of $\sigma_{\mu\nu}$. The expression
\eqn{T-xi} is then easily seen to be traceless and conserved.

A similar analysis can be made for the terms involving ${\bar \eta}$.
One would first have to compute the ${\bar \eta}$ dependence of all the
fields using the superconformal supersymmetry transformations, along the
same lines as in section 7.1. For present illustrative purposes we do not
need it. Having the four GCC in the energy-momentum we can saturate the
$\xi$ and $\bar\eta$ measure by computing the four-point function. The
$\mu$ and $\bar\mu$ coordinates are integrated out in the same way as was
done in the previous section and finally we get the result
\begin{eqnarray}
\label{T4}
&&\langle
{\mit\Theta}_{\mu_1\nu_1} (x_1)
\dots
{\mit\Theta}_{\mu_4\nu_4} (x_4)
\rangle
=
\widetilde C_N e^{- \left( \frac{8 \pi^2}{\g^2} - i \theta \right)}
\int\, d^4x_0 \, \frac{d\rho}{\rho^5}
\int\prod_{A = 1}^4 d^2 \xi^A \, d^2 \bar\eta^A
\prod_{j = 1}^{4}
\frac{\rho^4}{[ (x_j - x_0)^2 + \rho^2 ]^6} \nonumber\\
&&\qquad\qquad\qquad\qquad\times\epsilon_{ABCD}
\Big\{ (x_j - x_0)_{\rho_j} (x_j - x_0)_{\sigma_j}
\bigg( \xi^A\sigma_{\mu_j\rho_j}
\xi^{B} \bigg) \bigg( \xi^C \sigma_{\nu_j\sigma_j} \xi^{D} \bigg)
 + \dots \Big\}\ ,
\end{eqnarray}
where the dots stand for terms proportional to ${\bar \eta}$.
The normalization constant $\widetilde C_N$ reads in the large $N$ limit
\begin{equation}
\label{T4coeff}
\widetilde C_N =
{\rm const} \cdot \g^0 \, \frac{2^{-2 N} (2 N - 2)!}{(N - 1)! (N - 2)!}
\to \g^0 \sqrt{N}\ ,
\end{equation}
up to an $N$-independent constant. Thus in the large $N$
limit the four-point correlation function of energy-momentum
tensors has the same scaling in $N$ as the sixteen fermion correlator
discussed in the previous section.

%%%%%%%%%%%%%%%%%%%%%%%%%%%%%%%%%%%%%%%%%%%%%%%%%%%%%%%%%%%%%%%%%%%%%
\section{D-instantons in IIB supergravity.}
\setcounter{equation}{0}
%%%%%%%%%%%%%%%%%%%%%%%%%%%%%%%%%%%%%%%%%%%%%%%%%%%%%%%%%%%%%%%%%%%%%

After having studied instantons in (supersymmetric) Yang-Mills theories,
we will now discuss instantons in a (particular) supergravity theory,
which is related to YM theories via the AdS/CFT correspondence
\cite{ads/cft}. We consider IIB supergravity in ten dimensions, where the
bosonic fields are given by the ten-dimensional metric $g_{\mu\nu}$, the
dilaton $\phi$ and axion $a$, scalar and pseudoscalar respectively, and
some tensor fields which we shall specify later \cite{Sch83}. The
ten-dimensional action for these fields is
\begin{equation}
S_M^{\rm boson}
= \int \, d^{10}x \sqrt{g}
\left\{ \cR
- \ft12 \left( \partial_\mu\phi \right) \left( \partial^\mu\phi \right)
- \ft12 e^{2\phi} \left( \partial_\mu a \right)
\left( \partial^\mu a \right) \right\} \ .
\end{equation}
This action is written down in Einstein frame with minkowskian
signature. Our goal is now to discuss instantons in this system, which
requires the euclidean formulation of IIB supergravity. 
In distinction to $\cN = 4$ SYM, IIB supergravity can not be
obtained by dimensional reduction (over the time coordinate) of yet another
theory in higher dimensions. In fact, in euclidean space, there is no action
which is supersymmetric and real at the same time. This is
in agreement with the fact that there is no real form of the supersymmetry
algebra $OSp(1|32)$ with signature $(10,0)$, see e.g.\ \cite{BVP}.
The way to proceed then is to make all the fields complex, keeping the action
formally the same as in minkowskian space-time. This action will not be
real or hermitean, but depends holomorphically on all the fields.

To obtain (the bosonic part of) euclidean IIB supergravity one must flip
the sign in front of the kinetic energy for the axion field.
This sign change is explained by the argument \cite{Nie96} that
a pseudoscalar receives
a factor of $i$ after the Wick rotation from minkowskian to euclidean space,
$t\rightarrow \tau = - i t$. Since pseudoscalars can always be written in terms 
of scalars $S_i$, $i = 0, \dots, 9$ as
\begin{equation}
a = \epsilon^{\mu_0 \mu_1 \dots \mu_9} \left( \partial_{\mu_0} S_0 \right)
\dots \left( \partial_{\mu_9} S_{9} \right) \ .
\end{equation}
It becomes clear that the derivative associated with the time coordinate picks up 
a factor of $i$ after the Wick rotation. As a consequence, in the euclidean
theory, the sign in front of the axion kinetic Lagrangian changes,
\begin{equation}
\label{euclIIB}
- S_E^{\rm boson}
= \int \, d^{10}x \sqrt{g}
\left\{ \cR - \ft12 \left( \partial_\mu \phi \right)
\left( \partial^\mu \phi \right)
+ \ft12 e^{2\phi} \left( \partial_\mu a \right)
\left( \partial^\mu a \right) \right\} \ .
\end{equation}
This prescription is consistent with the procedure of making all the fields
complex. The sign change is then explained by taking only the real part of
the dilaton and the imaginary part of the axion to be non-zero.

The field equations that follow from \eqn{euclIIB} are (neglecting the
fermionic sector)
\begin{equation}
\cR_{\mu\nu} = \ft12
\left( \partial_\mu \phi \right) \left( \partial_\nu \phi \right)
- \ft12 e^{2\phi} \left( \partial_\mu a \right)
\left( \partial_\nu a \right) \ ,
\qquad
\nabla_\mu \left( e^{2\phi} \partial^\mu a \right) = 0 \ ,
\qquad
\nabla^2 \phi + e^{2\phi}
\left( \partial_\mu a \right) \left( \partial^\mu a \right) = 0 \ .
\end{equation}
The action contains further two-tensors $A_{\mu\nu} = - A_{\nu\mu}$
and a selfdual rank-five field strength, $F_{\mu_1\dots\mu_5}$. The latter
contributes only to the first of the above field equations,
\begin{equation}
\label{RF}
\cR_{\mu\nu} = \ft12 \left( \partial_\mu \phi \right)
\left( \partial_\nu \phi  \right)
- \ft12 e^{2\phi} \left( \partial_\mu a \right)
\left( \partial_\nu a \right)
+ \ft16 F_{\mu\mu_1\dots\mu_4} {F^{\mu_1\dots\mu_4}}_\nu\ .
\end{equation}
As we will see in subsection \ref{AdSinst}, this $F_{\mu_1\dots\mu_5}$  
becomes relevant when discussing instanton solutions. Again, in euclidean 
space, one must take notice of the fact that there exist
no real selfdual five-forms in ten dimensions, so we take the
strategy of working with a complex field strength.

The aim is now to discuss solutions of these equations by choosing a
particular background for the ten dimensional space-time. We will discuss
two examples which preserve the maximal number of supersymmetries, the
first one will be flat ${\bf R}^{10}$, and the other one
contains anti-de Sitter (AdS) space, namely $AdS_5 \times S^5$. The
D-instanton solution was found by Gibbons et al.\ in
\cite{Dinst}, see also Green and Gutperle in \cite{GG}, on which the
remainder of this section is heavily based. We will however concentrate
purely on the bosonic sector of the theory. An analysis of the fermionic
zero modes can be found in \cite{Dinst,GG,BGKR}.

%%%%%%%%%%%%%%%%%%%%%%%%%%%%%%%%%%%%%%%%%%%%%%%%%%%%%%%%%%%%%%%%%%%%%
\subsection{D-instantons in flat ${\bf R}^{10}$.}
%%%%%%%%%%%%%%%%%%%%%%%%%%%%%%%%%%%%%%%%%%%%%%%%%%%%%%%%%%%%%%%%%%%%%

In flat euclidean space, we set the rank-five field
strength equal to zero. The field equations then become
\begin{equation}
\left( \partial_\mu \phi \right)
\left( \partial_\nu \phi \right)
= e^{2\phi} \left( \partial_\mu a \right) \left( \partial_\nu a \right)\ ,
\qquad
\partial_\mu \left( e^{2\phi} \partial_\mu a \right) = 0\ ,
\qquad
\partial^2 \phi = - e^{2\phi}
\left( \partial_\mu a \right) \left( \partial_\mu a \right)\ .
\end{equation}
By taking the trace of the first equation and comparing it with the
third one, we find
\begin{equation}
\label{DinstEq}
\left( \partial_\mu \phi \right)^2
= - \partial^2 \phi
\qquad\Rightarrow\qquad \partial^2 \left( e^\phi \right) = 0\ .
\end{equation}
A spherically symmetric solution to this equation is
\begin{equation}
\label{dil-eqn}
e^\phi = e^{\phi_\infty} + \frac{c}{r^8}\ .
\end{equation}
Here, $\g_s \equiv \exp(\phi_{\infty})$ and $c$ are integration
constants, the first one being identified with the string coupling,
which is the value of the dilaton at infinity. Obviously \eqn{DinstEq}
is a solution everywhere except of the origin where it solves
(\ref{DinstEq}) with a delta function source, $\delta^{(10)} (x)$. One may
add another term to the action of the form $\int d^{(10)} x\, 
\delta^{(10)} (x) e^{- \phi}$. This term cancels the singularity with 
$\delta^{(10)} (x)$ in the field equations. The physical meaning of this 
term is that we now have a new object in the theory at $x = 0$, namely
the ``D-instanton''. This is different from YM-instantons, where we require 
regularity of the solution everywhere. D-instantons, or D-branes in general, 
as solutions of the supergravity equations of motion typically have delta 
function sources, but these singularities may be resolved in string theory.

The solution for the axion field equation $\partial_\mu a = \mp \partial_\mu
\exp\left( - \phi \right)$ is
\begin{equation}
\label{Axion}
a - a_\infty = \mp \left( e^{-\phi} - e^{-\phi_{\infty}} \right)\ ,
\end{equation}
where $a_{\infty}$ is again an integration constant, namely the value of
the axion at infinity. The plus or minus sign refers to the D-instanton
and D-anti-instanton respectively. One can actually write down a more general 
solution for the dilaton equation
\begin{equation}
\label{Dinst}
e^\phi = e^{\phi_\infty} + \frac{c}{|x - x_0|^8}\ ,
\end{equation}
and similarly for the axion field. The coordinates $x^\mu$ are just the
coordinates of ${\bf R}^{10}$. Then there are
ten bosonic collective coordinates
$x_0^\mu$, denoting the position of the D-instanton in ${\bf R}^{10}$.

Now we want to determine the instanton action. By plugging the solution given 
above into the action, one immediately sees that it gives zero (also for the 
$\delta$-function term), hence the instanton action vanishes. This is usually 
not the case for instantons, since they have finite but non-zero action. The
resolution is that we should have taken a boundary term
into account, of the form 
\begin{equation}
S_{\rm surf} = \int\, d^{10}x \, \partial_\mu \left\{ 
\sqrt{g} g^{\mu\nu} \exp 
\left( 2\phi \right) \left( a \partial_\nu a
\right) \right\} \ .
\end{equation}
Its origin was discussed in \cite{Dinst,GG} and we will
not repeat it here. This surface term is non-zero when evaluated in the
instanton background, and is proportional to the constant $c$, namely $S_{\rm cl} 
= 8 c {\rm Vol}\ S^9$. There is a conserved current $j_\mu = e^{2 \phi} 
\partial_\mu a$, whose charge is $Q = \int j_\mu d {\mit\Omega}^\mu = \mp 8 c 
\g_s {\rm Vol}\ S^9$. Hence $S = |Q|/\g_s$. Since D-instantons are  -1 
$p$-branes (which correspond to antisymmetric tensors with $p + 1$ indices),
their duals (which are field strengths with 9 indices) define gauge
potentials with 8 indices. These can couple to curls of D7-branes. Considering
the ``magnetic'' D7-brane in the presence of the ``electric'' D-instanton
one obtains the usual Dirac quantization condition
\begin{equation}
Q = 2 \pi k , \qquad k \in {\bf Z} \ .
\end{equation}
With this quantization condition, the action becomes
\begin{equation}
S_{\rm D-inst} = \frac{2 \pi |k|}{\g_s}\ .
\end{equation}
This is the same as for the YM-instanton action (without the $\theta$ angle)
upon identifying
\begin{equation}
\g_s=\frac{\g^2}{4\pi}\ ,
\end{equation}
with YM-coupling constant $\g$.

%%%%%%%%%%%%%%%%%%%%%%%%%%%%%%%%%%%%%%%%%%%%%%%%%%%%%%%%%%%%%%%%%%%%%
\subsection{D-instantons in $AdS_5\times S^5$.}
\label{AdSinst}
%%%%%%%%%%%%%%%%%%%%%%%%%%%%%%%%%%%%%%%%%%%%%%%%%%%%%%%%%%%%%%%%%%%%%

In this section we discuss D-instantons in IIB supergravity in a
background different from flat ten-dimensional space. Instead, we
choose the space $AdS_5 \times S^5$, and we want to solve the equations
of motion for the dilaton and axion in this background.

We start with some elementary facts about $AdS_5$, which is defined
as the hypersurface embedded in six-dimensional flat space-time by the
equation
\begin{equation}
\label{ads}
- \left( X^0 \right)^2 + \left( X^1 \right)^2
+ \left( X^2 \right)^2 + \left( X^3 \right)^2
+ \left( X^4 \right)^2 - \left( X^5 \right)^2
= - R^2\ .
\end{equation}
This surface defines a five-dimensional non-compact space with ``radius''
$R$. It has an isometry group $SO(4,2)$ which is the same as the conformal
group in four dimensional Minkowski space-time. The $AdS_5$ metric can be
obtained from the six dimensional flat metric with signature
\begin{equation}
ds^2 = - \left( dX^0 \right)^2 + \left( dX^1 \right)^2
+ \left( dX^2 \right)^2 + \left( dX^3 \right)^2
+ \left( dX^4 \right)^2 - \left( dX^5 \right)^2\ ,
\end{equation}
upon using the constraint \eqn{ads}. Defining the coordinates
\begin{equation}
U = X^4 + X^5\ , \qquad x^\mu = \frac{X^\mu R}{U}\ ,
\qquad\mbox{with}\qquad \mu = 0,1,2,3\ ,
\end{equation}
the $AdS_5$ metric can be written as
\begin{equation}
ds^2 = \frac{U^2}{R^2} \eta_{\mu\nu} d x^\mu d x^\nu
+ \frac{R^2}{U^2} \left( d U \right)^2 \ .
\end{equation}
In this expression, we have used the minkowskian metric $\eta_{\mu\nu}$
to contract the indices. But in fact, since instantons live in euclidean
space, we should take the euclidean version of the above metric and
replace $\eta_{\mu\nu}$ by $\delta_{\mu\nu}$. Defining the variable $\rho
\equiv R^2/U$, the metric takes the form
\begin{equation}
ds^2 = \frac{R^2}{\rho^2}\left\{ \delta_{\mu\nu}
dx_\mu dx_\nu + d\rho^2 \right\}\ .
\end{equation}
Notice that now there is just an overall factor in front of a
five-dimensional flat metric. Spaces with such a metric are called
conformally flat. We can now construct the invariant volume element of
$AdS_5$, which is
\begin{equation}
\int\, d^4x d\rho {\sqrt g} = R^5 \int\,d^4x \frac{d\rho}{\rho^5}\ .
\end{equation}
This is precisely the same expression (up to the prefactor $R^5$) as
obtained from the collective coordinate measure in $\cN = 4$ SYM theory
after integrating out the $\mu^A, {\bar\mu}^A$ fermionic collective
coordinates, combining \eqn{bos-meas}, \eqn{ferm-meas} and \eqn{recursion}.

The metric on the full ten-dimensional $AdS_5\times S^5$ space is
\begin{eqnarray}
\label{adss5metric}
d s^2\!\!\!&=&\!\!\!
\frac{R^2}{\rho^2} \left\{ dx_\mu dx_\mu + d \rho^2 \right\}
+ R^2 \, d S^5
= \frac{R^2}{\rho^2} dx_\mu dx_\mu
+ \frac{R^2}{\rho^2} \left\{ d\rho^2 + \rho^2 dS^5 \right\} \nonumber\\
&=&\!\!\! \frac{R^2}{\rho^2}
\left\{ dx_\mu dx_\mu + d y^i d y^i \right\}\ ,
\end{eqnarray}
which is conformally equivalent to ${\bf R}^{10}$.
We have taken the radius of the five-sphere to be the same as that of
$AdS_5$, and in the last equation we have introduced coordinates
$\{ y^i; i = 1,\dots,6 \}$ on ${\bf R}^6$, with $\rho^2= y^i y^i$.

Computing the Ricci tensor of $AdS_5\times S^5$, one finds that the only
non-zero components are given by
\begin{equation}
\cR_{MN}
= - \frac{4}{R^2} g_{MN}\ , \qquad \cR_{mn} = \frac{4}{R^2} g_{mn}\ ,
\end{equation}
where $M, N =1, \dots, 5$ run over the coordinates $(x^\mu,\rho)$, and
$m,n = 1,\dots,5$ label the angular coordinates of the five-sphere.
Our aim is to solve the field equations in this background. This is
different from the case of flat ${\bf R}^{10}$ in the sense that
now the Ricci tensor
does not vanish in the field equations. However, as mentioned above,
there is also a rank-five selfdual field strength $F_{\mu_1\dots\mu_5}$
which can compensate this effect. Indeed, if we choose the non-vanishing
components of this tensor to be (with flat indices to avoid having to write
determinants of vielbeins)
\begin{equation}
F_{MNPQS} = i \frac{1}{R} \epsilon_{MNPQS}\ ,
\qquad
F_{mnpqs} = \frac{1}{R} \epsilon_{mnpqs}\ ,
\end{equation}
we find a cancellation between the Ricci tensor and this five-form in
\eqn{RF}. Notice the factor $i$ in the AdS part, which is due to the Wick
rotation of the time coordinate $X^0$.

The field equations now reduce to
\begin{equation}
\left( \partial_\mu \phi \right) \left( \partial_\nu \phi \right)
= e^{2\phi} \left( \partial_\mu a \right) \left( \partial_\nu a \right)\ ,
\qquad
\nabla_\mu \left( e^{2\phi} \partial^\mu a \right) = 0\ ,
\qquad
\nabla^2 \phi + e^{2\phi}
\left( \partial_\mu a \right) \left( \partial^\mu a \right) = 0\ ,
\end{equation}
where now all indices are contracted with the ten-dimensional metric
\eqn{adss5metric}. Taking the trace of the first equation and combining
it with the third one, we get
\begin{equation}
g^{\mu\nu} \nabla_\mu \partial_\nu e^\phi 
= \frac{1}{\sqrt{g}} \partial_\mu 
\left( \sqrt{g} g^{\mu\nu} \partial_\nu e^\phi \right) 
= 0\ .
\end{equation}
The solution to this equation is a rescaled version of the
one in flat space, namely \cite{BGKR}
\begin{equation}
\label{AdSdilaton}
e^\phi = e^{\phi_\infty}
+ \frac{\rho_0^4 \rho^4}{R^8} \frac{c}{| X - X_0 |^8}\ ,
\end{equation}
where now $X = \left( x^\mu, y^i \right)$ and the collective coordinates 
$X_0 = \left( x_0^\mu, y_0^i \right)$
denote the position of the D-instanton in $AdS_5\times S^5$ and
$\rho_0^2 \equiv y^i_0 y^i_0$. To find the solution for the axion,
one proceeds along the same lines as in the flat case. It is given by
\eqn{Axion} with the dilaton profile from \eqn{AdSdilaton}. Finally
one has to determine the instanton action. Again the contribution will
come from a surface term and the instanton action is the same as in
${\bf R}^{10}$ \cite{BGKR}. Although we have not discussed the fermionic
sector of the theory, it turns out that the D-instanton in
$AdS_5\times S^5$ preserves half of the supersymmetries, hence this
background is on equal footing with the Minkowski background.

%%%%%%%%%%%%%%%%%%%%%%%%%%%%%%%%%%%%%%%%%%%%%%%%%%%%%%%%%%%%%%%%%%%
\subsection{Supergravity scattering amplitudes and AdS/CFT.}
%%%%%%%%%%%%%%%%%%%%%%%%%%%%%%%%%%%%%%%%%%%%%%%%%%%%%%%%%%%%%%%%%%%

In the context of the D-branes there an exact
correspondence was conjectured
between type IIB string theory on $AdS_5 \times S^5$
space and four-dimensional $\cN = 4$ SYM theory living on the boundary
of the anti-de Sitter space \cite{ads/cft}. It is expressed as an
equivalence between certain scattering amplitudes in ten-dimensional
superstring theory on $AdS_5 \times S^5$ and Green's functions of
composite operators on the field theory side,
\begin{equation}
\exp \left( - S_{\rm IIB} \left[ {\mit\Phi} (J) \right] \right)
= \int \left[ d\varphi \right] \exp
\left(
- S_{\rm SYM} \left[ \varphi \right] + \cO \left[ \varphi \right] \cdot J
\right) \ .
\end{equation}
Here $S_{\rm IIB}$ is the type IIB superstring effective action
with ${\mit\Phi}$ being solutions of the massless sugra or the massive 
Kaluza-Klein field equations with boundary values $J$. On the SYM side 
the latter couple to composite
operators $\cO$ which are functions of the quantum fields $\varphi$.
For instance, the graviton couples to the YM energy-momentum tensor
and one can compare correlations functions of the stress tensors with
multi-graviton scattering amplitudes.

The precise dictionary between the coupling constants in IIB sugra and 
SYM theories is
\begin{equation}
\g^2 = 4 \pi \g_s = 4 \pi e^{\phi_\infty} \ ,
\qquad
\theta = 2 \pi a_\infty \ ,
\qquad
\frac{R^2}{\alpha'} = \g \sqrt{N} \ ,
\end{equation}
where $\alpha'$ is the string tension and appears in front of the
supergravity action in the string frame, see below. From this it follows
that large 't~Hooft coupling $\g^2 N$ corresponds to a large
radius $R$ and hence small curvature of the AdS background. This is
precisely the regime where the supergravity approximation is valid.

Particular terms in the supergravity effective action which are of order
$\left( \alpha' \right)^3$ relative to the leading Einstein-Hilbert term
were derived in \cite{GG}. In the string frame they read
\begin{equation}
S_{\rm IIB} = \left( \alpha' \right)^{- 4} \int d^{10} x \sqrt{g}
\Bigg\{
e^{- 2 \phi} \cR
+ \frac{\left( \alpha' \right)^3}{2^{11}\pi^7}
f_4 \left( \tau, \bar\tau \right) e^{- \phi/2} \cR^4
+ \left( \alpha' \right)^3
f_{16} \left( \tau, \bar\tau \right) e^{- \phi/2} \Lambda^{16}
+ {\rm h.c.}
\Bigg\} \ ,
\end{equation}
where $\cR^4$ in the second term is a particular contraction of Riemann
tensors and $\Lambda$ is a complex chiral $SO(9,1)$ spinor (the dilatino).
The functions $f_i$ are $SL(2,{\bf Z})$ modular forms in the complex
parameter $\tau \equiv i e^{- \phi} + a$. They have the following
weak coupling $\g_s = e^{\phi}$ expansion \cite{GG,KehPar}
\begin{eqnarray}
f_m \!\!\!&=&\!\!\! a_m \zeta (3) e^{- 3\phi/2} + b_m e^{\phi/2} \\
&+&\!\!\! e^{\phi/2}
\sum_{k = 1}^{\infty} \left( \sum_{d|k} \frac{1}{d^2} \right)
\left( k e^{- \phi} \right)^{m - 7/2}
\exp\left( -2 \pi k \left( e^{- \phi} - i a \right) \right)
\left(
1 + \sum_{j = 1}^{\infty} c^k_{j,m} \left( k e^{- \phi } \right)^{- j}
\right)
\ . \nonumber
\end{eqnarray}
The first two terms with coefficients $a_m$ and $b_m$ correspond to
tree and one-loop results in target space. The last term can be viewed as the
$k$-anti-instanton contribution. There is a similar term coming from
$k$-instantons. Both terms contain the perturbative fluctuations
around the D-instanton with certain coefficients $c^k_{j,m}$. For our
present purposes we concentrate only on the leading part (without the
fluctuations) in the one-anti-instanton sector.

This conjecture can now be checked non-perturbatively making use of the
results obtained for the two correlation functions discussed in the
previous section. Let us first concentrate on the $\langle
\Lambda^{16} \rangle$-correlator on the SYM side, evaluated in the
semi-classical approximation, i.e.\ in the weak coupling region.
According to \cite{ads/cft} the operator \eqn{Lambda} is identified with
the (boundary value of) dilatino on the supergravity side. The
bulk-to-boundary scattering amplitude obtained from the 16 dilatino
vertex should therefore correctly reproduce \eqn{16corr}. The dependence
on the coupling constants
\begin{equation}
\left( \alpha' \right)^{- 1} e^{- \phi/2} f_{16} \
\sim \ g^{-24} \sqrt{N} \ ,
\end{equation}
can be seen to match with the large $N$ limit of SYM
prediction \eqn{16constant}. Note that also
the instanton actions on both sides are equal. Moreover
the bulk-to-boundary propagators for the dilatinos can be shown to
coincide with \eqn{adsPropag} implying complete agreement between
the two pictures \cite{BGKR,DorKhoMatVan98}, despite of the fact that
we are not in the strongly coupled regime. This indicates that this
correlator is protected from quantum corrections \cite{GopGre99}.

Now let us turn to the $\langle {\mit\Theta}^4 \rangle$ correlator \eqn{T4}.
Again, the coupling constant dependence of the four-graviton scattering
amplitude matches the large $N$ behaviour of \eqn{T4coeff}, as can be seen
from
\begin{equation}
\left( \alpha' \right)^{- 1} e^{- \phi/2} f_4 \ \sim \ g^0 \sqrt{N} \ .
\end{equation}
As for the $x$-dependence, there is no obvious agreement with
bulk-to-boundary propagator for the graviton \cite{HokFre}. This is no to be
expected since this correlator is not protected, contrary to the previous
case, by any non-renormalization theorems.

%%%%%%%%%%%%%%%%%%%%%%%%%%%%%%%%%%%%%%%%%%%%%%%%%%%%%%%%%%%%%%%%%%%
\section{Discussion.}
\setcounter{equation}{0}
%%%%%%%%%%%%%%%%%%%%%%%%%%%%%%%%%%%%%%%%%%%%%%%%%%%%%%%%%%%%%%%%%%%

In these lectures we have reviewed the general properties of single
YM-instantons, and have given tools to compute non-perturbative effects
in (non-) supersymmetric gauge theories. As an application we have only
considered the ${\cal N} = 4$ SYM theory in relation to the AdS/CFT
correspondence, but our analysis can be used for a much larger class of
models.

On the other hand, because of lack of space and time, we have omitted a
few topics which are relevant for instanton applications in more realistic 
models like QCD \cite{tHooft86,ShuSch97,Hooft3} and spontaneously broken gauge 
theories such as the Standard Model \cite{Hooft1}. Let us address some of 
these issues here.
\begin{itemize}
\item Constrained instantons: in spontaneously broken gauge theories,
or any other non-conformal model with a scale, exact instanton solutions
to the equations of motion do not exist. This can be proved by means of
Derrick's theorem \cite{Der64}. The way to proceed in these cases is to
construct approximate solutions which are still dominating the path
integral \cite{Affleck}. The technique to find these configurations is
somewhat similar to the method we described in section 7.2, but instead
of expanding in Grassmann collective coordinates, constrained instantons
are  expanded in the dimensionless parameter $v^2\rho^2 $, where $v$ is
the vacuum expectation value of the scalar fields. For a recent detailed
discussion on this procedure, see \cite{niels}. As was shown by 't Hooft
long ago \cite{Hooft1}, the integration over the size of the instanton
collective coordinate diverges for large $\rho$. In electroweak theories
this divergence can be cured by the Higgs fields of the constrained
instanton, providing a factor $ - \rho^2 v^2$ in the classical action 
which cuts the $\rho$ integral off above $\rho^2 
= 1/v^2$. With this cut-off mechanism one can compute correlation functions 
of fermions along the same lines as in section 8. Constrained instantons 
are also relevant for ${\cal N} = 1, 2$ SYM theories, as mentioned in 
the introduction. They can also be studied in the context of 
topological YM theories, as was recently discussed in \cite{BFT}.

\item{Vacuum tunneling:} instantons can be used to describe tunneling
processes between different vacua of the Minkowski theory. We have not
really discussed the structure of the vacuum in QCD-like theories and
how tunneling occurs. This is in particular related to the presence of
the theta angle term, and consequences thereof (CP violation, instanton
anti-instanton interactions etc.). For a discussion on this, we refer
to the original literature \cite{CDG,JR} or to other reviews \cite{Rev}.

\item{Perturbation theory around the instanton:} the methods described
here enable us to compute non-perturbative effects in the semi-classical
approximation where the coupling constant is small. It is in many cases
important to go beyond this limit, and to study subleading corrections
that arise from higher order perturbation theory around the instanton
\cite{TwoLoop1,TwoLoop2}.
Apart from a brief discussion about the one-loop determinants in section
5, we have not really addressed these issues. Not unrelated to this,
subleading corrections also arise from treating the fermionic zero modes
exactly, as explained in section 7. We have there indicated how one might
compute such corrections, but a detailed investigation remains to be done
\cite{BelVanNie00a}.

\item{Multi-instantons:} we have completely omitted a discussion on
multi-instantons. These can be constructed using the ADHM formalism
\cite{ADHM}. The main difficulty lies in the explicit construction of
the instanton solution and of the measure of collective coordinates beyond
instanton number $k=2$. However, it was recently demonstrated that certain
simplifications occur in the large $N$ limit of ${\cal N}=4$ SYM theories
\cite{DorHolKhoMatVan99}, where one can actually sum over all
multi-instantons to get exact results for certain correlation functions.
The same techniques were later applied for ${\cal N}=2,1$ SYM
\cite{N=2largeN,N=1largeN}, and it would be interesting to study the
consequences of multi-instantons for large $N$ non-supersymmetric theories.

\end{itemize}

\vspace{0.5cm}

{\bf Acknowledgement.}
These lecture notes grew out of a course on ``Solitons and Instantons''
given by S.V.\ and P.v.N.\ at the C.N.\ Yang Institute for Theoretical
Physics, SUNY, Stony Brook, together with results obtained in collaboration
with A.B. They were also presented by S.V.\ at the TMR School ``Quantum
aspects of gauge theories, supersymmetry and unification'', January 2000,
Turin, Italy, whose organizers he thanks for the invitation and for the
warm and pleasant atmosphere during that week. We thank Nick Dorey for
discussions. This work was supported by an NSF grant PHY9722101.

\appendix
%%%%%%%%%%%%%%%%%%%%%%%%%%%%%%%%%%%%%%%%%%%%%%%%%%%%%%%%%%%%%%%%%%%
\section{'t Hooft symbols and spinor algebra.}
\label{HooftSpinor}
\setcounter{equation}{0}
%%%%%%%%%%%%%%%%%%%%%%%%%%%%%%%%%%%%%%%%%%%%%%%%%%%%%%%%%%%%%%%%%%%

In this appendix we give a list of conventions and formulae useful
in instanton calculus.

Let us first discuss the structure of Lorentz algebra $so (3,1)$ in
Minkowski space-time. The generators can be represented by $L_{\mu\nu}
= - i (x_\mu \partial_\nu - x_\nu \partial_\mu)$ and form the algebra
$[L_{\mu\nu}, L_{\rho\sigma}] = i
\eta_{\mu\rho} L_{\nu\sigma} + i \eta_{\nu\sigma} L_{\mu\rho} - i
\eta_{\mu\sigma} L_{\nu\rho} - i \eta_{\nu\rho} L_{\mu\sigma}$, with
the signature $\eta_{\mu\nu} = {\rm diag} (-,+,+,+)$. The spatial rotations
$J_i \equiv \ft12 \epsilon_{ijk} L_{jk}$ and boosts $K_i \equiv L_{0i}$
are self-adjoint operators $J_i^\dagger = J_i$, $K_i^\dagger = K_i$, i.e.\
\begin{equation}
\int d^4 x\, {\mit\Psi}^\ast {\cal O} {\mit\Psi}
= \int d^4 x\, \left( {\cal O} {\mit\Psi} \right)^\ast {\mit\Psi} \ ,
\qquad\mbox{for}\qquad {\cal O} = J, K \ ,
\end{equation}
and upon introduction of combinations $N_i \equiv \ft12 (J_i + i K_i)$,
$M_i \equiv \ft12 (J_i - i K_i)$ they form two commuting $SU (2)$ algebras,
\begin{equation}
[N_i, M_j] = 0 \ , \qquad
[N_i, N_j] = i \epsilon_{ijk} N_k \ , \qquad
[M_i, M_j] = i \epsilon_{ijk} M_k  \ .
\end{equation}
However, the two $SU(2)$ algebras are related by complex conjugation
$\left( su (2)_N \right)^\ast = su (2)_M$, and spatial inversion
${\cal P} su (2)_N = su (2)_M$.

The situation differs for euclidean space ($\delta_{\mu\nu} = {\rm diag}
(+,+,+,+)$) with $SO (4)$ Lorentz group. For the present case the linear
combinations of $(ij)$ and $(4,i)$-plane rotations
\begin{equation}
N_i \equiv \ft12 (J_i + K_i) \ ,
\qquad
M_i \equiv \ft12 (J_i - K_i) \ ,
\end{equation}
where obviously $J_i \equiv \ft12 \epsilon_{ijk} L_{jk}$ and boosts $K_i
\equiv L_{4i}$, give the algebras of independent $SU (2)$ subgroups of
$SO (4) = SU (2) \times SU (2)$ in view of hermiticity $N_i^\dagger = N_i$,
$M_i^\dagger = M_i$. It is an easy exercise to check that
\begin{equation}
N_i = - \ft{i}2 \bar\eta_{i\mu\nu} x_\mu \partial_\nu \ ,
\qquad\mbox{and}\qquad
M_i = - \ft{i}2 \eta_{i\mu\nu} x_\mu \partial_\nu \ ,
\end{equation}
where we introduced 't Hooft symbols \cite{Hooft1}
\begin{eqnarray}
&&\eta_{a\mu\nu} \equiv \epsilon_{a\mu\nu} + \delta_{a\mu} \delta_{\nu4}
- \delta_{a\nu} \delta_{4\mu} \ , \nonumber\\
&&\bar\eta_{a\mu\nu} \equiv \epsilon_{a\mu\nu} - \delta_{a\mu} \delta_{\nu4}
+ \delta_{a\nu} \delta_{4\mu} \ ,
\end{eqnarray}
and $\bar\eta_{a\mu\nu} = (-1)^{\delta_{4\mu} + \delta_{4\nu}}
\eta_{a\mu\nu}$. They form a basis of anti-symmetric 4 by 4 matrices
and are (anti-)selfdual in vector indices ($\epsilon_{1234} = 1$)
\begin{equation}
\eta_{a\mu\nu} = \ft12 \epsilon_{\mu\nu\rho\sigma} \eta_{a\rho\sigma} \ ,
\qquad
\bar\eta_{a\mu\nu} = - \ft12 \epsilon_{\mu\nu\rho\sigma}
\bar\eta_{a\rho\sigma} \ .
\end{equation}
The $\eta$-symbols obey the following relations
\begin{eqnarray}
\label{eta-eta1}
&&\epsilon_{abc} \eta_{b\mu\nu} \eta_{c\rho\sigma}
= \delta_{\mu\rho} \eta_{a\nu\sigma}
+ \delta_{\nu\sigma} \eta_{a\mu\rho}
- \delta_{\mu\sigma} \eta_{a\nu\rho}
- \delta_{\nu\rho} \eta_{a\mu\sigma} \ , \\
&&\eta_{a\mu\nu} \eta_{a\rho\sigma}
= \delta_{\mu\rho} \delta_{\nu\sigma}
- \delta_{\mu\sigma} \delta_{\nu\rho}
+ \epsilon_{\mu\nu\rho\sigma} \ , \\
\label{eta-eta3}
&&\eta_{a\mu\rho} \eta_{b\mu\sigma}
= \delta_{ab} \delta_{\rho\sigma} + \epsilon_{abc} \eta_{c\rho\sigma} \ ,\\
&&\epsilon_{\mu\nu\rho\theta} \eta_{a\sigma\theta}
= \delta_{\sigma\mu} \eta_{a\nu\rho}
+ \delta_{\sigma\rho} \eta_{a\mu\nu}
- \delta_{\sigma\nu} \eta_{a\mu\rho} \ ,\\
&&\eta_{a\mu\nu} \eta_{a\mu\nu} = 12 \ ,\quad
\eta_{a\mu\nu} \eta_{b\mu\nu} = 4 \delta_{ab} \ ,\quad
\eta_{a\mu\rho} \eta_{a\mu\sigma} = 3 \delta_{\rho\sigma} \ .
\end{eqnarray}
The same holds for $\bar\eta$ except for
\begin{equation}
\bar\eta_{a\mu\nu} \bar\eta_{a\rho\sigma}
= \delta_{\mu\rho} \delta_{\nu\sigma}
- \delta_{\mu\sigma} \delta_{\nu\rho}
- \epsilon_{\mu\nu\rho\sigma} \ .
\end{equation}
Obviously $\eta_{a\mu\nu} \bar\eta_{b\mu\nu} = 0$ due to different
duality properties.

The spinor representation of the euclidean Lorentz algebra is defined by
\begin{equation}
\sigma_{\mu\nu} \equiv \ft12
[\sigma_\mu \bar\sigma_\nu - \sigma_\nu \bar\sigma_\mu] \ ,
\qquad
\bar\sigma_{\mu\nu} = \ft12 [\bar\sigma_\mu \sigma_\nu - \bar\sigma_\nu
\sigma_\mu] \ ,
\end{equation}
in terms of euclidean matrices
\begin{equation}
\sigma_\mu^{\alpha\beta'} = (\tau^a, i) \ , \qquad
\bar\sigma_{\mu\, \alpha'\beta} = (\tau^a, -i) \ , \qquad
\sigma_\mu^{\alpha\alpha'} = \bar\sigma_\mu^{\alpha'\alpha}
= \epsilon^{\alpha'\beta'} \bar\sigma_{\mu\, \beta'\beta}
\epsilon^{\alpha\beta} \ ,
\end{equation}
obeying the Clifford algebra $\sigma_\mu \bar\sigma_\nu + \sigma_\nu
\bar\sigma_\mu = 2 \delta_{\mu\nu}$. The spinor and vector representations
of the $su (2)$ algebra are related precisely via the 't Hooft symbols,
\begin{equation}
\bar\sigma_{\mu\nu} = i \eta_{a\mu\nu} \tau^a \ , \qquad
\sigma_{\mu\nu} = i \bar\eta_{a\mu\nu} \tau^a \ .
\end{equation}

Some frequently used identities are
\begin{eqnarray}
\bar\sigma_\mu \sigma_{\nu\rho} = \delta_{\mu\nu} \bar\sigma_\rho
- \delta_{\mu\rho} \bar\sigma_\nu - \epsilon_{\mu\nu\rho\sigma}
\bar\sigma_\sigma \ , \quad
\sigma_\mu \bar\sigma_{\nu\rho} = \delta_{\mu\nu} \sigma_\rho
- \delta_{\mu\rho} \sigma_\nu + \epsilon_{\mu\nu\rho\sigma}
\sigma_\sigma \ , \\
\sigma_{\mu\nu} \sigma_\rho = \delta_{\nu\rho} \sigma_\mu
- \delta_{\mu\rho} \sigma_\nu + \epsilon_{\mu\nu\rho\sigma}
\sigma_\sigma \ , \quad
\bar\sigma_{\mu\nu} \bar\sigma_\rho = \delta_{\nu\rho} \bar\sigma_\mu
- \delta_{\mu\rho} \bar\sigma_\nu - \epsilon_{\mu\nu\rho\sigma}
\bar\sigma_\sigma \ .
\end{eqnarray}
The Lorentz generators are antisymmetric in vector and symmetric in
spinor indices
\begin{equation}
\sigma_{\mu\nu\, \alpha\beta} = - \sigma_{\nu\mu\, \alpha\beta} \ , \qquad
\sigma_{\mu\nu\, \alpha\beta} = \sigma_{\mu\nu\, \beta\alpha} \ ,
\end{equation}
and obey the algebra
\begin{eqnarray}
\label{Sigma-Algebra}
[\sigma_{\mu\nu}, \sigma_{\rho\sigma}]
\!\!\!&=&\!\!\! - 2 \left\{
\delta_{\mu\rho} \sigma_{\nu\sigma} + \delta_{\nu\sigma} \sigma_{\mu\rho}
- \delta_{\mu\sigma} \sigma_{\nu\rho} - \delta_{\nu\rho} \sigma_{\mu\sigma}
\right\} \ , \\
\label{AntiComm}
\{\sigma_{\mu\nu}, \sigma_{\rho\sigma}\}
\!\!\!&=&\!\!\! - 2 \left\{ \delta_{\mu\rho} \delta_{\nu\sigma}
- \delta_{\mu\sigma} \delta_{\nu\rho} - \epsilon_{\mu\nu\rho\sigma}
\right\} \ .
\end{eqnarray}
The same relations hold for $\bar\sigma$ but with $+
\epsilon_{\mu\nu\rho\sigma}$. In spinor algebra the following
contractions are useful
\begin{equation}
\sigma_\mu^{\alpha\alpha'} \bar\sigma_{\mu\, \beta'\beta}
= 2 \delta_\alpha^{\phan{i}\beta} \delta_{\alpha'}^{\phan{i}\beta'} \ ,
\qquad
\sigma_{\rho\sigma\ \, \beta}^{\phan{ii}\alpha}
\sigma_{\rho\sigma\ \, \delta}^{\phan{ii}\gamma}
= 4 \left\{
\delta_\beta^{\phan{i}\alpha} \delta_\delta^{\phan{i}\gamma}
- 2
\delta_\delta^{\phan{i}\alpha} \delta_\beta^{\phan{i}\gamma}
\right\} \ .
\end{equation}
We use everywhere the north-west conventions for raising
and lowering the spinor indices
\begin{equation}
\epsilon^{\alpha\beta} \xi_\beta = \xi^\alpha \ ,
\qquad
\bar\xi^{\beta'} \epsilon_{\beta'\alpha'} = \bar\xi_{\alpha'} \ ,
\end{equation}
with $\epsilon_{\alpha\beta} = - \epsilon^{\alpha'\beta'}$,
$\epsilon_{\alpha\beta} = \epsilon^{\alpha\beta}$, so that
$\xi_{(1)}^\alpha \xi_{(2)\alpha} = \xi_{(2)}^\alpha
\xi_{(1)\alpha}$. For hermitean conjugation we define $\left(
\xi_{(1)}^\alpha \xi_{(2)\alpha} \right)^\dagger = \xi_{(2)\alpha}^\dagger
\xi_{(1)}^{\alpha\dagger}$ and the sigma matrices satisfy
\begin{equation}
\left( \sigma_{\mu}^{\alpha\beta'} \right)^\ast
= \sigma_{\mu\, \alpha\beta'} ,
\qquad
\left( \bar\sigma_{\mu\, \alpha'\beta} \right)^\ast
= \bar\sigma_{\mu}^{\alpha'\beta}\ .
\end{equation}

Throughout the paper we frequently use the following integral formula
\begin{equation}
\label{Integral}
\int d^4 x \frac{\left( x^2 \right)^n}{\left( x^2 + \rho^2 \right)^m}
= \pi^2 \left( \rho^2 \right)^{n - m + 2}
\frac{{\mit\Gamma} (n + 2){\mit\Gamma} (m - n - 2)}{{\mit\Gamma}(m)} \ ,
\end{equation}
which converges for $m - n > 2$.

%%%%%%%%%%%%%%%%%%%%%%%%%%%%%%%%%%%%%%%%%%%%%%%%%%%%%%%%%%%%%%%%%%%%%
\section{Winding number.}
\label{Winding}
\setcounter{equation}{0}
%%%%%%%%%%%%%%%%%%%%%%%%%%%%%%%%%%%%%%%%%%%%%%%%%%%%%%%%%%%%%%%%%%%%%

For a gauge field configuration with finite action the field strength must
tend to zero faster than $x^{-2}$ at large $x$. For vanishing $F_{\mu\nu}$,
the potential $A_\mu$ becomes a pure gauge, $A_\mu \stackrel{x \to \infty}{
\longrightarrow} U^{-1} \partial_\mu U$. All configurations of $A_\mu$
which become pure gauge at infinity fall into equivalence classes, where
each class has a definite winding number. As we now show, the
winding number is given by

\begin{equation}
k = - \frac{1}{16 \pi^2} \int d^4 x\,
{\rm tr}\, {^\ast F}_{\mu\nu} F_{\mu\nu} \ ,
\end{equation}
where the generators $T_a$ satisfy $\tr\, T_a T_b = - \ft12 \delta_{ab}$.
This is the normalization we adopt for the fundamental representation.
The key observation is that ${^\ast F}_{\mu\nu} F_{\mu\nu}$ is a
total derivative of a gauge variant current\footnote{Note that
${^\ast F}_{\mu\nu} F_{\mu\nu}$ equals to $2 \epsilon_{\mu\nu\rho\sigma}
\left\{ \partial_\mu A_\nu \partial_\rho A_\sigma + 2 \partial_\mu A_\nu
A_\rho A_\sigma + A_\mu A_\nu A_\rho A_\sigma \right\}$ but the last term
vanishes in the trace due to cyclicity of the trace.}
\begin{equation}
{\rm tr}\, {^\ast F}_{\mu\nu} F_{\mu\nu}
= 2 \partial_\mu {\rm tr}\,
\epsilon_{\mu\nu\rho\sigma}
\left\{
A_\nu \partial_\rho A_\sigma
+ \ft23 A_\nu A_\rho A_\sigma
\right\} \ .
\end{equation}
According to Stokes' theorem, the space-time integral becomes an integral
over the three-di\-men\-si\-onal boundary at infinity if one uses the
regular gauge in which there are no singularities at the orgin. Since
$A_\mu$ becomes a pure gauge at large $x$, one obtains
\begin{equation}
\label{k-charge}
k = \frac{1}{24 \pi^2} \oint_{S^3({\rm space})} d {\mit\Omega}_\mu
\epsilon_{\mu\nu\rho\sigma} {\rm tr}\,
\left\{
\left( U^{-1} \partial_\nu U \right)
\left( U^{-1} \partial_\rho U \right)
\left( U^{-1} \partial_\sigma U \right)
\right\} \ ,
\end{equation}
where the integration is over a large three-sphere, $S^3({\rm space})$,
in Euclidean space. To each point $x^\mu$ on this large three-sphere in
space corresponds a group element $U$ in the gauge group $G$. If $G =
SU (2)$, the group manifold is also a three-sphere\footnote{The elements
of $SU (2)$ can be written in the fundamental representation as $U =
a_0 \1 + i \sum_k a_k \tau_k$ where $a_0$ and $a_k$ are real coefficients
satisfying the condition $a_0^2 + \sum_k a_k^2 = 1$. This defines a sphere
$S^3({\rm group})$.} $S^3({\rm group})$. Then $U (x)$ maps $S^3({\rm
space})$ into $S^3({\rm group})$, and as we now show, $k$ is an integer
which counts how many times $S^3({\rm space})$ is wrapped around
$S^3({\rm group})$.

Choose a parametrization of the group elements of $SU (2)$ in terms of
group parameters\footnote{For example, Euler angles, or Lie
parameters $U = a_0 \1 + i \sum_k a_k \tau_k$ with $a_0 = \sqrt{1 -
\sum_k a_k^2}$.} $\xi^a (x)$ ($a = 1,2,3$). Hence the functions
$\xi^a (x)$ map $x$ into $SU (2)$. Consider a small surface element
of $S^3({\rm space})$. According to the chain rule
\begin{equation}
{\rm tr}\,
\left\{
\left( U^{-1} \partial_\nu U \right)
\left( U^{-1} \partial_\rho U \right)
\left( U^{-1} \partial_\sigma U \right)
\right\}
= \frac{\partial \xi^i}{\partial x_\nu}
\frac{\partial \xi^j}{\partial x_\rho}
\frac{\partial \xi^k}{\partial x_\sigma}
{\rm tr}\,
\left\{
\left( U^{-1} \partial_i U \right)
\left( U^{-1} \partial_j U \right)
\left( U^{-1} \partial_k U \right)
\right\} \ .
\end{equation}
Using\footnote{For example, if the surface element points in the
$x$-direction we have ${\mit\Delta\Omega} = {\mit\Delta}\tau
{\mit\Delta} y {\mit\Delta} z$.}
\begin{equation}
{\mit\Delta\Omega}_\mu = \ft16 \epsilon_{\mu\nu\rho\sigma}
{\mit\Delta} x_\nu
{\mit\Delta} x_\rho
{\mit\Delta} x_\sigma
\end{equation}
we obtain, from $\ft16 \epsilon_{\mu\nu\rho\sigma}
\epsilon_{\mu\alpha\beta\gamma} = \delta^{\nu\rho\sigma}_{
[\alpha\beta\gamma]}$ and $d \xi^{[i} d \xi^j d \xi^{k]} = \epsilon^{ijk}
d^3 \xi$, for the contribution ${\mit\Delta} k$ of the small surface
element to $k$
\begin{equation}
{\mit\Delta} k = \frac{1}{24\pi^2} \epsilon^{ijk}
{\rm tr}\,
\left\{
\left( U^{-1} \partial_i U \right)
\left( U^{-1} \partial_j U \right)
\left( U^{-1} \partial_k U \right)
\right\} d^3 \xi \ ,
\end{equation}
with $k = \oint_{S^3({\rm space})} {\mit\Delta} k$. The elements $\left(
U^{-1} \partial_i U \right)$ lie in the Lie algebra, and define the group
vielbein $e_i^a$ by
\begin{equation}
\left( U^{-1} \partial_i U \right) = e_i^a T_a \ .
\end{equation}
With $\epsilon^{ijk} e_i^a e_j^b e_k^c = \left( \det e \right) \,
\epsilon^{abc}$, we obtain for the contribution to $k$ from a surface
element ${\mit\Delta} {\mit\Omega}_\mu$
\begin{equation}
{\mit\Delta} k = \frac{1}{24 \pi^2}
\left( \det e \right)
\tr \left( \epsilon^{abc} T_a T_b T_c \right) d^3 \xi
= - \frac{1}{16 \pi^2}
\left( \det e \right) d^3 \xi \ .
\end{equation}

As we demonstrated, the original integral over the physical space is
reduced to one over the group with measure $\left( \det e \right) d^3 \xi$.
The volume of a surface element of $S^3({\rm group})$ with coordinates
$d \xi^i$ is proportional to $\left( \det e \right) d^3 \xi$. Since this
expression is a scalar in general relativity, we know that the value of
the volume does not depend on which coordinates one uses except for an
overall normalization. We fix this overall normalization of the group
volume such that near $\xi = 0$ the volume is $d^3 \xi$. Since $e_i^a =
\delta_i^a$ near $\xi = 0$, we have the usual euclidean measure $d^3 \xi$.
Each small patch on $S^3({\rm space})$ corresponds to a small patch on
$S^3({\rm group})$. Since the $U$'s fall into homotopy classes, integrating
once over $S^3({\rm space})$ we cover $S^3({\rm group})$ an integer number
of times. To check the proportionality factor in ${\mit\Delta} k \sim
{\rm Vol}\, \left( d^3 \xi \right)$, we consider the fundamental map
\begin{equation}
\label{Rotation}
U (x) = x_\mu \sigma_\mu / \sqrt{x^2} \ ,
\qquad
U^{-1} (x) = x_\mu \bar\sigma_\mu / \sqrt{x^2} \ .
\end{equation}
This is clearly a one-to-one map from $S^3({\rm space})$ to
$S^3({\rm group})$ and should yield $|k| = 1$. Direct calculation
gives
\begin{equation}
U^{-1} \partial_\mu U = - \bar\sigma_{\mu\nu} x_\nu /x^2 \ ,
\end{equation}
and substitution into \eqn{k-charge} leads to $k = - \frac{1}{2\pi^2} \oint
d {\mit\Omega}_\mu x_\mu/x^4 = -1$ making use of \eqn{Sigma-Algebra} and
\eqn{AntiComm}. To obtain $k = 1$ one has to make the change $\sigma 
\leftrightarrow \bar\sigma$
 or $x \leftrightarrow - x$ in Eq.\ (\ref{Rotation}).

Let us comment on the origin of the winding number of the instanton in
the singular gauge. In this case $A^{\rm sing}_\mu$ vanishes fast at
infinity, but becomes pure gauge near $x = 0$ . In the region between
a small sphere in the vicinity of $x = 0$ and a large sphere at $x =
\infty$ we have an expression for $k$ in terms of the total derivative,
but now for $A^{\rm sing}_\mu$ the only contribution to the topological
charge comes from the boundary near $x = 0$:
\begin{equation}
k = - \frac{1}{24 \pi^2} \oint_{S_{x \to 0}^3({\rm space})}
d {\mit\Omega}_\mu
\epsilon_{\mu\nu\rho\sigma} {\rm tr}\,
\left\{
\left( U^{-1} \partial_\nu U \right)
\left( U^{-1} \partial_\rho U \right)
\left( U^{-1} \partial_\sigma U \right)
\right\} \ .
\end{equation}
The extra minus sign is due to the fact that the normal to the
$S^3({\rm space})$ at $x = 0$ points inward. Furthermore, $A^{\rm sing}_\mu
\sim U^{-1} \partial_\mu U = - \bar\sigma_{\mu\nu} x_\nu / x^2$ near
$x = 0$, while $A^{\rm reg}_\mu \sim U \partial_\mu U^{-1} = -
\sigma_{\mu\nu} x_\nu / x^2$ for $x \sim \infty$. There is a second minus
sign in the evaluation of $k$ from the trace of Lorentz generators in both
solutions. As a result $k_{\rm sing} = k_{\rm reg}$, as it should be since
$k$ is a gauge invariant object. The gauge transformation which maps
$A^{\rm reg}_\mu$ to $A^{\rm sing}_\mu$ transfers the winding from a large
to a small $S^3({\rm space})$.

%%%%%%%%%%%%%%%%%%%%%%%%%%%%%%%%%%%%%%%%%%%%%%%%%%%%%%%%%%%%%%%%%%%%%
\section{The volume of the gauge orientation moduli space.}
\label{Volume}
\setcounter{equation}{0}
%%%%%%%%%%%%%%%%%%%%%%%%%%%%%%%%%%%%%%%%%%%%%%%%%%%%%%%%%%%%%%%%%%%%%

The purpose of this appendix is to prove the equation \eqn{VolCoset}.
Let us consider an instanton in an $SU (N)$ gauge theory.
Deformations of this configuration which are still self-dual and not
a gauge transformation are parametrized by collective coordinates.
Constant gauge transformations $A_\mu \to U^{-1} A_\mu U$ preserve
self-duality and transversality, $\partial_\mu A_\mu = 0$, but not all
constant $SU (N)$ matrices $U$ change $A_\mu$. Those $U$ which keep
$A_\mu$ fixed form the stability subgroup $H$ of the instanton, hence
we want to determine the volume of the coset space $SU (N)/H$.

If the instanton is embedded in the lower-right $2 \times 2$ submatrix
of the $N \times N$ $SU (N)$ matrix, then $H$ contains the $SU (N - 2)$
in the left-upper part, and the $U (1)$ subgroup with elements $\exp
\left( \theta A \right)$ where $A$ is the diagonal matrix
\begin{equation}
A = i \sqrt{\frac{N - 2}{N}} {\rm diag}
\left( \frac{2}{2 - N}, \dots, \frac{2}{2 - N}, 1, 1 \right) \ .
\end{equation}
All generators of $SU (N)$ (only for the purposes of the present appendix,
and also all generators of $SO (N)$ discussed below) are normalized
according to ${\rm tr} \, T_a T_b = - 2 \delta_{ab}$ in the defining
$N \times N$ matrix representation. Note that in the main text we use
$\tr \, T_a T_b = - \ft12 \delta_{ab}$.

At first sight one might expect the range of $\theta$ to be such that
all entries cover the range $2\pi$ an integer number of times. However,
this is incorrect: only for the last two entries of $\exp \left( \theta
A \right)$ we must require periodicity, because whatever happens in the
other $N - 2$ diagonal entries is already contained in the $SU (N - 2)$
part of the stability subgroup. Thus all elements $h$ in $H$ are of the
form \cite{Bernard}
\begin{equation}
h = e^{\theta A} g,
\qquad\mbox{with}\qquad
g \in SU (N - 2)
\qquad\mbox{and}\qquad
0 \leq \theta \leq \theta_{\rm max} = 2 \pi \sqrt{\frac{N}{N - 2}} \ .
\end{equation}
For $N = 3, 4$ this range of $\theta$ corresponds to periodicity of all
entries, but for $N \geq 5$ the range of $\theta$ is less than required for
periodicity. Thus $H \neq SU(N) \times U (1)$  for $N \geq 5$. The
first $N - 2$ entries of $\exp \left( k \theta_{\rm max} A \right)$
with integer $k$ are given by $\exp \left( - i k \frac{4\pi}{N - 2}
\right)$ and lie therefore in the center $Z_N$ of $SU (N - 2)$. Note
that the $SU (N)$ group elements $h = \exp\left( \theta A \right) g$
with $0 \leq \theta \leq \theta_{\rm max}$ form a subgroup. We shall
denote $H$ by $SU (N - 2) \times ``U (1)"$ where $``U (1)"$ denotes
the part of the $U (1)$ generated by $A$ which lies in $H$.

We now use three theorems to evaluate the volume of $SU (N)/H$:
\begin{eqnarray}
({\rm I})   && {\rm Vol}\, \frac{SU (N)}{SU (N - 2) \times ``U (1)"}
= \frac{{\rm Vol} \, \left( SU (N)/SU (N - 2) \right)
}{{\rm Vol} \, ``U (1)"} \ ,
\nonumber\\
({\rm II})  && {\rm Vol} \, \frac{SU (N)}{SU (N - 2)}
= {\rm Vol} \, \frac{SU (N)}{SU (N - 1)}
{\rm Vol} \, \frac{SU (N - 1)}{SU (N - 2)} \ , \\
({\rm III}) && {\rm Vol}\, \frac{SU (N)}{SU (N - 1)}
= \frac{{\rm Vol} \, SU (N)}{{\rm Vol} \, SU (N - 1)} \ .
\nonumber
\end{eqnarray}
It is, in fact, easiest to first compute ${\rm Vol} \left( SU (N)/
SU (N - 1) \right)$ and then to use this result for the evaluation of
${\rm Vol}\ G/H$ (with ${\rm Vol}\ SU (N)$ as a bonus).

In general the volume of a coset manifold $G/H$ is given by $V = \int
\prod_{\mu} dx^\mu\, \det\ e^m_\mu (x)$ where $x^\mu$ are the
coordinates on the coset manifold and $e^m_\mu (x)$ are the coset
vielbeins. One begins with ``coset representatives'' $L (x)$ which are
group elements $g \in G$ such that every group element can be decomposed as
$g = k h$ with $h \in H$. We denote the coset generators by $K_\mu$
and the subgroup generators by $H_i$. Then $L^{-1} (x) \partial_\mu L (x)
= e^m_\mu (x) K_m + \omega_\mu^i (x) H_i$. Under a general coordinate
transformation from $x^\mu$ to $y^\mu (x)$, the vielbein transforms as
a covariant vector with index $\mu$ but also as a contra-covariant vector
with index $m$ at $x = 0$. Hence $V$ does (only) depend on the choice
of the coordinates at the origin. Near the origin, $L (x) = \1 + dx^\mu
K_\mu$ , and we fix the normalization of $T_a = \left\{ K_\mu, H_i
\right\}$ by ${\rm tr}\, T_a T_b = - 2 \delta_{ab}$ for $T_a \in SU (N)$.

To find the volume of $SU (N)/SU (N - 1)$  we note that the group elements
of $SU (N)$ have a natural action on the space ${\rm\bf C}^N$ and map a
vector $\left( z^1, \dots, z^N \right) \in {\rm\bf C}^N$ on the complex
hypersphere $\sum_{i = 1}^{N}\left| z^i \right|^2 = 1$ into another vector
on the complex hypersphere. The ``south-pole'' $(0, \dots, 0, 1)$ is kept
invariant by the subgroup $SU (N - 1)$, and points on the complex
hypersphere are in one-to-one correspondence with the coset representatives
$L (z)$ of $SU (N)/SU (N - 1)$. We use as generators for $SU (N)$ the
generators for $SU (N - 1)$ in the upper-left block, and further the
following coset generators: $N - 1$ pairs $T_{2 k}$ and $T_{2k + 1}$
each of them containing only two non-zero elements
\begin{equation}
\left(
\begin{array}{cccc}
0      & \ldots &        & 0 \\
\vdots &        &        & 1      \\
       &        & \ddots & \vdots  \\
0      &  - 1   & \ldots &   0    \\
\end{array}
\right) \ ,
\qquad\qquad
\left(
\begin{array}{cccc}
0      & \ldots &        & 0 \\
\vdots &        &        & i      \\
       &        & \ddots & \vdots  \\
0      &    i   & \ldots &   0    \\
\end{array}
\right) \ ,
\end{equation}
and further one diagonal generator
\begin{equation}
T_{N^2 - 1} = i \sqrt{\frac{2}{N (N - 1)}}
{\rm diag} \left( - 1, \dots, - 1, N - 1 \right) .
\end{equation}
(For instance, for $SU (3)$ there are two pairs, the usual $\lambda_4$
and $\lambda_5$ and $\lambda_6$ and $\lambda_7$ and the diagonal hypercharge
generator $\lambda_8$.)

The idea now is to establish a natural one-to-one correspondence between
points in ${\bf C}^N$ and points in ${\bf R}^{2N}$, namely we write all
points $(x^1, \dots, x^{2N})$ in ${\rm\bf R}^{2N}$ as points in
${\rm\bf C}^N$ as $(i x^1 + x^2, \dots, i x^{2N - 1} + x^{2N})$. In
particular the south pole in ${\rm\bf R}^{2N}$ corresponds to the south
pole in ${\rm\bf C}^{N}$ and the sphere $\sum_{i = 1}^{2N} (x^i)^2 = 1$
in ${\bf R}^{2N}$ corresponds to the hypersphere $\sum_{i = 1}^N |z^i|^2
= 1$. Points on the sphere $S^{2N - 1}$ in ${\bf R}^{2N}$ correspond
one-to-one to coset generators of $SO (2N)/ SO (2N - 1)$. The coset
generators of $SO(2N)/SO(2N - 1)$ are antisymmetric $2N \times 2N$
matrices $A_I$ $(I = 1, \dots, 2N - 1)$ with the entry $+1$ in the last
column and $-1$ in the last row. The coset element $1 + \delta g = 1 + dt^I
A_I$ maps the south pole $s = (0, \dots, 0, 1)$ in ${\rm\bf R}^{2N}$ to a
point $s + \delta s$ in ${\rm\bf R}^{2N}$ where $\delta s = (dt^1, \dots ,
dt^{2N - 1}, 0)$. In ${\rm\bf C}^N$ the action of this same coset element
is defined as follows: it maps $s = (0, \dots, 0, 1)$ to $s + \delta s$
with $\delta s = (i dt^1 + dt^2, \dots, i dt^{2N - 1})$. The coset generators
of $SU (N)/SU (N - 1)$ also act in ${\rm\bf C}^N$, but $g = 1 + dx^\mu
K_\mu$ maps $s$ to $s + \delta s$ where now $\delta s = ( i dx^1 +
dx^2, \dots, i \sqrt{\ft{2(N - 1)}{N}}dx^{2N - 1} )$.

We can cover $S^{2N - 1}$ with small patches. Each patch can be brought
by the action of a suitable coset element to the south pole, and then we
can use the inverse of this group element to map this patch back into
the manifold $SU (N)/SU (N - 1)$. In this way both $S^{2N - 1}$ and
$SU (N)/SU (N - 1)$ are covered by patches which are in a one-to-one
correspondence. Each pair of patches has the same ratio of volumes since both
patches can be brought to the south pole by the same group element and
at the south pole the ratio of their volumes is the same. To find the ratio
of the volumes of $S^{2N - 1}$ and $SU (N)/SU (N - 1)$, it is then
sufficient to consider a small patch near the south pole.

Consider then a small patch at the south pole of $S^{2N - 1}$ with
coordinates $\left( dt^1, \dots, dt^{2N - 1} \right)$ and volume
$dt^1 \dots dt^{2N - 1}$. The corresponding patch at the south pole of
${\bf C}^N$ has coordinates $\left( i dt^1 + dt^2, \dots, i dt^{2N - 1}
\right)$ and the same volume $dt^1 \dots dt^{2N - 1}$. The same patch at
the south pole in ${\bf C}^N$ has coordinates $d x^\mu$ where $\left(
i dt^1 + dt^2, \dots, i dt^{2N - 1}\right) = \left( i dx^1 + dx^2, \dots,
i \sqrt{\frac{2 (N - 1)}{N}} dx^{2N - 1} \right)$. The volume of a patch
in $SU (N)/SU (N - 1)$ with coordinates $dx^1, \dots, dx^{2N - 1}$
is $dx^1 \dots dx^{2N - 1}$. It follows that the volume of $SU (N)/SU
(N - 1)$ equals the volume of $S^{2N - 1}$ times $\sqrt{\frac{N}{2 (N
- 1)}}$,
\begin{equation}
\label{SUNtoN1}
{\rm Vol} \ \frac{SU (N)}{SU (N - 1)}
= \sqrt{\frac{N}{2(N - 1)}}\ {\rm Vol}\ S^{2N - 1} \ .
\end{equation}

From here the evaluation of ${\rm Vol}\ SU(N)/H$  is straightforward.
Using
\begin{equation}
{\rm Vol}\ S^{2N - 1} = \frac{2 \pi^N}{(N - 1)!} \ ,
\end{equation}
we obtain
\begin{equation}
\label{VolSUN}
{\rm Vol}\ SU(N) = \sqrt{N} \prod_{k = 2}^{N}
\frac{\sqrt{2}\pi^k}{(k - 1)!} ,
\end{equation}
and
\begin{eqnarray}
&&{\rm Vol}\ H = {\rm Vol}\ SU (N - 2) {\rm Vol}\ ``U (1)" \ ,
\qquad
{\rm Vol}\ ``U(1)" = 2 \pi \sqrt{\frac{N}{N - 2}} \ , \nonumber\\
&&{\rm Vol}\ SU (N)/H = \frac{\pi^{2N - 2}}{(N - 1)!(N - 2)!} \ .
\end{eqnarray}

As an application and check of this analysis let us demonstrate a few
relations between the volumes of different groups. Let us check that
${\rm Vol}\ SU (2) = 2 {\rm Vol}\ SO (3)$, ${\rm Vol}\ SU (4) = 2
{\rm Vol}\ SO (6)$ and ${\rm Vol}\ SO (4) = \ft12 \left( {\rm Vol}\
SU (2) \right)^2$ (the latter will follow from $SO (4) = SU (2) \times
SU (2)/Z_2$).

We begin with the usual formula for the surface of a sphere with unit
radius (given already above for odd $N$)
\begin{equation}
{\rm Vol}\ S^N = \frac{2 \pi^{(N + 1)/2}}{{\mit\Gamma}
\left( \frac{N + 1}{2} \right)} \ .
\end{equation}
In particular
\begin{eqnarray}
&&{\rm Vol}\ S^2 = 4 \pi \ , \quad
{\rm Vol}\ S^3 = 2 \pi^2 \ , \quad
{\rm Vol}\ S^4 = \ft83 \pi^2 \ , \nonumber\\
&&{\rm Vol}\ S^5 = \pi^3, \quad
{\rm Vol}\ S^6 = \ft{16}{15} \pi^3 \ , \quad
{\rm Vol}\ S^7 = \ft13 \pi^4 \ .
\end{eqnarray}
Furthermore ${\rm Vol}\ SO (2) = 2 \pi$ since the $SO (2)$ generator
is $T = \left( {0\, \ 1\atop - 1\, 0} \right)$ and $\exp (\theta T)$ is an
ordinary rotations $\left( {\cos \theta\, \ \sin \theta \atop - \sin \theta
\, \cos \theta} \right)$ for which $0 \leq \theta \leq 2 \pi$.
Using ${\rm Vol}\, SO (N) = {\rm Vol} \, S^{N - 1} {\rm Vol} \, SO(N - 1)$
we obtain
\begin{eqnarray}
&&{\rm Vol}\ SO (2) = 2 \pi \ , \quad
{\rm Vol}\ SO (3) = 8 \pi^2 \ , \quad
{\rm Vol}\ SO (4) = 16 \pi^4 \ , \nonumber\\
&&\qquad\qquad{\rm Vol}\ SO (5) = \ft{128}{3} \pi^6 \ , \quad
{\rm Vol}\ SO (6) = \ft{128}{3} \pi^9 \ .
\end{eqnarray}

Now consider $SU (2)$. In the normalization $T_1 = - i \tau_1$,
$T_2 = - i \tau_2$ and $T_3 = - i \tau_3$ (so that ${\rm tr}\
T_a T_b = - 2 \delta_{ab}$) we find by direct evaluation using Euler
angles ${\rm Vol}\ SU (2) = 2 \pi^2$. This also agrees with Eq.\ (\ref{SUNtoN1})
for $N = 2$, using ${\rm Vol}\ SU (1) = 1$. For higher $N$ we get
\begin{equation}
{\rm Vol}\ SU (2) = 2 \pi^2 \ ,
\qquad
{\rm Vol}\ SU (3) = \sqrt{3} \pi^5 \ ,
\qquad
{\rm Vol}\ SU (4) = \ft{\sqrt{2}}{3} \pi^9 \  .
\end{equation}
The group elements of $SU (2)$ can be written as $g = x^4  + i
\vec\tau \cdot \vec x$ with $\left( x^4 \right)^2 + \left( \vec x
\right)^2 = 1$ which defines a sphere $S^3$. Since near the unit
element $g \approx 1 + \vec\tau \cdot \delta\vec x$, the normalization
of the generators is as before, and hence for this parametrization
${\rm Vol}\ SU (2) = 2 \pi^2$. This is indeed equal to ${\rm Vol}\ S^3$.

However, ${\rm Vol}\ SU (2)$ is not yet equal to $2 {\rm Vol}\ SO (3)$.
The reason is that in order to compare properties of different groups
we should normalize the generators such that the structure constants
are the same (the Lie algebras are the same, although the group volumes
are not). In other words, we should use the normalization that the adjoint
representations have the same ${\rm tr} \ T_a T_b$.

For $SU (2)$ the generators which lead to the same commutators as the
usual $SO (3)$ rotation matrices (with entries $+1$ and $-1$) are
$T_a = \left\{ -\ft{i}2 \tau_1, -\ft{i}2 \tau_2, -\ft{i}2
\tau_3 \right\}$. Then ${\rm tr}\ T_a T_b = - \ft12 \delta_{ab}$.
In this normalization, the range of each group coordinate is multiplied by
2, leading to ${\rm Vol}\ SU (2) = 2^3 \cdot 2 \pi^2 = 16 \pi^2$. Now
indeed ${\rm Vol}\ SU (2) = 2{\rm Vol}\ SO (3)$.

For $SU (4)$ the generators with the same Lie algebra as $SO (6)$ are the
generators $\ft14 ( \gamma_m \gamma_n$ $- \gamma_n \gamma_m )$, $i \gamma_m$
and $i \gamma_5$, where $\gamma_m$ and $\gamma_5$ are the $4 \times 4$
matrices obeying the Clifford
algebra $\{ \gamma_m, \gamma_n \} = 2 \delta_{mn}$. Now, ${\rm tr}\ T_a T_b
= - \delta_{ab}$ (for example, $\tr \left\{ \left( \ft12 \gamma_1 \gamma_2
\right) \left( \ft12 \gamma_1 \gamma_2 \right) \right\} = - 1$). Recall that
originally we had chosen the normalization
${\rm tr}\ T_a T_b = - 2 \delta_{ab}$. We must thus multiply the range of
each coordinate by a factor $\sqrt{2}$, and hence we must multiply our original
result for ${\rm Vol}\ SU (4)$ by a factor $\left( \sqrt{2} \right)^5$. We
find that indeed the relation ${\rm Vol}\ SU (4) = 2 {\rm Vol}\ SO (6)$ is
fulfilled.

Finally, we consider the relation $SO (4) = SU (2) \times SU (2)/ Z_2$.
(The vector representation of $SO (4)$ corresponds to the representation
$\left( \ft12, \ft12 \right)$ of $SU (2) \times SU (2)$, but
representations like $\left( \ft12, 0 \right)$ and $\left( 0, \ft12
\right)$ are not representations of $SO (4)$ and hence we must divide by
$Z_2$. The reasoning is the same as for $SU (2)$ and $SO (3)$, or
$SU (4)$ and $SO (6)$.) We choose the generators of $SO (4)$
as follows
\begin{equation}
T_1^{(+)} = \frac{1}{\sqrt{2}} \left( L_{14} + L_{23} \right) \ , \quad
T_2^{(+)} = \frac{1}{\sqrt{2}} \left( L_{31} + L_{24} \right) \ , \quad
T_3^{(+)} = \frac{1}{\sqrt{2}} \left( L_{12} + L_{34} \right) \ ,
\end{equation}
and the same but with minus sign denoted by $T_i^{(-)}$. Here $L_{mn}$ equals
$+1$ in the $m^{\rm th}$ column and $n^{\rm th}$ row, and is antisymmetric.
Clearly $\tr\ T_a T_b = - 2 \delta_{ab}$.
The structure constants follow from
\begin{equation}
\left[
\frac{1}{\sqrt{2}} \left( L_{12} + L_{34} \right),
\frac{1}{\sqrt{2}} \left( L_{14} + L_{23} \right),
\right]
= - \left( L_{31} + L_{24} \right) \ ,
\end{equation}
thus
\begin{equation}
\label{CRSO4}
\left[ T_i^{(+)}, T_j^{(+)} \right]
= - \sqrt{2} \epsilon_{ijk} T_k^{(+)} \ , \qquad
\left[ T_i^{(-)}, T_j^{(-)} \right]
= - \sqrt{2} \epsilon_{ijk} T_k^{(-)} \ , \qquad
\left[ T_i^{(+)}, T_j^{(-)} \right] = 0 \ .
\end{equation}
We choose for the generators of $SU (2) \times SU (2)$ the representation
\begin{equation}
T^{(+)}_i = \frac{i \tau_i}{\sqrt{2}} \otimes \1 \ , \qquad
T^{(-)}_i = \1 \otimes \frac{i \tau_i}{\sqrt{2}} \ .
\end{equation}
Then we get the same commutation relations as for $SO (4)$ generators
(\ref{CRSO4}); however, the generators are normalized differently,
namely ${\rm tr}\ T_a T_b = - 2 \delta_{ab}$ for $SO (4)$ but ${\rm tr}\
T_a T_b = - \delta_{ab}$ for $SU (2)$. With the normalization $\tr\ T_a
T_b = - 2 \delta_{ab}$ we found ${\rm Vol}\ SU (2) = 2 \pi^2$. In the \
present normalization we find ${\rm Vol}\ SU (2) = 2 \pi^2 \left( \sqrt{2}
\right)^3$. The relation ${\rm Vol}\ SO (4)
= \ft12 \left( {\rm Vol}\ SU (2) \right)^2$ is now indeed satisfied
\begin{equation}
{\rm Vol}\ SO (4) = 16 \pi^4
= \ft12 \left( {\rm Vol}\ SU (2) \right)^2
= \frac{1}{2}\left( 2 \pi^2 \left( \sqrt{2} \right)^3 \right)^2 \ .
\end{equation}

%%%%%%%%%%%%%%%%%%%%%%%%%%%%%%%%%%%%%%%%%%%%%%%%%%%%%%%%%%%%%%%%%%%%%%
%%%%%%%%%%%%%%%%%%%%%%%%%%%%%%%%%%%%%%%%%%%%%%%%%%%%%%%%%%%%%%%%%%%%%%

\end{document}